\documentclass [10pt,twocolumn]{IEEEtran}
\pdfoutput=1
\usepackage{algorithm,algorithmic,amsbsy,amsmath,amssymb,epsfig,bbm,bm,mathrsfs,multirow,amsthm}
\usepackage[T1]{fontenc}
\usepackage[latin9]{inputenc}
\usepackage{amsmath}
\usepackage{algorithm}
\usepackage{array,multirow,graphicx}
\usepackage{color}
\usepackage{makecell}
\usepackage[hidelinks]{hyperref}
\usepackage[flushleft]{threeparttable}

\begin{document}

\title{Ambient Backscatter Communications: \\ A Contemporary Survey}
\author{\IEEEauthorblockN{Nguyen Van Huynh\IEEEauthorrefmark{1}, Dinh Thai Hoang\IEEEauthorrefmark{1}, Xiao Lu\IEEEauthorrefmark{2}, Dusit Niyato\IEEEauthorrefmark{1}, Ping Wang\IEEEauthorrefmark{1}, and Dong In Kim\IEEEauthorrefmark{3}}
	
	\IEEEauthorblockA{\IEEEauthorrefmark{1}School of Computer Science and Engineering, Nanyang Technological University, Singapore}	\\
	\IEEEauthorblockA{\IEEEauthorrefmark{2}Department of Electrical and Computer Engineering, University of Alberta, Canada}	\\
	\IEEEauthorblockA{\IEEEauthorrefmark{3}School of Information and Communication Engineering, Sungkyunkwan University (SKKU), Korea}
}

\maketitle
\begin{abstract}
Recently, ambient backscatter communications has been introduced as a cutting-edge technology which enables smart devices to communicate by utilizing ambient radio frequency (RF) signals without requiring active RF transmission. This technology is especially effective in addressing communication and energy efficiency problems for low-power communications systems such as sensor networks. It is expected to realize numerous Internet-of-Things (IoT) applications. Therefore, this paper aims to provide a contemporary and comprehensive literature review on fundamentals, applications, challenges, and research efforts/progress of ambient backscatter communications. In particular, we first present fundamentals of backscatter communications and briefly review bistatic backscatter communications systems. Then, the general architecture, advantages, and solutions to address existing issues and limitations of ambient backscatter communications systems are discussed. Additionally, emerging applications of ambient backscatter communications are highlighted. Finally, we outline some open issues and future research directions.
\end{abstract}

\section{Introduction}
\label{sec:Intro}
\begin{table*}
	\footnotesize
	\centering
	\caption{\footnotesize LIST OF ABBREVIATIONS} \label{tab:abbreviation} 
	\begin{tabular}{|>{\raggedright\arraybackslash}m{1.5cm}|>{\raggedright\arraybackslash}m{6cm}|>{\raggedright\arraybackslash}m{1.5cm}|>{\raggedright\arraybackslash}m{6cm}|}
		\hline
		\multicolumn{1}{|>{\centering\arraybackslash}m{1.5cm}|}{\textbf{Abbreviation}} & \multicolumn{1}{>{\centering\arraybackslash}m{6cm}|}{\textbf{Description}} & \multicolumn{1}{>{\centering\arraybackslash}m{1.5cm}|}{\textbf{Abbreviation}} & \multicolumn{1}{>{\centering\arraybackslash}m{6cm}|}{\textbf{Description}} \\ \hline		\hline
		MBCSs & Monostatic backscatter communications systems & UHF& Ultra high frequency  \\ \hline
		SHF & Super high frequency  & UWB &Ultra-wideband \\ \hline
		NRZ & Non-return-to-zero  & OSTBC & Orthogonal space-time block code \\ \hline
		ASK & Amplitude shift keying & FSK & Frequency-shift keying \\ \hline
		PSK & Phase shift keying & BPSK & Binary phase shift keying \\ \hline
		QPSK & Quadrature phase shift keying & FDMA & Frequency division	multiple access\\ \hline
		QAM & Quadrature amplitude modulation & OOK& On-off-keying\\ \hline
		BBCSs & Bistatic backscatter communications systems & IoT & Internet of Things \\ \hline
		MCU & Micro-controller unit& CFO & Carrier frequency offset \\ \hline
		BER & Bit-error-rate & SNR & Signal-to-noise ratio \\ \hline
		TDM & Time-division multiplexing & CMOS & Complementary metal-oxide-semiconductor \\ \hline
		CDMA & Code division multiple access & ABCSs & Ambient backscatter communications systems \\ \hline
		AP & Access point & D2D & Device-to-device\\ \hline
		ML & Maximum-likelihood & OFDM & Orthogonal frequency division multiplexing \\ \hline
		WPCNs & Wireless powered communication networks & CRN & Cognitive radio network \\ \hline
	\end{tabular}
\end{table*}

Modulated backscatter technique was first introduced by Stockman in 1948~\cite{StockmanModulated} and quickly became the key technology for low-power wireless communication systems. In modulated backscatter communications systems, a backscatter transmitter modulates and reflects received RF signals to transmit data instead of generating RF signals by itself~\cite{Vann2008},~\cite{Blet2008Anti},~\cite{Kimionis2013Bistatic}. As a result, this technique has found many useful applications in practice such as radio-frequency identification (RFID), tracking devices, remote switches, medical telemetry, and low-cost sensor networks~\cite{Blet2009Anti},~\cite{Griffin2009Complete}. However, due to some limitations~\cite{RFID2006survey}--\cite{Klair2010Survey}, conventional backscatter communications cannot be widely implemented for data-intensive wireless communications systems~\cite{Zhang2012BLINK}. First, traditional backscatter communications require backscatter transmitters to be placed near their RF sources, and hence they may not be suitable for dense deployment scenarios. Second, in conventional backscatter communications, the backscatter receiver and the RF source are located in the same device, i.e., the reader, which can cause the interference between receive and transmit antennas, thereby reducing the communication performance. Moreover, conventional backscatter communications systems operate passively, i.e., backscatter transmitters only transmit data when inquired by backscatter receivers. Thus, they are only adopted by some limited applications. 

Recently, ambient backscatter~\cite{Liu2013Ambient} has been emerging as a promising technology for low-energy communication systems which can address effectively the aforementioned limitations in conventional backscatter communications systems. In ambient backscatter communications systems (ABCSs), backscatter devices can communicate with each other by utilizing surrounding signals broadcast from ambient RF sources, e.g., TV towels, FM towels, cellular base stations, and Wi-Fi access points (APs). In particular, in an ABCS, the backscatter transmitter can transmit data to the backscatter receiver by modulating and reflecting surrounding ambient signals. Hence, the communication in the ABCS does not require dedicated frequency spectrum which is scarce and expensive. Based on the received signals from the backscatter transmitter and the RF source or carrier emitter, the receiver then can decode and obtain useful information from the transmitter. By separating the carrier emitter and the backscatter receiver, RF components are minimized at backscatter devices and the devices can operate actively, i.e., backscatter transmitters can transmit data anytime without initiation from receivers. This capability allows the ABCSs to be adopted widely in many practical applications.

Although ambient backscatter communications has a great potential for future low-energy communication systems, especially Internet-of-Things (IoT), they are still facing many challenges. In particular, unlike conventional backscatter communications systems, the transmission efficiency of an ABCS much depends on the ambient source such as the type, e.g., TV signal or Wi-Fi signal, RF source location, and environment, e.g., indoor or outdoor. Therefore, ABCS has to be designed specifically for particular ambient sources. Furthermore, due to the dynamic of ambient signals, data transmission scheduling for backscatter devices to maximize the usability of ambient signals is another important protocol design issue. Additionally, using ambient signals from licensed sources, the communication protocols of ABCSs have to guarantee not to interfere the transmissions of the licensed users. Therefore, considerable research efforts have been reported to improve the ABCS in various aspects. This paper is the first to provide a comprehensive overview on the state-of-the-art research and technological developments on the architectures, protocols, and applications of emerging ABCSs. The key features and objectives of this paper are 
\begin{itemize}
	\item To provide a fundamental background for general readers to understand basic concepts, operation methods and mechanisms, and applications of ABCSs, 
	\item To summarize advanced design techniques related to architectures, hardware designs, network protocols, standards, and solutions of the ABCSs, and 
	\item To discuss challenges, open issues, and potential future research directions.
\end{itemize}

The rest of this paper is organized as follows. Section~\ref{sec:PrincipleModulatedBack} provides fundamental knowledge about modulated backscatter communications including operation mechanism, antenna design, channel coding, and modulation schemes. Section~\ref{sec:Bistatic} and~\ref{sec:AmbientBack} describe general architectures of bistatic backscatter communication systems (BBCSs) and ABCSs, respectively. We also review many research works in the literature aiming to address various existing problems in ABCSs, e.g., network design, scheduling, power management, and multiple-access. Additionally, some potential applications are also discussed in Section~\ref{sec:Bistatic} and~\ref{sec:AmbientBack}. Then, emerging backscatter communications systems are reviewed in Section~\ref{sec:Emerging}. Section~\ref{sec:ChallengesandFuture} discusses challenges and future directions of ABCSs. Finally, we summarize and conclude the paper in Section~\ref{sec:Conclusion}. The abbreviations used in this article are summarized in Table~\ref{tab:abbreviation}.

\section{Ambient Backscatter Communications: An Overview}
\label{sec:PrincipleModulatedBack}

In this section, we first provide an overview of backscatter communications systems and fundamentals of modulated backscatter communications. Then, key features in designing antennas for ABCSs are highlighted. Finally, typical modulation and channel coding techniques used in ABCSs are discussed.

\subsection{Backscatter Communications Systems}

\begin{figure*}[tbh]
	\centering
	\includegraphics[scale=0.33]{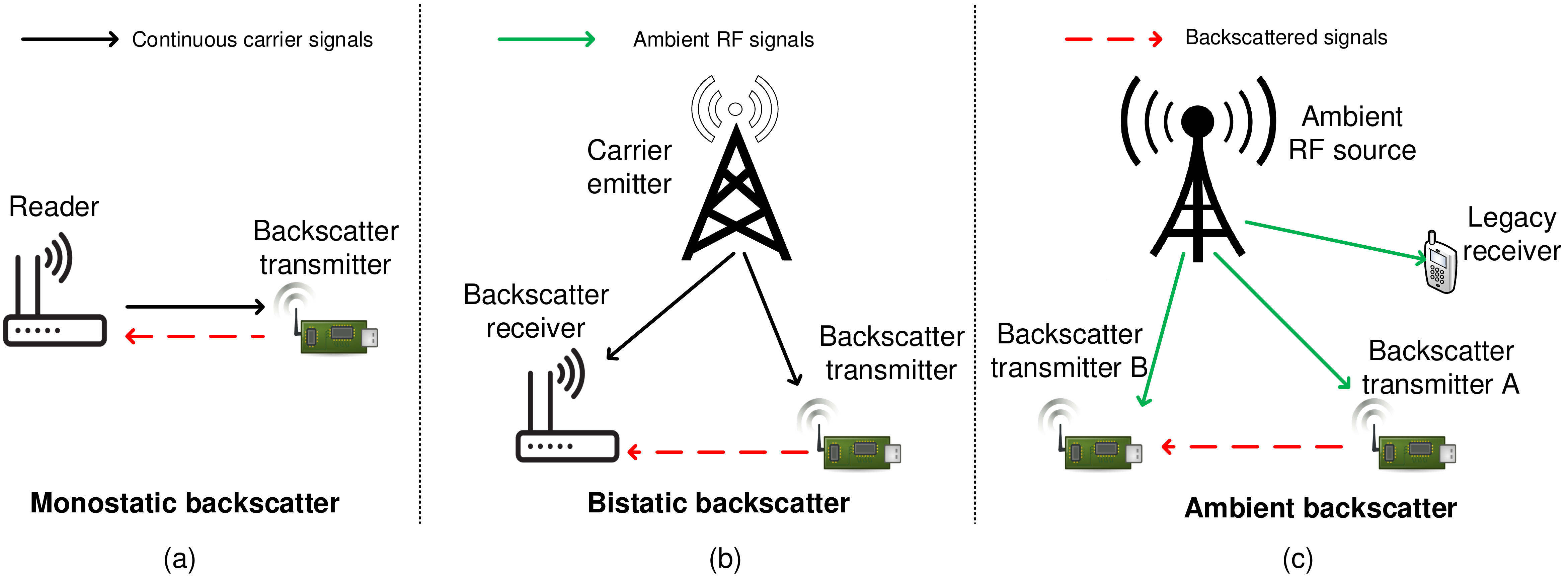}
	\caption{Paradigms for backscatter communications.}
	\label{fig:chap2_paradigms_backscatter}
\end{figure*}

Backscatter communications systems can be classified into three major types based on their architectures: monostatic backscatter communications systems (MBCSs), BBCSs, and ABCSs as shown in Fig.~\ref{fig:chap2_paradigms_backscatter}.

\subsubsection{Monostatic Backscatter Communications Systems}
In an MBCS, e.g., an RFID system, there are two main components: a backscatter transmitter, e.g., an RFID tag, and a reader as shown in Fig.~\ref{fig:chap2_paradigms_backscatter}(a). The reader consists of, in the same device, an RF source and a backscatter receiver. The RF source generates RF signals to activate the tag. Then, the backscatter transmitter modulates and reflects the RF signals sent from the RF source to transmit its data to the backscatter receiver. As the RF source and the backscatter receiver are placed on the same device, i.e., the tag reader, the modulated signals may suffer from a round-trip path loss~\cite{Kimionis2014Increased}. Moreover, MBCSs can be affected by the doubly near-far problem. In particular, due to signal loss from the RF source to the backscatter transmitter, and vice versa, if a backscatter transmitter is located far from the reader, it can experience a higher energy outage probability and a lower modulated backscatter signal strength~\cite{Choi2015Backscatter}. The MBCSs are mainly adopted for short-range RFID applications.

\subsubsection{Bistatic Backscatter Communications Systems}

Different from MBCSs, in a BBCS, the RF source, i.e., the carrier emitter, and the backscatter receiver are separated as shown in Fig.~\ref{fig:chap2_paradigms_backscatter}(b). As such, the BBCSs can avoid the round-trip path loss as in MBCSs. Additionally, the performance of the BBCS can be improved dramatically by placing carrier emitters at optimal locations. Specifically, one centralized backscatter receiver can be located in the field while multiple carrier emitters are well placed around backscatter transmitters. Consequently, the overall field coverage can be expanded. Moreover, the doubly near-far problem can be mitigated as backscatter transmitters can derive RF signals sent from nearby carrier emitters to harvest energy and backscatter data~\cite{Choi2015Backscatter}. Although carrier emitters are bulky and their deployment is costly, the manufacturing cost for carrier emitters and backscatter receivers of BBCSs is cheaper than that of MBCSs due to the simple design of the components~\cite{Lu2017Ambient}.

\subsubsection{Ambient Backscatter Communications Systems}
Similar to BBCSs, carrier emitters in ABCSs are also separated from backscatter receivers. Different from BBCSs, carrier emitters in ABCSs are available ambient RF sources, e.g., TV towers, cellular base stations, and Wi-Fi APs instead of using dedicated RF sources as in BBCSs. As a result, ABCSs have some advantages compared with BBCSs. First, because of using already-available RF sources, there is no need to deploy and maintain dedicated RF sources, thereby reducing the cost and power consumption for ABCSs. Second, by utilizing existing RF signals, there is no need to allocate new frequency spectrum for ABCSs, and the spectrum resource utilization can be improved. However, because of using ambient signals for backscatter communications, there are some disadvantages in ABCSs compared with BBCSs. First, ambient RF signals are unpredictable and dynamic, and thus the performance of an ABCS may not be as stable as that of the BBCS. Second, since ambient RF sources of ABCSs are not controllable, e.g., transmission power and locations, the design and deployment of an ABCS to achieve optimal performance is often more complicated than that of an BBCS.

\subsection{Fundamentals of Modulated Backscatter Communications}
\begin{figure*}[tbh]
	\centering
	\includegraphics[scale=0.27]{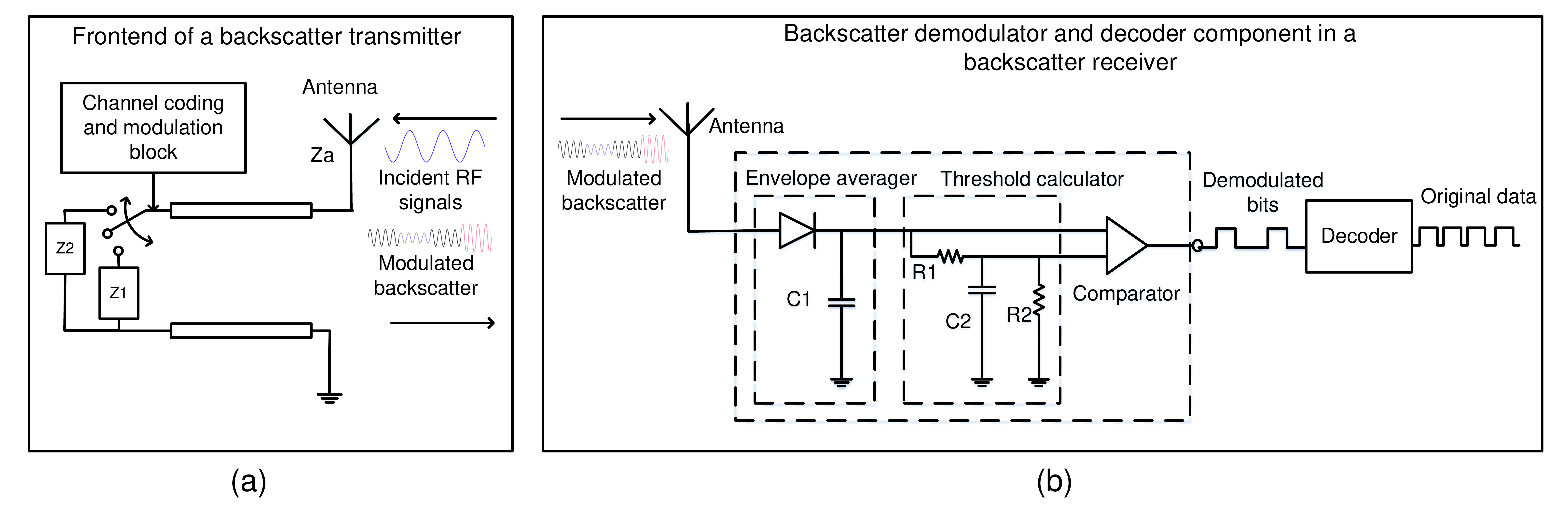}
	\caption{The main components of (a) a backscatter transmitter and (b) a backscatter receiver in a backscatter communications system~\cite{Lu2017Ambient}.}
	\label{fig:chap3_modulated_backscatter}
\end{figure*}

Despite differences in configurations, MBCSs, BBCSs, and ABCSs share the same fundamentals. In particular, instead of initiating their own RF transmissions as conventional wireless systems, a backscatter transmitter can send data to a backscatter receiver just by tuning its antenna impedance to reflect the received RF signals. Specifically, the backscatter transmitter maps its bit sequence to RF waveforms by adjusting the load impedance of the antenna. The reflection coefficient of the antenna is computed by~\cite{Griffin2009Complete},~\cite{Loo2008Chipimpedance},~\cite{Zhang2017Areview},~\cite{Nikitin2008Antenna}:
\begin{equation} \label{reflection-coefficient}
\Gamma_i=\frac{Z_i-Z_a^*}{Z_i+Z_a},
\end{equation}
where $Z_a$ is the antenna impedance, $*$ is the complex-conjugate operator, and $i = 1,2$ represents the switch state. In general, the number of states can be greater than 2, e.g., 4 or 8 states. However, in backscatter communications systems, the two-state modulation is typically used because of its simplicity. By switching between two loads $Z_1$ and $Z_2$ as shown in Fig.~\ref{fig:chap3_modulated_backscatter}(a), the reflection coefficient can be shifted between absorbing and reflecting states, respectively. In the absorbing state, i.e., impedance matching, RF signals are absorbed, and this state represents bit `0'. Conversely, in the reflecting state, i.e., impedance mismatching, the RF signals are reflected, and this state represents bit `1'. This scheme is known as the \textit{load modulation}.
There are two ways to decode the modulated signals sent from the backscatter transmitter: (i) using analog-to-digital converter (ADC)~\cite{ADC} and (ii) using an averaging mechanism.

The ADC has been commonly used in backscatter communications systems, especially for RFID systems. The procedures of using the ADC to decode modulated signals are as follows. The backscatter receiver samples the received signals at the Nyquist-information rate of the ambient signals, e.g., TV signals. The received samples, i.e., $y[n]$, at the backscatter receiver are expressed as follows:
\begin{equation} \label{receive_signals}
y[n] = x[n] + \alpha B[n]x[n] + w[n],
\end{equation}
where $x[n]$ are the samples of the TV signals received by the backscatter receiver, $w[n]$ is the noise, $\alpha$ is the complex attenuation of the backscattered signals relative to the TV signals, and $B[n]$ are the bits which are transmitted by the backscatter transmitter.
Then, the average powers of $N$ received samples are calculated by the backscatter receiver as follows:
\begin{equation} \label{receive_power}
\frac{1}{N}\sum_{i=1}^{N} {|y[n]|}^{2} = \frac{1}{N}\sum_{i=1}^{N} {|x[n]+\alpha Bx[n] + w[n]|}^2,
\end{equation}
where $B$ takes a value of `0' or `1' depending on the non-reflecting and reflecting states, respectively. As $x[n]$ is uncorrelated with the noise $w[n]$, (\ref{receive_power}) can be expressed as follows:
\begin{equation} \label{receive_data}
\frac{1}{N}\sum_{i=1}^{N} {|y[n]|}^{2} = \frac{{|1+\alpha B|}^2}{N} \sum_{i=1}^{N} {|x[n]|}^2 + \frac{1}{N} \sum_{i=1}^{N} {w[n]}^2.
\end{equation}
Denote $P$ as the average power of the received TV signals, i.e., $P=\frac{1}{N}\sum_{i=1}^{N}{|x[n]|}^2$. Ignoring the noise, the average power at the backscatter receiver is ${|1+\alpha|}^2 P$ and $P$ when the backscatter transmitter is at the reflecting ($B=1$) and non-reflecting ($B=0$) states, respectively. Based on the differences between the two power levels, i.e., ${|1+\alpha|}^2 P$ and $P$, the backscatter receiver can decode the data from the backscattered signals with a conventional digital receiver.

However, the ADC component consumes a significant amount of power, and thus may not be feasible to use in ultra-low-power systems. Therefore, the authors in~\cite{Liu2013Ambient} propose the averaging mechanism to decode the modulated signals without using ADCs and oscillators. The averaging mechanism requires only simple analog devices, i.e, an envelope average and a threshold calculator, at the backscatter receiver as shown in Fig.~\ref{fig:chap3_modulated_backscatter}(b). By averaging the received signals, the envelope circuit first smooths these signals. Then, the threshold calculator computes the threshold value which is the average of the two signal levels, and compares with the smoothed signals to detect bits `1' and `0'. After that, demodulated bits are passed through a decoder to derive the original data. In backscatter communications systems, the backscatter transmitter and backscatter receiver do not require complex components such as oscillator, amplifier, filter, and mixer, which consume a considerable amount of energy. Thus, the backscatter communications systems have low-power consumption, low implementation cost, and thus are easy to implement and deploy.

\subsection{Antenna Design}
In a backscatter communications system, an antenna is an essential component used to receive and backscatter signals. Thus, the design of the antenna can significantly affect the performance of the backscatter communications system. The maximum practical distance between the backscatter transmitter and the RF source\footnote{Depending on the type of the backscatter communications system, the RF source can be a reader in RFID systems, a carrier emitter in BBCSs, or an ambient RF source in ABCSs.}, of the system can be calculated by the Friis equation~\cite{Rao2015Antenna} as follows:
\begin{equation} \label{Friis}
r=\frac{\lambda}{4\pi}\sqrt{\frac{P_tG_t(\theta, \varphi)G_r(\theta, \varphi)p\tau}{P_{th}}},
\end{equation}
where $\lambda$ is the wavelength, $P_t$ is the power transmitted by the RF source, $G_t(\theta, \varphi)$ and $G_r(\theta, \varphi)$ are the gain of the transmit antenna and the gain of the receive antenna on the angles $(\theta, \varphi)$, respectively. $P_{th}$ is the minimum threshold power that is necessary to provide sufficient power to the backscatter transmitter chip attached to the antenna of the backscatter receiver, $p$ is the polarization efficiency, and $\tau$ is the power transmission coefficient based on antenna impedance and chip impedance of the backscatter transmitter. Accordingly, it is important to adjust and set these parameters to achieve optimal performance of the backscatter communications system.

\subsubsection{Operating Frequency} In a backscatter communications system, the operating frequency of an antenna varies in a wide range depending on many factors such as local regulations, target applications, and the transmission protocols~\cite{Loo2008Chipimpedance},~\cite{RFID_fre_regulations}. For example, RFID systems operate at the frequency ranging from the low frequency band, i.e., 125 kHz - 134.2 kHz, and the high frequency band, i.e., 13.56 MHz, to the ultra high frequency (UHF) band, i.e., 860 MHz - 960 MHz, and the super high frequency (SHF) band, i.e., 2.4 GHz - 2.5 GHz and 5.725 GHz - 5.875 GHz~\cite{RFID_fre_regulations},~\cite{Lehpamer2012RFID}. Most of the recent RFID systems adopt EPC Global Class 1 Gen 2 and ISO 18000-6c as standard regulations for designs in UHF. However, the deployed frequency is dissimilar in different regions, e.g., 866.5 MHz in Europe, 915 MHz in North America, and 953 MHz in Asia~\cite{Loo2008Chipimpedance},~\cite{Lehpamer2012RFID}.

It is important to note that increasing the operating frequency results in a higher power consumption and a more complicated design for active RF circuits~\cite{Shirane2015RFpowered}. However, in a backscatter communications system, the backscatter transmitter antenna does not contain active RF circuits, and thus the power consumption may negligibly increase for higher frequencies. Therefore, several works in the literature suggest that backscatter communications systems have some benefits when operating in the SHF band as follows:
\begin{itemize}
	\item By backscattering SHF signals, backscatter communications systems can be compatible with billions of existing Bluetooth and Wi-Fi devices~\cite{Ensworth2015Every}. Hence, it is highly potential to capitalize the ubiquitous characteristics of conventional wireless systems to support low-cost, low-power backscatter communications systems~\cite{Ensworth2015BLE},~\cite{Kim2003Measurments},~\cite{Kellogg2014WiFiIn}.
	\item As the operating frequency of the backscatter transmitters increases, the half-wave dipole, i.e., a half of wavelength, is reduced, e.g., 16 cm at 915 MHz, 6 cm at 2.45 GHz, and 2.5 cm at 5.79 GHz~\cite{Griffin2009Complete}. Hence, the size of the antenna can be greatly shrunk at the SHF band~\cite{Shirane2015RFpowered}. Thus, this increases the antenna gain and object immunity~\cite{Griffin2009HighDiss}. As the antenna is smaller, the backscatter transmitter size becomes smaller, thereby reducing the backscatter receiver's size and making it possible to be embedded on mobile and hand-held readers~\cite{Griffin2009HighDiss}.
	\item As the SHF band has more available bandwidth than that of the UHF band, backscatter communications systems are able to operate on the spread spectrum to increase the data rate~\cite{Griffin2009HighDiss}.
\end{itemize}

Recently, ultra-wideband (UWB) backscatter technology has been introduced~\cite{Dardari2010Ultrawide},~\cite{Guidi2010Backscatter}. The UWB system can operate with instantaneous spectral occupancy of 500 MHz or a fractional bandwidth of more than 20\%~\cite{UWB}. The key idea of the UWB system is that the UWB signals are generated by driving the antenna with very short electrical pulses, i.e., one nanosecond or less. As such, the bandwidth of transmitted signals can increase up to one or more GHz. Hence, the UWB avoids the multi-path fading effect, thereby increasing the robustness and reliability of backscatter communications systems. Furthermore, as the UWB system operates at baseband, it is free of sine-wave carriers and does not require intermediate frequency processing. This can reduce the hardware complexity and power consumption.
\subsubsection{Impedance Matching}
The impedance matching (mismatching) between the chip impedance, i.e., the load impedance, and the antenna impedance is required to ensure that most of the RF signals are absorbed (reflected) in the absorbing (reflecting) state. Thus, finding suitable values of the antenna impedance and the chip impedance is critical in the antenna design.

The complex chip impedance and antenna impedance are expressed as follows~\cite{Rao2015Antenna},~\cite{Yeoman2014Impedance}:
\begin{equation}
\begin{split}
Z_c= R_c+jX_c,\\
Z_a= R_a+jX_a,
\end{split}
\end{equation}
where $R_c$ and $R_a$ are the chip and antenna resistances, respectively, and $X_c$ and $X_a$ are the chip and antenna reactances, respectively. The chip impedance $Z_c$ is hard to change due to technological limitations~\cite{Grosinger2012Diss}. This stems from the fact that $Z_c$ is a function of the operating frequency and the power received by the chip $P_c$~\cite{Loo2008Chipimpedance}. As a result, changing the antenna impedance is more convenient in performing the impedance matching. $P_c$ can be represented by the power received at the antenna $P_a$ and the power transmission coefficient $\tau$ as $P_c=P_a\tau$. Here, $\tau$ is expressed as follows~\cite{Rao2015Antenna},~\cite{Yeoman2014Impedance}:
\begin{equation}\label{perfect_impedance}
\tau=\frac{4R_cR_a}{|Z_c+Z_a|^2}.
\end{equation}
The closer $\tau$ to 1, the better the impedance matching between the backscatter transmitter chip and antenna. The impedance matching will be perfect when $\tau=1$. Thus, based on~(\ref{perfect_impedance}), the antenna impedance can be easily determined to achieve the perfect impedance matching, i.e., $\tau=1$ when $Z_a=Z_c^*$.

\subsubsection{Antenna Gain}
Antenna gain is the amount of power transmitted in the direction of peak radiation to an isotropic source~\cite{AntennaDesign}. In general, the higher antenna gain leads to the longer range of transmission. Thus, it is important to calculate the antenna gain based on the target communication distance when designing the antenna~\cite{AntennaGain}. However, as the price of a high-gain antenna is more expensive and its size is larger than that of a low-gain antenna, the high-gain antenna is not always a feasible and economical choice for implementation. In particular, for the scenario in which the backscatter transmitters are not far away from the backscatter receiver, or information about the direction of incoming signals is not available, low-gain antennas are more preferred~\cite{AntennaGain},~\cite{6Factors}. Another important factor in designing the antenna is the \textit{on-object gain penalty}, i.e., the gain penalty loss. This loss represents the reduction of antenna gain due to the material attachment~\cite{Lehpamer2012RFID},~\cite{Griffin2009HighDiss}. The on-object gain penalty depends on different factors such as material properties, object geometry, frequency, and antenna types. Hence, it is difficult to directly calculate the on-object gain penalty. Currently, a common and effective method to determine the on-object gain penalty is through simulations and measurements~\cite{Griffin2009HighDiss}.

\subsubsection{Polarization}
Polarization, also known as \textit{orientation}, is the curve traced by an end point of the vector to represent the instantaneous electric field~\cite{AntennaTheory}. In other words, it describes how the direction and magnitude of the field vector change over time. According to the shape of the trace, the polarization is classified into linear, circular, and elliptical groups. The power received at the antenna is maximized when the polarization of the incident wave is matched to that of the antenna. Thus, orientations of the backscatter receiver and the backscatter transmitter can significantly affect the received power and the range of the transmission. For example, when the antennas of backscatter receiver and backscatter transmitter are placed parallelly, the received power at the antennas is maximized. Otherwise, if the backscatter transmitter's antenna is displaced by $90^\circ$, i.e., complete polarization mismatch, it is unable to communicate with the backscatter receiver. This is known as the polarization mismatch problem~\cite{Lehpamer2012RFID}.

The polarization mismatch problem is an important issue, which needs to be carefully considered when designing the antenna, as an orientation of the backscatter transmitter is usually arbitrary~\cite{Griffin2009HighDiss}. Several works aim to solve this problem. One of the effective solutions is transmitting a circularly polarized wave from the reader in the monostatic system~\cite{Nikitin2008Antenna},~\cite{Wu2013circularpolarization},~\cite{Deavours2009circularpolarization},~\cite{Nikitin2006Reply}. In this way, the uplink polarization mismatch and downlink polarization mismatch are both equal to 3dB~\cite{Nikitin2006Reply}. Thereby the backscatter transmitter is able to communicate with the backscatter receiver regardless of their orientation. In~\cite{Griffin2009Complete}, the authors implement two linearly-polarized antennas, which are oriented at $45^\circ$ with respect to each other, on the backscatter transmitter. By doing this, the complete polarization mismatch problem can be largely avoided.

\subsection{Channel Coding and Decoding}
Channel coding, i.e., \textit{coding in the baseband}, is a process that matches a message and its signal representation to the characteristics of the transmission channel. The main purpose of the coding process is to ensure reliable transmissions by protecting the message from interference, collision, and intentional modification of certain signal characteristics~\cite{RFIDHandbook}. At the backscatter receiver, the encoded baseband signals are decoded to recover the original message and detect any transmission errors. 

In backscatter communications systems, many conventional coding techniques can be adopted such as non-return-to-zero (NRZ), Manchester, Miller, and FM0~\cite{RFIDHandbook},~\cite{RFIDFundamentals}.

\begin{itemize}
	\item \textit{NRZ code:} Bit `1' is represented by \textit{high signals} and bit `0' is represented by \textit{low signals}. 
	\item \textit{Manchester code:} Bit `1' is represented by a negative transition, i.e., from a high level to a low level, in the middle of the bit period. Bit `0' is represented by a positive transition, i.e., a low level to a high level, at the start of the clock. 
	\item \textit{Miller code:} Bit `1' is represented by a transition of either high to low levels or low to high levels in the half-bit period, while bit `0' is represented by the continuance of the bit `1' level over the next bit period~\cite{RFIDHandbook}.
	\item \textit{FM0 code:} The phases of the baseband signals are all inverted at the beginning of each symbol. Bit `0' has a transition in the middle of the clock. In contrast, bit `1' has no transition during the symbol period~\cite{Griffin2009Fundamentals},~\cite{Lalitha2014CodingReview}.
\end{itemize}
NRZ and Manchester are the two simple channel coding techniques and widely adopted in backscatter communications systems, especially in RFID systems~\cite{RFIDHandbook},~\cite{RFIDFundamentals}. However, the NRZ code has a limitation when the transmitted data has a long string of bits `1' or `0' and the Manchester code requires more bits to be transmitted than that in the original signals. Thus, existing backscatter communications systems, i.e., UHF Class 1 Gen 2 RFID, BBCSs, and ABCSs, usually adopt the Miller and FM0 channel coding techniques due to their advantages such as enhanced signal reliability, reduced noise, and simplicity~\cite{Liu2013Ambient},~\cite{Lalitha2014CodingReview},~\cite{Kim2016Selfpowered},~\cite{Kim2017Implementation}.

Nonetheless, as backscatter communications systems are emerging rapidly in terms of the application, technology, and scale, the conventional channel coding techniques may not meet the emerging requirements such as high data rates, long communication range, and robustness. Hence, several novel coding techniques are proposed. In~\cite{Boyer2013Spacetime}, the authors introduce an orthogonal space-time block code (OSTBC) to improve the data rate and reliability of RFID systems. The key idea of the OSTBC is to transmit data through multiple orthogonal antennas, i.e., multiple-input multiple-output technology. In particular, this channel coding scheme transmits several symbols simultaneously which are spread into block codes over space and time. As such, the OSTBC achieves a maximum diversity order with linear decoding complexity, thereby improving the performance of the system. In~\cite{Durgin2017Improved}, the authors highlight that the FM0 coding used in ISO 18000-6C standard for UHF RFID tags is simple, but may not achieve maximum throughput. The authors then propose a \emph{balanced block code} to increase the throughput while maintaining the simplicity of the system. To do so, the balanced block code calculates the frequency spectrum for each of the resulting balanced codewords\footnote{Each codeword contains the same number of bits `1' and `0'. Specific codewords for different input bit sequences are described in~\cite{Durgin2017Improved} and~\cite{Durgin2017Abetter}.}. Then, the codewords with the deepest spectral nulls at direct current are selected and assigned to a
Grey-coded ordered set of the input bits. If the Hamming distance between the codeword and its non-adjacent neighbor is lower than that between the codeword and its adjacent neighbor, the current codeword and its adjacent neighbor are swapped. As a result, the current codeword is placed next to its non-adjacent neighbor. This procedure achieves a local optimum that minimizes the bit errors. The experimental results demonstrate that the balanced block code increases the throughput by 50\% compared to the conventional channel coding techniques, e.g., FM0.

In BBCSs, to deal with the interleaving of backscatter channels, an efficient encoding technique, namely \emph{short block-length cyclic channel code}, is developed~\cite{Nikos2015Coherent}. In particular, based on the principle of the cyclic code~\cite{cycliccode}, this technique encodes data by associating the code with polynomials. Thus, this short block-length cyclic channel code can be performed efficiently by using a simple shift register. The experimental results demonstrate that the proposed encoding technique can support communication ranges up to 150 meters. In~\cite{Park2014Turbocharging}, the authors introduce $\mu code$, a low-power encoding technique, to increase the communication range and ensure concurrent transmissions for ABCSs. Instead of using a pseudorandom chip sequence, $\mu code$ uses a periodic signal to represent the information. In this way, the transmitted signals can be detected at the backscatter receiver without any phase synchronization when the receiver knows the frequency of the sinusoidal signals. The authors also note that the backscatter transmitter cannot transmit sine waves as it supports only two states, i.e., absorbing and reflecting states. Hence, a periodic alternating sequence of bits ``0'' and ``1'' is adopted. With no need for the synchronization, $\mu code$ reduces the energy consumption as well as the complexity of the backscatter receiver. Through the experiments, the authors demonstrate that $\mu code$ enables long communication ranges, i.e., 40 times more than that of conventional backscatter communications systems, and also support multiple concurrent transmissions.

\subsection{Modulation and Demodulation}
Modulation is a process of varying one or more properties, i.e., frequency, amplitude, and phase, of carrier signals. At a backscatter receiver, by analyzing the characteristics of the received signals, we can reconstruct the original data by measuring the changes in reception phase, amplitude, or frequency, i.e., demodulation. Table~\ref{tab_sec3:summarymodulation} summaries the principle, advantages, disadvantages along with references of popular modulation schemes in backscatter communications systems. 

\begin{table*}
	\footnotesize
	\centering
	\caption{\footnotesize SUMMARY OF MODULATION SCHEMES} \label{tab_sec3:summarymodulation} 
	\begin{tabular}{|>{\raggedright\arraybackslash}m{1.5cm}|>{\raggedright\arraybackslash}m{4cm}|>{\raggedright\arraybackslash}m{2.5cm}|>{\raggedright\arraybackslash}m{2.0cm}|>{\raggedright\arraybackslash}m{1.5cm}|>{\raggedright\arraybackslash}m{1.5cm}|>{\raggedright\arraybackslash}m{1.5cm}|}
		\hline
		\multicolumn{1}{|>{\centering\arraybackslash}m{1.5cm}|}{\multirow{2}{*}{\textbf{Modulation}}} & \multicolumn{1}{>{\centering\arraybackslash}m{4cm}|}{\multirow{2}{*}{\textbf{Principle}}} & \multicolumn{1}{>{\centering\arraybackslash}m{2.5cm}|}{\multirow{2}{*}{\textbf{Advantages}}} & \multicolumn{1}{>{\centering\arraybackslash}m{2.0cm}|}{\multirow{2}{*}{\textbf{Disadvantages}}} & \multicolumn{3}{c|}{\textbf{References}} \\ 
		\cline{5-7}
		& & & & RFID & BBCSs & ABCSs \\
		\hline		
		\hline
		ASK & Represent the binary data in the form of variations in the amplitude levels, i.e., high and low voltage, of RF carrier signals & Provide continuous power to backscatter transmitters and enables relatively simple backscatter receiver design~\cite{RFIDHandbook}  & Very sensitive to noise and interference~\cite{ASK_advantges_disadvantages} &\cite{Molina2013BAT},~\cite{Kuester2012Baseband} & \cite{Kimionis2012Bistatic} &\makecell[l]{\cite{Zhang2016HitchHike},~\cite{EkhoNet2014Zhang}\\} \\ \hline

		FSK & The frequency $F_c$ of the carrier signals is switched between two frequencies $f_1$ and $f_2$ according to the digital signal changes, i.e., bits `1' and bits `0', respectively & Resilient to the noise and signal strength variations 	& Require more spectrum &  \cite{Molina2013BAT}&\makecell[l]{\cite{Kimionis2013Bistatic},~\cite{Kimionis2014Increased},\\\cite{Choi2015Backscatter},~\cite{Ensworth2015Every},\\\cite{Nikos2015Coherent},~\cite{Tountas2015Bistatic},\\\cite{Kampi2013Backscatter}--\cite{Alevizos2017Scatter}}&\cite{Liu2017Backscatter},~\cite{Wang2017FMBackscatter} \\ \hline
		
		PSK & The phase of carrier signals varies to represent bits `1' and `0'. Based on the number of phases, there are several forms of PSK such as binary PSK (BPSK), quadrature PSK (QPSK), and 16-PSK. & Allow backscatter transmitters to backscatter data in a smaller number of radio frequency cycles resulting in a higher data transmission rate & The recovery process is more complicated than other schemes. & \cite{Kuester2012Baseband},~\cite{Griffin2008Gains} & 
		&\makecell[l]{\cite{Liu2017Backscatter},~\cite{Darsena2017Modeling},\\\cite{Iyer2016Inter}--\cite{Kim2017Optimum}}\\ \hline
		
		QAM & Convey two analog message signals, i.e., two digital bit streams, by changing the amplitudes of two carrier waves, i.e., two-dimensional signaling  Support several forms of QAM such as 2-QAM, 4-QAM, 8-QAM, and 32-QAM & Increase the efficiency of transmissions  & Susceptible to noise, require power-hungry linear amplifiers~\cite{Wang2012Efficient}  & \cite{Boyer2012codedQAM} & \cite{Correia2016Design},~\cite{Lee2017Determination}&\cite{Kim2017Optimum} \\ \hline
	\end{tabular}
\end{table*}

In general, there are three basic modulation schemes corresponding to the changes of the amplitude, frequency, and phase in the carrier signals, i.e., amplitude shift keying (ASK), frequency shift keying (FSK), and phase shift keying (PSK). These modulation schemes are commonly adopted in backscatter communications systems~\cite{RFIDHandbook},~\cite{Kimionis2012Bistatic},~\cite{Zhang2016HitchHike},~\cite{Kampi2013Backscatter},~\cite{Liu2017Backscatter},~\cite{Iyer2016Inter}. In BBCSs, FSK is more favorable. In particular, as several backscatter transmitters in BBCSs may communicate with the backscatter receiver simultaneously, there is a need for a multiple access mechanism. Hence, several works choose FSK and frequency-division multiple access (FDMA) for BBCSs since the characteristics of FSK are perfectly fit with FDMA~\cite{Choi2015Backscatter},~\cite{Ensworth2015Every},~\cite{Kampi2013Backscatter},~\cite{Vannucci2007Implementing}. Furthermore, FSK is resilient to noise and signal strength variations~\cite{FSK_advantages_disadvantages}. On the contrary, PSK is mainly adopted in ambient backscatter systems~\cite{Liu2017Backscatter},~\cite{Iyer2016Inter},~\cite{Bharadia2015BackFi}. Specifically, PSK can support high data rate transmissions since it transmits data in a small number of radio frequency cycles. In~\cite{Darsena2016Performance}, the authors compare the performance between PSK and ASK in different angle $\phi$~\footnote{In~\cite{Darsena2016Performance}, the authors use a model including one backscatter node, i.e., $BS$, and two legendary nodes, i.e., $L1$ and $L2$. The backscatter node can transmit data to $L1$ and $L2$ by using RF signals from $L1$. $\phi$ is the angle between $BS-L1$ and $L1-L2$ paths.}. The numerical results show that the quadrature PSK (QPSK) modulation with $\phi = \pi/18$ achieves the highest capacity while the 4-ASK modulation with $\phi = \pi/3$ offers the lowest capacity. In~\cite{Shen2016Phase}, a multi-phase backscatter technique is proposed for ASK and PSK to reduce the phase cancellation problem. The phase cancellation problem occurs when there is a relative phase difference between the carrier and backscattered signals received at the backscatter receiver~\cite{Shen2016Phase}. Through the simulation and experimental results, the authors note that the performance of PSK is better than that of ASK. The reason is that the phase cancellation can be theoretically avoided completely if the difference between the phases of the two pair of impedances takes a value between $0$ and $\frac{\pi}{2}$. This can be easily achieved by using the PSK modulation scheme.

Some other modulation schemes are also adopted in backscatter communications systems. In~\cite{Shirane2015RFpowered}, by using the n-quadrature amplitude modulation (QAM) scheme, i.e., 32-QAM, a passive RF-powered backscatter transmitter operating at 5.8 GHz can achieve 2.5 Mbps data rate at a distance of ten centimeters. Nevertheless, the n-QAM modulation is susceptible to noise, thereby resulting in the normalized power loss~\cite{Bharadia2015BackFi}. In~\cite{Boyer2012codedQAM}, the authors measure the normalized power loss by analyzing the use of higher dimensional modulation schemes, e.g., 4-QAM or 8-QAM. The numerical results show that the normalized power loss is significantly increased from 2-QAM to 4-QAM. Therefore, the authors propose a novel QAM modulation scheme to combine QAM with \emph{unequal error protection} to minimize the normalized power loss. Unequal error protection protects bits at different levels. In particular, bits that are more susceptible to errors will have more protection, and vice versa. Through the numerical results, the authors demonstrate that the normalized power loss is greatly reduced by using the proposed QAM modulation scheme. In~\cite{Vannucci2007Implementing}, the authors introduce minimum-shift keying (MSK), i.e., a special case of FSK, to minimize interference at the backscatter receiver. The principle of MSK is that signals from the backscatter transmitter will be modulated at different sub-carrier frequencies. Through the experimental results, the authors demonstrate that the MSK modulation scheme can significantly minimize the collision at the backscatter receiver.

At the backscatter receiver, there is a need to detect modulated signals from the backscatter transmitter. Many detection mechanisms have been proposed in the literature. Among them, the noncoherent detection is most commonly adopted because of its simplicity and effectiveness~\cite{Kimionis2013Bistatic},~\cite{Darsena2017Modeling},~\cite{Qian2016Noncoherent},~\cite{Boyer2014Backscatter}. In particular, the noncoherent detection does not need to estimate the carrier phase, thereby reducing the complexity of the backscatter receiver circuit. This detection mechanism is suitable for the ASK and FSK modulation schemes. However, the noncoherent detection offers only a low bitrate~\cite{Noncoherent}. Therefore, some works adopt the coherent detection to increase the bitrate~\cite{Tountas2015Bistatic},~\cite{Nikos2015Trans}. Different from the noncoherent detection, the coherent detection requires knowledge about the carrier phase resulting in a more complicated backscatter receiver circuit. The PSK modulation usually prefers the coherent detection since 
its phases are varied to modulate signals. It is also important to note that in ambient backscatter communications systems, as the ambient RF signals are indeterminate or even unknown, many existing works assume that the ambient RF signals follow zero-mean circularly symmetric complex Gaussian distributions. Then, maximum-likelihood (ML) detectors~\cite{Proakis2007DigitalcommunicationBook} can be adopted to detect the modulated signals at the backscatter receiver~\cite{Wang2015Uplink},~\cite{Qian2017SemiCoherent},~\cite{Lu2015Signal}.

\subsection{Backscatter Communications Channels}
In the following, we describe general models of backscatter communication channels. Then, theoretical analyses and experimental measurements for the backscatter channels are discussed.

\subsubsection{Backscatter Communications Channels}
\label{subsec:PopularModel}

\begin{figure*}[tbh]
	\centering
	\includegraphics[scale=0.3]{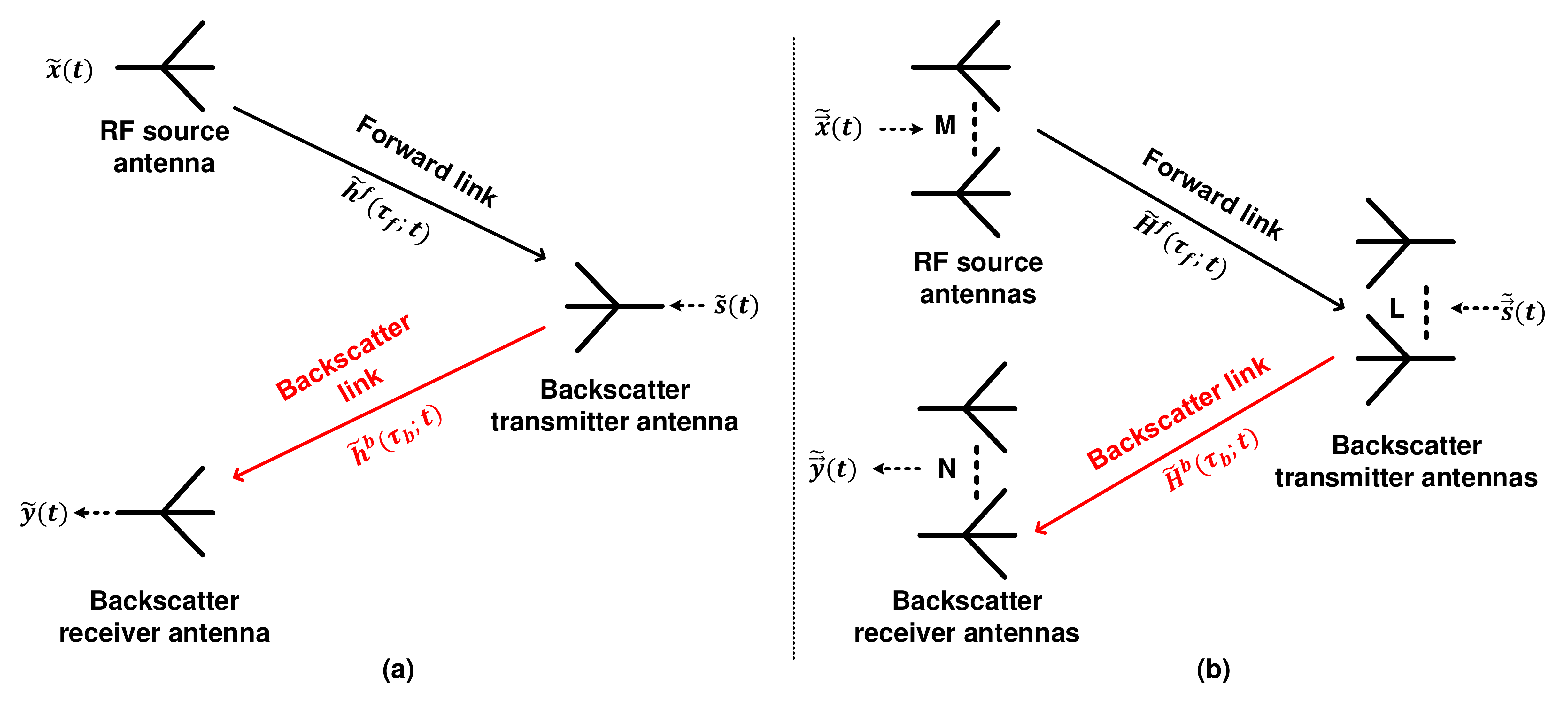}
	\caption{(a) Basic backscatter channel and (b) Dyadic backscatter channel~\cite{Griffin2009HighDiss},~\cite{Griffin2009Fundamentals},~\cite{Griffin2008Gains}.}
	\label{fig:chap4_backscatter_channel}
\end{figure*}

\paragraph{Basic Backscatter Channel}
A general system model of a backscatter communications system consists of three main components: (i) an RF source, (ii) a backscatter receiver, and (iii) a backscatter transmitter as shown in Fig.~\ref{fig:chap4_backscatter_channel}(a). Note that the RF source and the backscatter receiver can be in the same device, i.e., a reader, in the monostatic systems, or in different devices in BBCSs and ABCSs.

To transmit signals to the backscatter receiver, the backscatter transmitter modulates the carrier signals, which are transmitted from the RF source through the forward link. Then, the modulated signals is transmitted to the backscatter receiver through the backscatter link. The modulated signals received at the backscatter receiver are expressed as follows~\cite{Griffin2009Fundamentals}:
\begin{equation}
\begin{split}
\widetilde{y}(t) = \frac{1}{2}\int_{-\infty}^{+\infty}\int_{-\infty}^{+\infty}\widetilde{h}^b(\tau_b;t)\widetilde{s}(t)\widetilde{h}^f(\tau_f;t)\\
\times \widetilde{x}(t-\tau_b-\tau_f)d\tau_bd\tau_f+\widetilde{n}(t),
\end{split}
\end{equation}
where $\widetilde{h}^b(\tau_b;t)$ is the baseband channel impulse of the backscatter link, i.e., the link between the backscatter transmitter and the backscatter receiver, $\widetilde{h}^f(\tau_f;t)$ is the baseband channel impulse of the forward link, i.e., the link between the RF source and the backscatter transmitter, $\widetilde{s}(t)$ is the information signals transmitted from the backscatter transmitter, $\widetilde{x}(t)$ is the carrier signals transmitted from the RF source, and $\widetilde{n}(t)$ is the noise.

\paragraph{Dyadic Backscatter Channel}
Recently, a dyadic backscatter channel model is derived to characterize the multiple antenna backscatter channels~\cite{Griffin2009Complete},~\cite{Griffin2009HighDiss},~\cite{Griffin2008Gains}. As shown in Fig.~\ref{fig:chap4_backscatter_channel}(b), multiple antennas are employed, i.e., $M$ antennas at the RF source, $L$ antennas at the backscatter transmitter, and $N$ antennas at the backscatter receiver. Hence, the dyadic backscatter channel is also known as the $M \times L \times N$ backscatter channel. Similar to the basic backscatter channel, the received signals at the backscatter receiver are expressed as follows~\cite{Griffin2009Complete},~\cite{Griffin2009HighDiss},~\cite{Griffin2008Gains}:
\begin{equation}
\begin{split}
\widetilde{\vec{y}}(t) = \frac{1}{2}\int_{-\infty}^{+\infty}\int_{-\infty}^{+\infty}\widetilde{H}^b(\tau_b;t)\widetilde{S}(t)\widetilde{H}^f(\tau_f;t)\\
\times \widetilde{\vec{x}}(t-\tau_b-\tau_f)d\tau_bd\tau_f+\widetilde{\vec{n}}(t),
\end{split}
\end{equation}
where $\widetilde{\vec{y}}(t)$ is an $N\times1$ vector of received complex baseband signals, $\widetilde{H}^b(\tau_b;t)$ is the $N \times L$ complex baseband channel impulse response matrix of the backscatter link, and $\widetilde{H}^f(\tau_f;t)$ is the $L \times M$ complex baseband channel impulse response matrix of the forward link. $\widetilde{S}(t)$ is the backscatter transmitter's narrow band $L \times L$ signaling matrix, $\widetilde{\vec{x}}(t)$ is an $M \times 1$ vector of the signals transmitted from the RF source antennas, and $\widetilde{\vec{n}}(t)$ is an $N \times 1$ vector of noise components. The term \textit{dyadic} represents for the two-fold nature of a two-way channel and the matrix form of the modulated signals. In~\cite{Ingram2001Transmite}, this channel is investigated in the context of semi-passive backscatter transmitters to achieve diversity and spatial multiplexing. The authors demonstrate that by using multiple antennas at both the backscatter transmitter and backscatter receiver, the communication range is significantly extended. The reason is that in the $M \times L \times N$ backscatter channel, small-scale fading effects can be reduced~\cite{Ingram2001Transmite},~\cite{Griffin2007Reduced}, thereby improving the performance of backscatter communications systems~\cite{Griffin2007Link}.

\paragraph{Link Budgets for Backscatter Channels}
In a backscatter communications system, there are two major link budgets, i.e., the forward link and the backscatter link budgets, that affects performance of the system (Fig.~\ref{fig:chap4_backscatter_channel}). In particular, the forward link budget is defined as the amount of power received by the backscatter transmitter, and the backscatter link budget is the amount of power received by the backscatter receiver~\cite{Griffin2009HighDiss}. The forward link budget is calculated as follows:
\begin{equation}
P_t=\frac{P_TG_TG_t\lambda^2X\tau}{(4 \pi r_f)^2 \Theta BF_p},
\end{equation}
where $P_t$ is the power coupled into the backscatter transmitter, $P_T$ is the transmit power of the RF source, $G_T$ and $G_t$ are the antenna gains of the RF source and the backscatter transmitter, respectively. $\lambda$ is the frequency wavelength, $X$ is the polarization mismatch, $\tau$ is the power transmission coefficient, $r_f$ is the distance between the RF source and the backscatter transmitter, $\Theta$ is the backscatter transmitter's antennas on-object gain penalty, $B$ is the path blockage loss, and $F_p$ is the forward link fade margin.

The backscatter link budget is calculated as follows:
\begin{equation}
P_R= \frac{P_TG_RG_TG_t^2\lambda^4X_fX_bM}{(4 \pi)^4r_f^2r_b^2 \Theta^2B_fB_bF},
\end{equation}
where $M$ is the modulation factor, $r_b$ is the distance between the backscatter transmitter and the backscatter receiver, $X_f$ and $X_b$ are the forward link and backscatter link polarization mismatches, respectively. $B_f$ and $B_b$ are the forward link and backscatter link path blockage losses, respectively, and $F$ is the backscatter link fade margin. 

The link budgets will take different forms depending on the configurations, i.e., MBCSs, BBCSs, and ABCSs. However, the detail is beyond the scope of this survey. The more information can be found in~\cite{Griffin2009HighDiss} and~\cite{Griffin2009Fundamentals}.

\subsubsection{Theoretical Analyses and Experimental Measurements}
\begin{table*}
	\footnotesize
	\centering
	\caption{\footnotesize BER Performance of Backscatter Communications Systems} \label{tab_sec4:Experimental} 
	\begin{threeparttable}
	\begin{tabular}{|>{\raggedright\arraybackslash}m{1.5cm}|>{\raggedright\arraybackslash}m{2.0cm}|>{\raggedright\arraybackslash}m{6.0cm}|>{\raggedright\arraybackslash}m{1.5cm}|>{\raggedright\arraybackslash}m{1.5cm}|}
		\hline
		\multicolumn{1}{|>{\centering\arraybackslash}m{1.5cm}|}{\textbf{References}} &\multicolumn{1}{>{\centering\arraybackslash}m{2.0cm}|}{\textbf{Configurations}} & \multicolumn{1}{>{\centering\arraybackslash}m{6.0cm}|}{\textbf{Strategies}} &\multicolumn{1}{>{\centering\arraybackslash}m{1.5cm}|}{\textbf{Achieved SNR}} & \multicolumn{1}{>{\centering\arraybackslash}m{1.5cm}|}{\textbf{Achieved BER}}\\ 
		\hline		
		\hline
		\cite{Kimionis2013Bistatic}& Bistatic backscatter & Adopt the FSK modulation scheme & 9 dB& $\sim10^{-2}$ \\ \hline
		\cite{Kimionis2012Bistatic}&Bistatic backscatter & Adopt a ML detector & 10 dB& $\sim10^{-3}$ \\ \hline
		\cite{Qian2016Noncoherent}&Ambient backscatter & Adopt the 8-PSK modulation scheme at the backscatter transmitter and a noncoherent detector at the backscatter receiver& 20 dB& $\sim10^{-4}$ \\ \hline
		\cite{Lee2017Determination}& Ambient backscatter & Use 8 antennas at the backscatter transmitter & 50 dB& $\sim10^{-5}$ \\ \hline
		
		\cite{Kang2017Signaldetection}& Ambient backscatter & Use 2 antennas at the backscatter transmitter and adopt an energy detector\tnote{*} at the backscatter receiver. & 50 dB & $\sim10^{-4}$ \\ \hline
		\cite{Smietanka2012Modeling}& Monostatic backscatter & Use 2 antennas at the backscatter transmitter and adopt the Miller-8 encoding technique & 5.3 dB& $\sim10^{-4}$ \\ \hline
		\cite{He2010Gains}& Monostatic backscatter & Use 2 antennas at the backscatter transmitter and adopt the space-time block coding& 18 dB & $\sim10^{-4}$ \\ \hline
	
	\end{tabular}
	\begin{tablenotes}
	\item[*] This detector uses the difference in energy levels of the received signals to detect bit `1' and `0'.
	\end{tablenotes}
	\end{threeparttable}
\end{table*}
Based on the above models, many works focus on measuring and evaluating performance of backscatter channels. In~\cite{Griffin2009Complete}, by adopting different antenna materials, e.g., cardboard sheet, aluminum slab, or pine plywood, under the three configurations, the performance of the backscatter communications system is measured in terms of the link budgets. In particular, the authors demonstrate that reducing antenna impedance results in a small power transmission coefficient that may prevent the backscatter transmitter from turning on. It is also shown that the object attachment and multi-path fading may have significant effects on the performance of the system in terms of the communication range and bitrate between the backscatter transmitter and the backscatter receiver. The authors suggest that using multiple antennas operating at high frequencies provides many benefits such as increasing antenna gain and object immunity, and reducing small-scale fading to facilitate backscatter propagation. In~\cite{Kim2003Measurments} and~\cite{Kim2001Small}, the path loss and small-scale fading of backscatter communications systems are extensively investigated in an indoor environment. The authors demonstrate that the small-scale fading of the backscatter channel can be modeled as two uncorrelated traditional one-way fades, and the path loss of the backscatter channel is twice that of the one-way channel.

The multiple-antenna backscatter channels are investigated in references~\cite{Griffin2007Link},~\cite{Griffin2011Fading},~\cite{Griffin2010Multipath},~\cite{Griffin2009Multipath}. By using the cumulative distribution functions to determine the multi-path fading of backscatter channels, the authors in~\cite{Griffin2010Multipath} and~\cite{Griffin2009Multipath} demonstrate that multi-path fading on the modulated backscatter signals can be up to 20 dB and 40 dB with line-of-sight and non-line-of-sight backscatter channels, respectively. However, this multi-path fading can be significantly reduced by using multiple antennas at the backscatter transmitter to modulate data~\cite{Griffin2009Multipath}. Furthermore, in~\cite{Griffin2011Fading}, the authors suggest that the dyadic backscatter channel with two antennas at the backscatter transmitter can improve the reliability of the system and increase the communication range by 78\% with a bit-error rate (BER) of $10^{-4}$ compared with basic backscatter channels. Another link budget that needs to be considered is the link envelope correlation. In particular, the link envelope correlation may have negative effects on the performance of the system by coupling fading in the forward and backscatter links even if fading in each link is uncorrelated. In~\cite{Griffin2007Link}, the authors adopt probability density functions to analyze the link envelope correlation of the dyadic backscatter channel. The theoretical results show that using multiple antennas at the backscatter transmitter can reduce the link envelope correlation effect, especially for the system in which the RF source and the backscatter receiver are separated, i.e., BBCSs and ABCSs.

Different from all the aforementioned works, many works focus on measuring and analyzing the BER of backscatter communications~\cite{Kang2017Signaldetection}. Table~\ref{tab_sec4:Experimental} shows the summary of BER versus SNR in different system setups. Obviously, many factors can affect the BER performance such as antenna configurations, detectors, channel coding, and modulation schemes. In general, using multiple antennas at the backscatter transmitter to modulate data can significantly improve the BER performance. For example, in~\cite{Lee2017Determination}, by using 8 antennas at the backscatter transmitter, the BER of $10^{-5}$ can be achieved at 50 dB of SNR. However, this may increase the complexity of the backscatter transmitter. Thus, many works reduce BER at the backscatter receiver through novel channel coding and modulation as well as detection schemes. In~\cite{Qian2016Noncoherent}, by adopting the 8-PSK modulation and a noncoherent detector, the authors can reduce BER to $10^{-4}$ at 20 dB of SNR. Furthermore, by using the Miller-8 encoding technique with 2 antennas at the backscatter transmitter, the work in~\cite{Smietanka2012Modeling} achieves the BER of $10^{-4}$ at 5.3 dB of SNR.

In this section we have provided the principles of modulated backscatter with regard to the fundamentals, antenna design, channel coding and modulation schemes as well as backscatter channel models. In the following, we review various designs and techniques developed for BBCSs and ABCSs.

\section{Bistatic Backscatter Communications Systems}
\label{sec:Bistatic}
In this section, we first describe the general architecture of BBCSs. Then, we review existing approaches which aim to enhance the performance for the BBCSs. Table~\ref{table_sec5_sum} provides the summary of BBCSs.
\begin{table*}
	\caption{Summary of Bistatic Backscatter Communications Systems} 
	\label{table_sec5_sum}
	\begin{centering}
		\begin{tabular}{|>{\raggedright\arraybackslash}m{1.0cm}|>{\raggedright\arraybackslash}m{3.5cm}|>{\raggedright\arraybackslash}m{1.5cm}|>{\raggedright\arraybackslash}m{1.5cm}|>{\raggedright\arraybackslash}m{2cm}|>{\raggedright\arraybackslash}m{3.0cm}|>{\raggedright\arraybackslash}m{2.0cm}|}
			\hline 
			\multicolumn{1}{|>{\centering\arraybackslash}m{1.0cm}|}{\multirow{4}{*}{\textbf{Article}}} & \multicolumn{1}{>{\centering\arraybackslash}m{3.5cm}|}{\multirow{4}{*}{\textbf{Design Goals}}} & \multicolumn{2}{c|}{\textbf{Modeling}} & \multicolumn{1}{>{\centering\arraybackslash}m{2.0cm}|}{\multirow{4}{*}{\textbf{Key idea}}} & \multicolumn{1}{>{\centering\arraybackslash}m{3.0cm}|}{\multirow{4}{*}{\textbf{Experiment setup}}} & \multicolumn{1}{>{\centering\arraybackslash}m{2.0cm}|}{\multirow{4}{*}{\textbf{Results}}}\\
			\cline{3-4}
			& & Channel Coding \& Modulation & Multiple Access Protocol & & &\\
			\hline 
			\hline 
			\cite{Kimionis2014Increased} & Increase the communication range (theoretical analysis, simulation, and experimental) & OOK \& FSK & TDM \& FDM & Implement CFO compensation block and noncoherent detectors & 13 dBm of emitter power at 867 MHz and 1 kbps bitrate, $d_{et}$ = 2-4 m & The distance between the backscatter transmitter and the backscatter receiver ($d_{tr}$) is 130 meters \\
			\cline{1-7}
			\cite{Choi2015Backscatter} & Increase the communication range (simulation and theoretical analysis) & FSK & TDM \& FDM & Use a H-AP as additional RF source & 25 dBm and 13 dBm of transmit power at H-AP and carrier emitter at 868 MHz, respectively, the distance between the carrier emitter and the backscatter transmitter ($d_{et}$) is 1 m & $d_{tr}$ = 120 meters \\
			\cline{1-7}
			\cite{Kimionis2012Bistatic} & Increase the communication range (theoretical analysis, simulation, and experimental) & OOK & N.~A. & Design near-optimal detectors to improve the BER performance & 30 dBm of carrier emitter power and 1 kbps bitrate & $d_{tr}$=60 meters \\
			\cline{1-7}
			\cite{Alevizos2014Channel} & Increase the communication range (theoretical analysis, simulation, and experimental) & Reed-Mullers code \& FSK & N.~A. & Employ channel coding and interleaving technique & 13 dBm of emitter power at 867 MHz and 1 kbps bitrate, $d_{et}$ = 2.8 m & $d_{tr}$ = 134 meters \\
			\cline{1-7}
			\cite{Varshney2016Loera} & Design a low-power backscatter transmitter \& increase the communication range (experimental and prototype) & OOK \& FSK & N.~A. & Use 2.4 GHz ISM band and design a backscatter transmitter with low-power components & 26 dBm of emitter power at 2.4 GHz and 2.6 kbps bitrate, $d_{et}$ = 1 m & $d_{tr}$ = 225 meters and the backscatter transmitter consumes 7.2 mW of power \\
			\cline{1-7}
			\cite{Vougiou2016Could} & Increase the communication range (experimental and prototype) & FSK & N.~A. & Design a backscatter transmitter & 13 dBm of emitter power at 868 MHz and 1.2 kbps bitrate, $d_{et}$ = 3 m & $d_{tr}$ = 269 meters \\
			\cline{1-7}
			\cite{Nikos2015Trans} & Increase the communication range (theoretical analysis, simulation, and experimental) & Short block-length cyclic \& FSK & N.~A. & Use channel coding \& coherent detectors & 20 mW of emitter power at 868 MHz and 1 kbps bitrate, $d_{et}$ = 10 m & $d_{tr}$ = 150 meters\\
			\cline{1-7}
			\cite{Konstan2016Converting} & Allow the backscatter transmitter to harvest energy from both the carrier emitter and plants in the field (experimental and prototype) & Frequency modulation & N.~A. & Harvest biologic energy from a plant & Two backscatter sensor transmitters occupy the frequency range at 868.016-868.021 KHz and 868.023-868.030 KHz, respectively & Harvest not less than 10.6 $\mu$W of power from the plant and the carrier emitter \\
			\cline{1-7}
			\cite{Kampianakis2014Wireless} & Increase the communication range (theoretical analysis, experimental and prototype) & Analog frequency modulation & TDM \& FDM & Propose a data smoothing technique & Less than 1 mW of power at 868 MHz and 100 kbps bitrate & $d_{tr}$ = 100 meters \\
			\cline{1-7}
			\cite{Daskalakis2016Soil} & Increase the communication range (theoretical analysis, experimental, and prototype) & Modulation pulses with 50\% of duty-cycle & FDM & Propose a modulation pulse with 50\% of duty-cycle & 13 dBm of emitter power at 868 MHz and sampling rate is 1 MHz, $d_{et}$ = 3 m & $d_{tr}$ = 250 meters \\
			\cline{1-7}
			\hline
		\end{tabular}
		\par\end{centering}
\end{table*}
\subsection{Overview of Bistatic Backscatter Communications Systems}
\label{subsec:overviewBBCSs}
BBCSs have been introduced for low-cost, low-power, and large-scale wireless networks. Due to the prominent characteristics, BBCSs have been adopted in many applications such as wireless sensor networks, IoT, and smart agriculture~\cite{Kimionis2012Bistatic},~\cite{Kamp2014Wireless},~\cite{Konstan2016Converting}.

\begin{figure*}[tbh]
		\centering
		\includegraphics[scale=0.35]{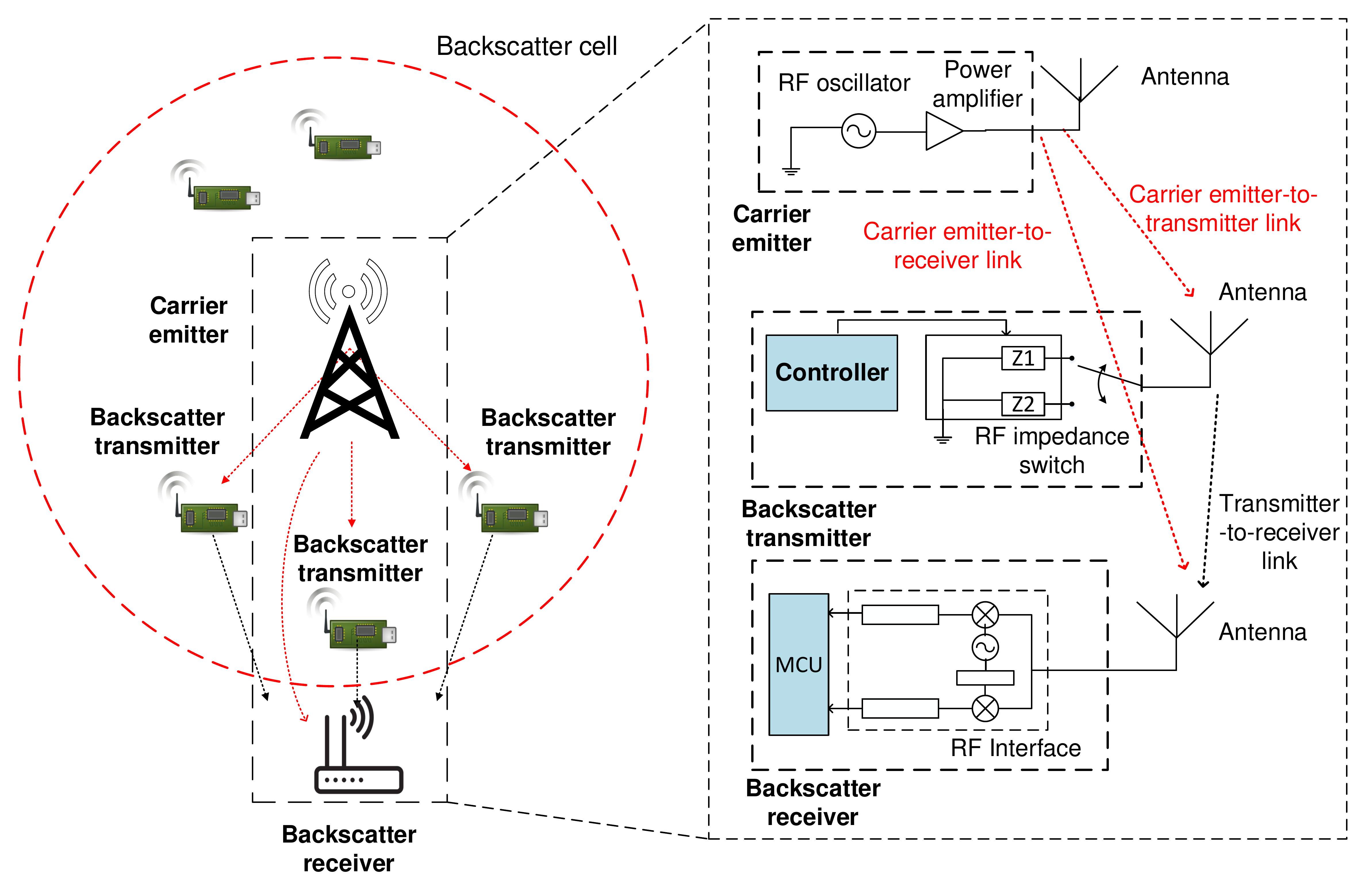}
		\caption{A general bistatic backscatter communications architecture.}
		\label{fig:chap5_bi_archs}
\end{figure*}

As shown in Fig.~\ref{fig:chap5_bi_archs}, there are three major components in the BBCS architecture: (i) backscatter transmitters, (ii) a backscatter receiver, and (iii) a carrier emitter, i.e., RF source. Unlike monostatic configuration, e.g., RFID, where the RF source and the backscatter receiver reside on the same device, i.e., the reader~\cite{Kimionis2013Bistatic}, in the bistatic systems, the carrier emitter and the backscatter receiver are physically separated. To transmit data to the backscatter receiver, the carrier emitter first transmits RF signals, which are produced by the RF oscillator, to a backscatter transmitter through the emitter's antenna which is connected to the power amplifier as shown in Fig.~\ref{fig:chap5_bi_archs}. Then, the backscatter transmitter harvests energy from the received signals to support its internal operation functions, such as data sensing and processing. After that, under the instruction of the backscatter transmitter's controller, the carrier signals are modulated and reflected by switching the antenna impedance with different backscatter rates~\cite{Choi2015Backscatter} through the RF impedance switch. The carrier emitter's signals and the transmitter's reflected signals are received at the antenna of the backscatter receiver and processed by the RF interface. First, the received signals are passed to the filters to recover the reflected signals from the backscatter transmitter. Then, the signals are demodulated by the demodulator and converted to bits by the converter to extract useful data. The extracted data is collected and processed by the micro-controller unit (MCU) inside the backscatter receiver. References~\cite{Choi2015Backscatter} and~\cite{Kimionis2012Bistatic} provide more comprehensive details for the standard models of bistatic systems.

The BBCSs have many advantages compared with conventional wireless communications systems as follows:
	\begin{itemize}
		\item {\textit{Low power consumption:} As the backscatter transmitters do not need to generate active RF signals, they have lower power consumption than those of conventional wireless systems. For example, a low-power backscatter transmitter, which consists of an HMC190BMS8 RF switch~\cite{HMC190BMS8} as the front-end and an MSP430~\cite{MSP430} as the controller, is introduced in~\cite{Varshney2016Loera}. The power consumptions of HMC190BMS8 and MSP430 are as little as 0.3 $\mu$W and 7.2 mW, respectively. Moreover, the backscatter transmitter used in~\cite{Konstan2016Converting}, consumes 10.6 $\mu$W for its operations. In~\cite{Vougiou2016Could}, a Silicon Laboratories SI1064 ultra-low power MCU~\cite{SI1064Datasheet} with an integrated transceiver is used as the backscatter receiver and another SI1064 is configured as the carrier emitter. The power consumption of the SI1064 is less than 10.7 mA RX and 18 mA TX at 10 dBm of output power. These power consumptions are significantly lower than that of conventional wireless systems' components. For example, a typical commercial RFID reader, i.e., the Speedway Revolution R420 from Impinj~\cite{SpeedwayR420}, consumes 15 W of power for its operations. Furthermore, in wireless sensor networks, an active RF sensor, named N6841A~\cite{N6841A}, may need as much as 30 W of power to operate, and a Sensaphone WSG30 gateway~\cite{SensaphoneWSG30} requires 120 V of alternating current power supply.}
		
		\item {\textit{Low implementation cost:} As the backscatter transmitters are battery-less, the size and the complexity of the electronic design are reduced. Therefore, the implementation cost can be significantly decreased especially in large-scale wireless systems, which typically involve a large number of backscatter transmitters. For example, the implementation cost for 100 backscatter sensor transmitters, which are introduced in~\cite{Konstan2016Converting}, is as little as \$10. In contrast, the price of an N6841A sensor~\cite{N6841A}, which is a typical active RF sensor, is about \$18. In RFID systems, an active RFID tag usually costs \$25 up to \$100, and a passive 96-bit EPC tag costs 7 to 15 U.S. cents~\cite{RFIDPrice}. In addition, by using off-the-shelf devices, the prices of the carrier emitter and the backscatter receiver are significantly reduced. In~\cite{Varshney2016Loera}, a CC2420 radio chip~\cite{CC2420Data} is used as the carrier emitter and a Texas Instruments CC2500~\cite{CC2500Data} is utilized as the backscatter receiver. Both CC2420 and CC2500 can work at frequencies in the range of UHF. CC2420 and CC2500 cost \$3.95~\cite{CC2420} and \$1.19~\cite{CC2500}, respectively, while an RFID reader costs under \$100 for low-frequency models, from \$200 to \$300 for high-frequency models, and from \$500 to \$2000 for UHF models~\cite{RFIDReaderCost}.}
		
		\item {\textit{Scalability:} In the bistatic systems, as the carrier emitters are separated from the backscatter receiver and deployed close to the backscatter transmitters, the emitter-to-transmitter path loss can be significantly reduced. Therefore, with more carrier emitters in the field and the backscatter transmitters being placed around, the transmission coverage of the systems can be extended~\cite{Kimionis2014Increased}. In addition, as bistatic backscatter radio is suitable for low-bit rate sensing applications~\cite{Nikos2015Coherent}, the backscatter transmitter occupies a narrow bandwidth. Thus, the number of backscatter transmitters in the systems can be increased in the frequency domain~\cite{Kimionis2013Bistatic}.}
	\end{itemize}

Compared with the monostatic systems, e.g., RFID, the communication ranges and transmission rates of bistatic systems are usually greater~\cite{Hoang2017Optimal}. However, the performances are still limited, especially compared with active radio communications systems. This is due to the fact that the backscatter transmitters are battery-less and hardware-constrained devices. Furthermore, important issues such as multiple access and energy management need to be addressed. Therefore, in the following, we review solutions to address the major challenges in the bistatic systems.

\subsection{Performance Improvement for Bistatic Backscatter Communications Systems}
\label{subsec: solution}
\subsubsection{Communication Improvement}

As mentioned above, a bistatic backscatter offers better communication range and transmission rate than those of MBCSs. Nevertheless, the performances need to be further improved to meet requirements of future wireless systems and their applications.

The authors in~\cite{Kimionis2012Bistatic} propose a backscatter receiver design to increase the communication range of BBCSs. One of the most important findings in this paper is that the time-varying carrier frequency offset (CFO) between a carrier emitter and a backscatter receiver can significantly reduce the communication range. The CFO often occurs when the local oscillator at the backscatter receiver does not synchronize with the carrier signals, i.e., oscillator inaccuracy. Thus, the authors eliminate the CFO by passing the received signals to an absolute operator at the backscatter receiver. This absolute operator can divide the received signals into noiseless and noise signals. Then, the backscatter receiver observes the amplitude of the noiseless signals which take two distinct values according to the binary modulation performed by the backscatter transmitter, and thus the CFO is removed. Furthermore, the near-optimal detectors are adopted in order to improve the BER performance, which also increases the transmitter-to-receiver distance. The experimental results show that the proposed backscatter receiver design can extend the communication range up to 60 meters at 1 kbps and 30 dBm emitter power in an outdoor environment.

In~\cite{Kampianakis2014Wireless}, the authors propose a data smoothing technique including two phases of filtering to increase the communication range. The first phase adopts the histogram filtering process that calculates the histogram of collected data and derives these data with the highest occurrence in a certain range. Then, the Savitzky-Golay filtering process~\cite{Schater2011Savitzky} is implemented in the second phase to exploit least-squares data smoothing on the measurements. The proposed two-phase filtering can significantly reduce an error that may occur in the transmission, and thus the signal-to-noise ratio (SNR) at the backscatter receiver can be increased. The experimental results show that the proposed technique can extend the communication range up to 100 meters with less than 1 mW of emitter power at 868 MHz and 100 kbps bitrate.

The study in~\cite{Kimionis2014Increased} introduces a system model for BBCSs taking into account the important microwave parameters such as CFO, BER, and SNR, which impact the transmitter-to-receiver communication performance. The authors in~\cite{Kimionis2014Increased} then design a non-conventional backscatter radio system architecture with a CFO compensation block and noncoherent detectors. It is shown that the proposed architecture can increase the communication range up to 130 meters at 13 dBm emitter power by using the FSK modulation scheme and 1 kbps bitrate. In~\cite{Alevizos2014Channel}, the authors indicate that employing channel coding can increase the communication range and the reliability of BBCSs. To do so, the codeword needs to be simple so that the backscatter transmitter and the backscatter receiver with limited power can process. Thus, the Reed-Mullers code~\cite{Forney2005BlockCode} is adopted because of its small length and sufficient error correction capability. Moreover, the authors show that the BBCSs can suffer from the interleaving of carrier-to-transmitter and transmitter-to-receiver channels, which results in the reduction of the communication range. To solve this problem, an interleaving technique is employed in conjunction with the block codes. The key idea is that the backscatter transmitter stores a block of codewords and transmits bits in the block in sequence. As a result, the burst errors affect bits of different codewords rather than bits of the same codeword. From the experimental results, the transmitter-to-receiver communication range can be extended up to 134 meters with 13 dBm emitter power and 1 kbps bitrate.

However, the interleaving technique incurs delay and requires additional memory at both the backscatter transmitter and the backscatter receiver. Therefore, the authors in~\cite{Nikos2015Trans} propose a more sophisticated method based on short block-length cyclic channel codes, named \textit{interleaved code}, to reduce the memory requirements. Specifically, the authors extend the work in~\cite{Kimionis2014Increased} by developing coherent detectors to estimate unknown parameters of channel and microwave such as CFO and interleaving. Both the simulation and experimental results show that the proposed solution can achieve the communication range of 150 meters with 20 mW emitter power and 1 kbps bitrate. The authors in~\cite{Varshney2016Loera} introduce a backscatter receiver, named LOREA, to increase the communication range of BBCSs. To achieve this, LOREA decouples the backscatter receiver from the carrier emitter in frequency and space domains by (i) using different frequencies for the carrier emitter and the backscatter receiver and (ii) locating them in different devices. Therefore, the self-interference can be significantly reduced. In addition, LOREA uses 2.4 GHz Industrial Scientific Medical (ISM) band for transmissions, which enables to utilize the signals from other devices, such as sensor nodes and Wi-Fi devices, for carrier signals. By applying LOREA, the communication range can be extended to 225 meters at 26 dBm emitter power and 2.6 kbps bitrate in line-of-sight (LOS) scenarios.

In~\cite{Daskalakis2016Soil}, the authors indicate that the backscattered power at the backscatter transmitter will be increased when the duty-cycle approaches 50\%~\cite{Daskalakis2014Soil},~\cite{Smith1997dutycycle}. The authors also note that square waves having the duty cycle different from 50\% occupy additional bandwidth. To achieve 50\% duty cycle, a backscatter sensor transmitter with an analog switch and a resistor ($R_2$) in the circuit is proposed. The duty cycle of the produced signals is calculated as $\frac{R_2}{2R_2}=50\%$. The experiments demonstrate that the communication range is significantly extended up to 250 meters with the sampling rate of 1 MHz and 13 dBm carrier emitter power.

\begin{figure*}[htb]
	\centering
	\includegraphics[scale=0.32]{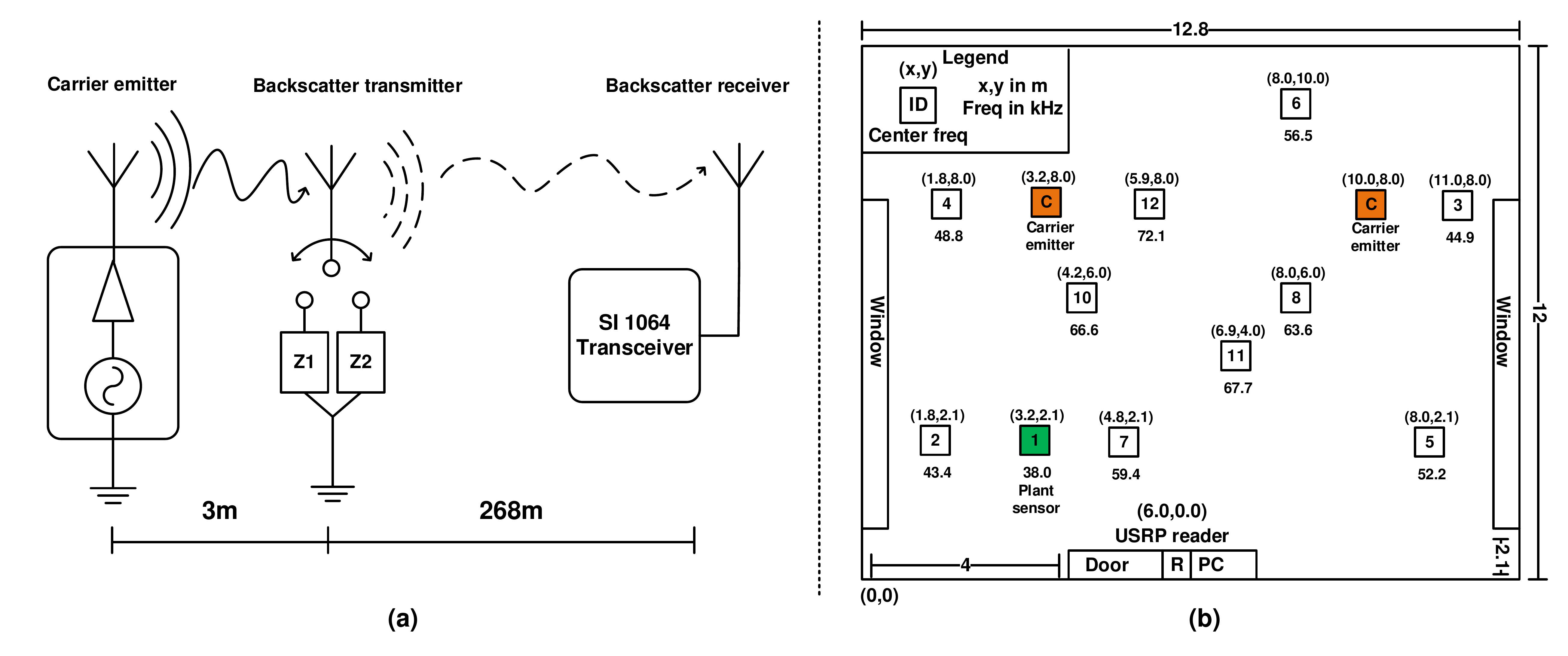}
	\caption{Experiment setup for measuring (a) communication range~\cite{Vougiou2016Could}, and (b) multiple access~\cite{Kampianakis2014Wireless}.}
	\label{fig:chap5_exp}
\end{figure*}
The authors in~\cite{Vougiou2016Could} introduce a backscatter transmitter circuit based on the Arduino development board, which uses a bit vector to form a packet consisting of 8-byte preamble, 4-byte sync, and 6-byte data. This packet is modulated by BFSK modulation and sent to the backscatter receiver. By selecting a long stream of preamble/sync bytes, the authors can minimize the effects of noise at the backscatter receiver. This leads to the reduction of packet error rate, and thus the communication range can be increased. To extract data in the packets from the backscatter transmitters, at the backscatter receiver, the authors deploy a Silicon Laboratories SI1064 ultra-low power MCU and an embedded TI 1101. The SI1064 MCU is integrated with a transceiver and configured to receive BFSK-modulated signals which are reflected from the backscatter transmitters. The TI 1101 is tested to verify the reception of these signals. Then, a prototype is implemented and based on the topology as shown in Fig.~\ref{fig:chap5_exp}(a). The carrier emitter produces RF signals at 868 MHz with 13 dBm of power, and the emitter-to-transmitter distance $d_{et}$ is 3 meters. The backscatter transmitter modulates data at 1.2 kbps using FSK modulation. The experimental results show that the communication range between the backscatter transmitter and the backscatter receiver $d_{tr}$ can be extended up to 268 meters.

\subsubsection{Multiple Access}
In the bistatic systems, the backscatter receiver may receive reflected signals from multiple backscatter transmitters simultaneously. Therefore, controlling the interference/collision among received signals is a challenge. There are several solutions in the literature to deal with this problem. FSK and On-off-keying (OOK) are the commonly used modulation schemes in BBCSs. Although FSK requires extra processing for CFO estimation compared with OOK, it outperforms OOK in terms of the BER performance~\cite{Kimionis2014Increased}. Furthermore, FSK and frequency-division multiplexing (FDM) are suitable for BBCSs. With FDM, since the reserved bandwidth for each backscatter transmitter is narrow, with a given frequency band, many backscatter transmitters can operate simultaneously. As the sub-carrier frequency reserved for each backscatter transmitter is unique, the collisions among the backscatter transmitters are eliminated~\cite{Kimionis2014Increased}. As a result, a majority of models in the literature, e.g.,~\cite{Kimionis2013Bistatic},~\cite{Kimionis2014Increased},~\cite{Alevizos2017Noncoherent},~\cite{Nikos2015Trans},~\cite{Kampianakis2014Wireless},~\cite{Daskalakis2016Soil}, use FDM as a multiple access scheme.

The authors in~\cite{Kampianakis2014Wireless} introduce an expression to estimate the operating sub-carrier frequency of a backscatter sensor transmitter for environment (humidity) monitoring applications. This expression is calculated by using the values of resistors and capacitors in the backscatter transmitter circuitry, i.e., a resistor-capacitor network. All backscatter sensor transmitters have different resistor-capacitor networks. Therefore, the center frequency and the spectrum band of each backscatter sensor transmitter are unique. By applying this expression, appropriate values of resistor-capacitor components can be chosen and the frequency for each individual transmitter can be calculated. To demonstrate the efficiency of the FDM scheme, the authors deploy the bistatic backscatter sensor system in a green house based on the topology as shown in Fig.~\ref{fig:chap5_exp}(b). The system consists of 10 environmental relative humidity backscatter sensor transmitters, and the topology of these transmitters is presented in Fig.~\ref{fig:chap5_exp}b. All the transmitters utilize different resistor-capacitor components in order to apply the FDM scheme as discussed above. To extend the coverage of the system, two carrier emitters with 20 mW emitter power are used. The experimental results show that the backscatter sensor transmitters are able to communicate with the backscatter receiver in a collision-free manner.

Similar to~\cite{Kampianakis2014Wireless}, the authors in~\cite{Daskalakis2016Soil} also define an expression to estimate the sub-carrier frequency for the backscatter sensor transmitters. However, the authors indicate that the outdoor temperature variations affect the circuit operation of the backscatter sensor transmitters, and thus the reserved sub-carrier frequency for each backscatter sensor transmitter may be drifted. Thus, in practice, the reserved bandwidth for each backscatter sensor transmitter is increased, and the number of backscatter sensor transmitters working on a given spectrum band will be reduced. It is also noted that the trade-off between scalability, i.e., the number of simultaneously operating backscatter sensor transmitters, and the environmental parameters for FDM scheme should be also analyzed.

In some cases, multiple BBCSs operate simultaneously at the same location, which can cause serious interference and reduce the performance of the whole network. To address this issue, the works in~\cite{Choi2015Backscatter},~\cite{Tountas2015Bistatic}, and~\cite{Kampianakis2014Wireless} adopt time-division multiplexing (TDM) to ensure that in each time frame there is only one active carrier emitter. In a single time frame, the carrier emitter transmits the carrier signals, based on which a certain backscatter transmitter backscatters their data to the backscatter receiver. Hence, the interference among emitters and the transmitters in the network is avoided.

\subsubsection{Energy Consumption Reduction}
The backscatter transmitters in BBCSs use energy harvested from an environment for their internal operations such as modulating and transmitting. However, the amount of the harvested energy is typically small. Therefore, several designs are proposed to use energy efficiently in BBCSs. In~\cite{Iyer2016Inter}, the authors design a backscatter transmitter in order to reduce the power consumption of BBCSs. The key idea is using 65 nm low power complementary metal-oxide-semiconductor (CMOS) technology which enables the backscatter transmitter to consume very small amount of energy in an idle state. The experimental results demonstrate that the power consumption of the backscatter transmitter is as little as 28 $\mu$W. Similarly, the authors in~\cite{Varshney2016Loera} introduce a backscatter transmitter design which consists of several low-power components such as MSP430~\cite{MSP430} for generating baseband signals and HMC190BMS8 RF switch~\cite{HMC190BMS8} for the backscatter front-end. Under this design, the backscatter transmitter consumes only 7.2 mW of power as shown in the experimental results.

In~\cite{Konstan2016Converting}, the authors propose a low-power backscatter sensor transmitter which can harvest energy from both the carrier emitter and the plant in the field. The authors note that the plant power-voltage characteristic varies in the range of 0.52-0.67 V, depending on the solar radiation and the ambient environmental temperature. This potential energy can be used to support internal operations of the backscatter sensor transmitter. Thus, an energy storage capacitor is employed to accumulate the biologic energy of the plant through a charging/discharging process. During the charging period, the operations of the backscatter sensor transmitter are suspended, and the biologic energy is harvested and stored in the capacitor. After the capacitor accumulates sufficient energy, the backscatter sensor transmitter is reactivated during the discharging period. The charging/discharging process is repeated constantly based on the transmission time interval of the backscatter sensor transmitter, which is controlled by a power management unit. The experimental results show that the proposed backscatter sensor transmitter consumes around 10.6 $\mu$W of power, and the harvested energy from the plant and the carrier emitter is sufficient for the operations. The authors also demonstrate that the capacitor of the backscatter sensor transmitter can have almost 0.7 V of biologic power after 1200 seconds of the charging time.

\subsection{Discussion}

In this section, we have provided an overview of BBCSs and reviewed several state-of-the-art approaches to enhance the designs and performances of the BBCSs. We summarize the approaches along with the references in Table~\ref{table_sec5_sum}. From the table, we observe that many existing works focus on improving the communication range of the BBCSs, while energy consumption reduction and multiple access are less studied. For the power management, some techniques such as eliminating the overheads of sensing, data handling, and communication~\cite{EkhoNet2014Zhang}, and increasing the efficiency of energy harvesting can be developed for the BBCSs. Furthermore, instead of FDMA and TDMA, several multiple access schemes, such as non-orthogonal multiple access, code division multiple access (CDMA), space division multiple access, and random access schemes such as ALOHA and CSMA, can be exploited to avoid transmission collisions.

Besides, as BBCSs use simple modulation and coding schemes, e.g., FSK and OOK, the systems may suffer from potential security attacks which adversely impact the performance and reliability of the systems. However, to the best of our knowledge, the security aspects of BBCSs are marginally studied in the existing works. Thus, this poses a need for effective schemes to prevent the vulnerability of security attacks.

In the following section, we give a comprehensive review of ABCSs.

\section{Ambient Backscatter Communications Systems}
\label{sec:AmbientBack}

In this section, we first present a general architecture of ABCSs. Then, existing approaches to improve performance of the ABCSs are discussed. Finally, we review emerging applications of ABCSs. Table~\ref{table_sec6_sum} provides the summary of ABCSs.
\subsection{Overview of Ambient Backscatter Communications Systems}
\subsubsection{Definition and Architecture}
\begin{table*}
	\caption{Summary of Ambient Backscatter Communications Systems}
	\label{table_sec6_sum}
	
	\begin{centering}
		\begin{tabular}{|>{\raggedright\arraybackslash}m{1.0cm}|>{\raggedright\arraybackslash}m{3.5cm}|>{\raggedright\arraybackslash}m{3.5cm}|>{\raggedright\arraybackslash}m{2.0cm}|>{\raggedright\arraybackslash}m{4.0cm}|}
			\hline 
			\multicolumn{1}{|>{\centering\arraybackslash}m{1.0cm}|}{\multirow{1}{*}{\textbf{Article}}} & 
			\multicolumn{1}{>{\centering\arraybackslash}m{3.5cm}|}{\multirow{1}{*}{\textbf{Design Goals}}} & \multicolumn{1}{>{\centering\arraybackslash}m{3.5cm}|}{\multirow{1}{*}{\textbf{Key idea}}} & \multicolumn{1}{>{\centering\arraybackslash}m{2.0cm}|}{\multirow{1}{*}{\textbf{RF source}}} & \multicolumn{1}{>{\centering\arraybackslash}m{4.0cm}|}{\multirow{1}{*}{\textbf{Results}}}\\
			\hline 
			\hline 
			\cite{Park2014Turbocharging}& Increase the communication range and bitrate (theoretical analysis, experiment, and prototype) & Design a multi-antenna backscatter transmitter and a low-power coding scheme & TV tower, 539 MHz & 1 Mbps at distances from 4 feet to 7 feet and 1 kbps at a distance of 80 feet \\
			\cline{1-5}
			\cite{Wang2017FMBackscatter}& Reduce the energy consumption of backscatter transmitters (experiment and prototype) & Deploy a backscatter transmitter consisting of low-power analog devices & FM tower, 91.5 MHz & 11.7 $\mu W$ at the backscatter transmitter  \\
			\cline{1-5}
			\cite{Bharadia2015BackFi}& Increase the communication range and bitrate (theoretical analysis, experiment, and prototype) & Design a self-interference cancellation technique & Wi-Fi AP, 2.4 GHz & 5 Mbps at a range of 1 meter and 1 Mbps at a range of 5 meters  \\
			\cline{1-5}
			\cite{Qian2016Noncoherent}& Improve BER performance and reduce the complexity of detectors (theoretical analysis and simulation) & Design a detector operates based on statistic variances of the received signals & N.~A. & Reduce the complexity while maintaining the BER performance as good as in~\cite{Wang2016Ambient} \\
			\cline{1-5}
			\cite{Shen2016Phase}& Reduce the phase cancellation problem (theoretical analysis, simulation, experiment, and prototype) & Design a multi-phase backscatter modulator & A signal generator operates at 915 MHz & Significantly reduce the phase cancellation problem, and thus increase the communication range and robustness \\
			\cline{1-5}
			\cite{Wang2016Ambient}& Minimize BER (theoretical analysis and simulation) & Design an ML detector with a threshold value to decode received signals without acknowledging the channel state information & N.~A. & $10^{-1}$ and $10^{-2}$ BER with 5 dB and 30 dB of transmit SNR, respectively \\
			\cline{1-5}
			\cite{Barott2014Coherent}& Increase the communication range (theoretical analysis) & Propose a passive coherent processing with four stages & TV tower, 626-632 MHz & 1 kbps at a range of 100 meters \\
			\cline{1-5}
			\cite{Zhang2016Enabling}& Increase the communication range and bitrate (experiment and prototype) & Design a frequency-shifted backscatter technique to reduce self-interference & Wi-Fi AP, 2.4 GHz & 50 kbps at a range of 3.6 meters with $10^{-3}$ BER  \\
			\cline{1-5}
			\cite{Penichet2016Do}& Increase the bitrate (theoretical analysis and simulation) & Encode multiple bits per symbol & N.~A. & Increase the bitrate while reducing the robustness \\
			\cline{1-5}
			\cite{Liu2014Enabling}& Improve performance of ABCSs in terms of BER, communication range, bitrate, reliability and energy consumption & Deploy a full-duplex backscatter backscatter transmitter consisting of low-power components & TV tower, 920 MHz & The backscatter transmitter consumes 0.25 $\mu W$ for TX and 0.54 $\mu W$ for RX\\
			\cline{1-5}
			\cite{Liu2017Coding}& Improve BER performance (theoretical analysis and simulation) & Implement a coding scheme & N.~A. & $10^{-3}$ BER with 15 dB of transmit SNR and $10^{-1}$ bits/s/Hz with 20 dB of transmit SNR \\
			\cline{1-5}
			\cite{Kellogg2016Passive}& Reduce the energy consumption of backscatter transmitters (experiment and prototype) & Deploy a backscatter transmitter consisting of low-power analog devices & Wi-Fi AP, 2.4 GHz & 14.5 $\mu W$ at 1 Mbps and 59.2 $\mu W$ at 11 Mbps \\
			\cline{1-5}
			\cite{Zhou2017An}& Address multiple access problem (theoretical analysis and simulation) & Design a backscatter transmitter selection technique & N.~A. & Up to 8 backscatter transmitters \\
			\cline{1-5}
			\cite{Munir2017Lowpower}& Increase the bitrate (simulation) & Design a relaying technique for full-duplex backscatter devices & TV tower, 539 MHz & 2 kbps between the backscatter transmitter and a relay node and 1 kbps between the relay node and the backscatter receiver \\
			\cline{1-5}
			\hline
		\end{tabular}
		\par\end{centering}
\end{table*}

The first ABCS is introduced in~\cite{Liu2013Ambient}, and it has quickly become an effective communication solution which can be adopted in many wireless applications and systems. Unlike BBCSs, ABCSs allow backscatter transmitters to communicate by using signals from ambient RF sources, e.g., TV towers, cellular and FM base stations, and Wi-Fi APs. As an enabler for device-to-device (D2D) communications, ABCSs have received a lot of attention from both academia and industry~\cite{Gollakota2013Emergence},~\cite{Penichet2016PhD}.

\begin{figure*}[tbh]
	\centering
	\includegraphics[scale=0.47]{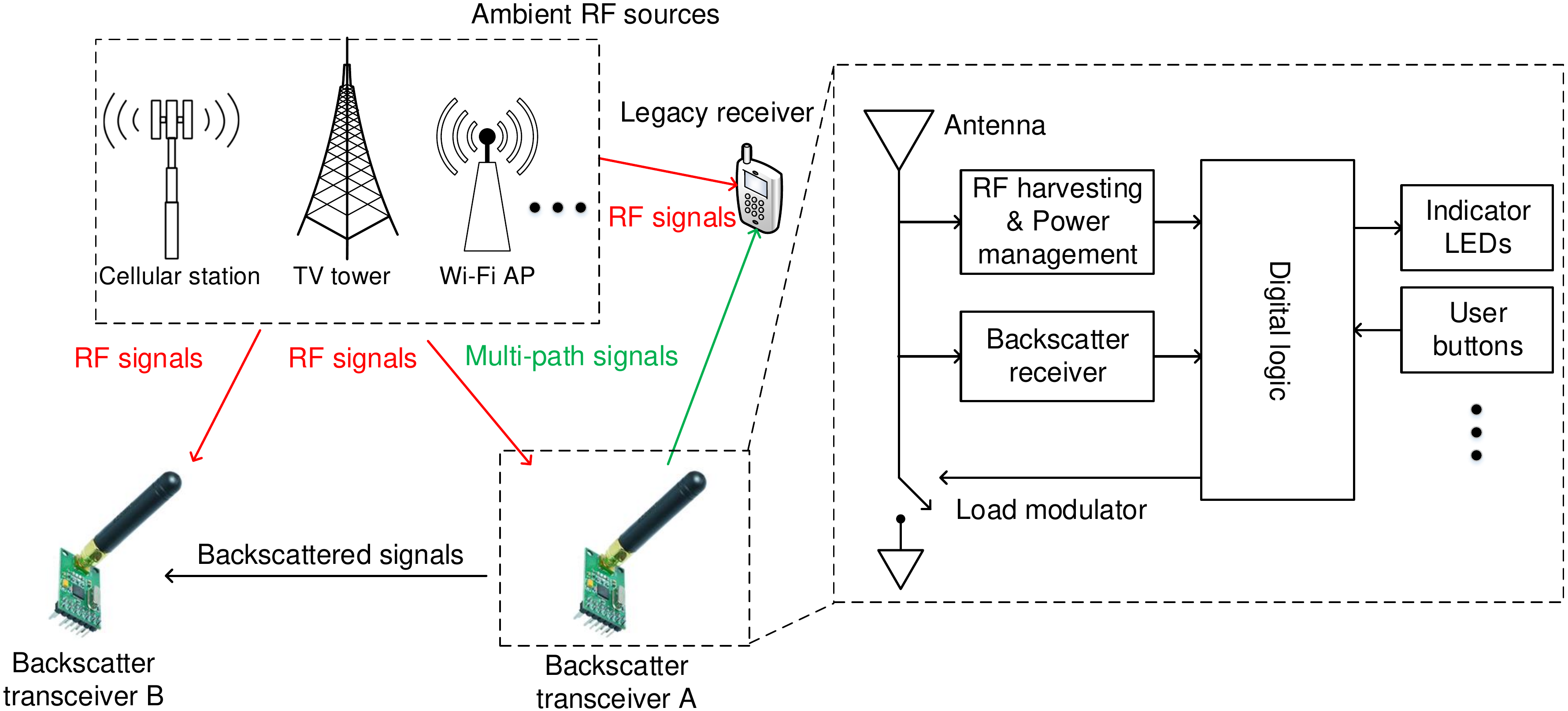}
	\caption{A general ambient backscatter communications architecture.}
	\label{fig:chap6_am_archs}
\end{figure*}

As shown in Fig.~\ref{fig:chap6_am_archs}, a general ABCS architecture consists of three major components: (i) RF sources, (ii) ambient backscatter transmitters, and (iii) ambient backscatter receivers. The ambient backscatter transmitter and receiver can be co-located and known as a transceiver. The ambient RF sources can be divided into two types, i.e., \textit{static} and \textit{dynamic} ambient RF sources~\cite{Xiao2015RFHarvest}. Table~\ref{table_sec6_ambient_source} shows the transmit power and RF source-to-transmitter distance of different RF sources. 

\begin{table*}
	\caption{Ambient RF sources} 
	\label{table_sec6_ambient_source}
	\begin{centering}
		\begin{tabular}{|>{\raggedright\arraybackslash}m{3cm}|>{\raggedright\arraybackslash}m{2cm}|>{\raggedright\arraybackslash}m{1.5cm}|>{\raggedright\arraybackslash}m{2cm}|>{\raggedright\arraybackslash}m{3cm}|>{\raggedright\arraybackslash}m{2.5cm}|}
			\hline 
			\multicolumn{1}{|>{\centering\arraybackslash}m{3.0cm}|}{\textbf{Type}} & \multicolumn{1}{>{\centering\arraybackslash}m{2.0cm}|}{\textbf{RF source}} & \multicolumn{1}{>{\centering\arraybackslash}m{1.5cm}|}{\textbf{Transmit power}} & \multicolumn{1}{>{\centering\arraybackslash}m{2.0cm}|}{\textbf{Frequency}} & \multicolumn{1}{>{\centering\arraybackslash}m{3.0cm}|}{\textbf{Transmission rate}} & \multicolumn{1}{>{\centering\arraybackslash}m{2.5cm}|}{\textbf{RF source-to-transmitter distance}}\tabularnewline
			\hline 
			\hline 
			\parbox[t]{2mm}{\multirow{4}{*}{Static RF source}} 
			& TV Tower & Up to 1 MW & 470-890 MHz	& 1 kbps at 539 MHz and 1 MW of transmit power~\cite{Liu2013Ambient} & Several kilometers	\tabularnewline 	\cline{2-6} 
			& FM base station & Up to 100 kW & 88-108 MHz & 3.2 kbps at 91.5 MHz and a received power at backscatter transmitters of -20 dBm~\cite{Wang2017FMBackscatter} & Several kilometers		\tabularnewline 	\cline{2-6} 
			& Cellular base station & Up to 10 W & 900 MHz (GSM 900) & N.~A. & Several hundred meters	\tabularnewline 	\cline{2-6} 
			\hline 
			\parbox[t]{2mm}{\multirow{1}{*}{Dynamic RF source}} 
			& Wi-Fi AP & Up to 0.1 W & 2.4 GHz & 1 kbps with 40 mW of transmit power~\cite{Kellogg2016Passive} & Several meters	\tabularnewline \cline{2-6} 
			\hline 
		\end{tabular}
		\par\end{centering}
\end{table*}

\begin{itemize}
	\item {\textit{Static ambient RF sources:} Static ambient RF sources are the sources which transmit RF signals constantly, e.g., TV towers and FM base stations. The transmit powers of these RF sources are usually high, e.g., up to 1 MW for TV towers~\cite{Xiao2015RFHarvest}. The transmitter-to-RF source distance can vary from several hundred meters to several kilometers~\cite{Liu2013Ambient},~\cite{Wang2017FMBackscatter}.}
	\item {\textit{Dynamic ambient RF sources:} Dynamic ambient RF sources are the sources which operate periodically or randomly with typically lower transmit power, e.g., Wi-Fi AP. The transmitter-to-RF source distance is often very short, e.g., 1-5 meters~\cite{Bharadia2015BackFi}.}
\end{itemize}

\subsubsection{Ambient Backscatter Design}

In~\cite{Liu2013Ambient}, the authors design an ambient backscatter transmitter which can act as a transceiver as shown in Fig.~\ref{fig:chap6_am_archs}. A transceiver consists of three main components: (i) the harvester, (ii) backscatter transmitter, and (iii) the backscatter receiver. The components are all connected to the same antenna. To transmit data, the harvester extracts energy from ambient RF signals to supply energy for the \textit{backscatter transceiver A}. Then, by modulating and reflecting the ambient RF signals, the \textit{backscatter transceiver A} can send data to \textit{backscatter transceiver B}. To do so, \textit{backscatter transceiver A} uses a switch which consists of a transistor connected to the antenna. The input of the \textit{backscatter transceiver A} is a stream of one and zero bits. When the input bit is zero, the transistor is off, and thus the \textit{backscatter transceiver A} is in the non-reflecting state. Otherwise, when the input bit is one, the transistor is on, and thus the \textit{backscatter transceiver A} is in the reflecting state. As such, \textit{backscatter transceiver A} is able to transfer bits to \textit{backscatter transceiver B}. Clearly, \textit{backscatter transceiver B} can also send data to \textit{backscatter transceiver A} in the same way.

In ABCSs, to extract data transferred from the ambient backscatter transmitter, an averaging mechanism is adopted at the ambient backscatter receiver. The main idea of the averaging mechanism is that the backscatter receiver can separate the ambient RF signals and the backscattered signals if the bitrates of these signals are significantly different. Therefore, the backscatter transmitter transmits the backscattered signals at a lower frequency than that of the ambient RF signals, and hence adjacent samples in the ambient RF signals are more likely uncorrelated than adjacent samples in the backscattered signals. As such, the backscatter receiver can remove the variations in the ambient RF signals while the variations in the backscattered signals remain. The backscatter receiver can decode data in the backscattered signals by using two average power levels of the ambient and backscattered signals.

It is important to note that the inputs of the averaging mechanism are digital samples. Hence, another challenge when designing the backscatter receiver is how to decode backscattered data without using an analog-to-digital converter which consumes a significant amount of energy. The authors in~\cite{Liu2013Ambient} thus design a demodulator as shown in Fig.~\ref{fig:chap3_modulated_backscatter}(b). First, at the receiver, the received signals are smoothed by an envelope circuit. Then, a threshold between the voltage levels of zero and one bits is computed by a compute-threshold circuit. After that, the comparator compares the average envelope signals with a predefined threshold to generate output bits.

\subsubsection{Advantages and Limitations}

In ABCSs, as the backscatter transmitters can be designed with low-cost and low-power components, the system costs as well as system power consumption can be significantly lowered~\cite{Liu2013Ambient}. For example, the ambient backscatter transceivers in~\cite{Liu2013Ambient} include several analog components such as MSP430~\cite{MSP430} as a micro-controller and ADG902~\cite{ADG902} as an RF switch. The power consumption of the analog components of this transceiver is as low as 0.25 $\mu$W TX and 0.54 $\mu$W RX, while the analog components of a traditional backscatter system, i.e., Wireless Identification and Sensing Platform (WISP)~\cite{WISP}, consume 2.32 $\mu$W TX and 18 $\mu$W RX. Furthermore, by using ambient RF signals, there is virtually no cost for deploying and maintaining RF sources, e.g., carrier emitters in BBCSs and readers in RFID systems. ABCSs also enable ubiquitous computing and allow direct D2D and multi-hop communications~\cite{Darsena2016Performance},~\cite{Wang2016Ambient}. Moreover, backscatter transmitters in ABCSs only modulate and reflect existing signals rather than actively transmit signals in the licensed spectrum. Consequently, their interference to the licensed users is almost negligible. Therefore, the ABCSs can be considered to be legal under current spectrum usage policies~\cite{Liu2013Ambient}, and they they do not require dedicated frequency spectrum to operate, thereby saving system cost further.

Nevertheless, ABCSs have some limitations. As backscatter transmitters use ambient RF signals for circuit operation and data transmission, it is typically not possible to control the RF sources in terms of quality-of-service such as transmit power, scheduling, and frequencies. In addition, ABCSs may potentially face several security issues since the backscatter transmitters are simple devices and the RF sources are not controllable. Moreover, as the harvested energy from the ambient RF signals is usually small~\cite{Hoang2017Ambient}, and these signals can be affected by fading and noise on the communication channels, the bitrate and communication range between the backscatter transmitters of ABCSs are limited.

In the following, we review existing solutions to address the aforementioned limitations of ABCSs.

\subsection{Performance Improvement for Ambient Backscatter Communications Systems}
\subsubsection{Communication Improvement}
Although ABCSs possess many advantages as mentioned above, their communication ranges and bitrates are very limited. In particular, for the first ABCS introduced in~\cite{Liu2013Ambient}, to achieve a target BER of $10^{-2}$, the backscatter receiver can receive at a rate of 1 kbps at distance up to 2.5 feet for an outdoor environment and up to 1.5 feet for an indoor environment. Thus, solutions to improve communication efficiency for the ABCSs need to be developed. 

\begin{figure}[htb]
	\centering
	\includegraphics[scale=0.23]{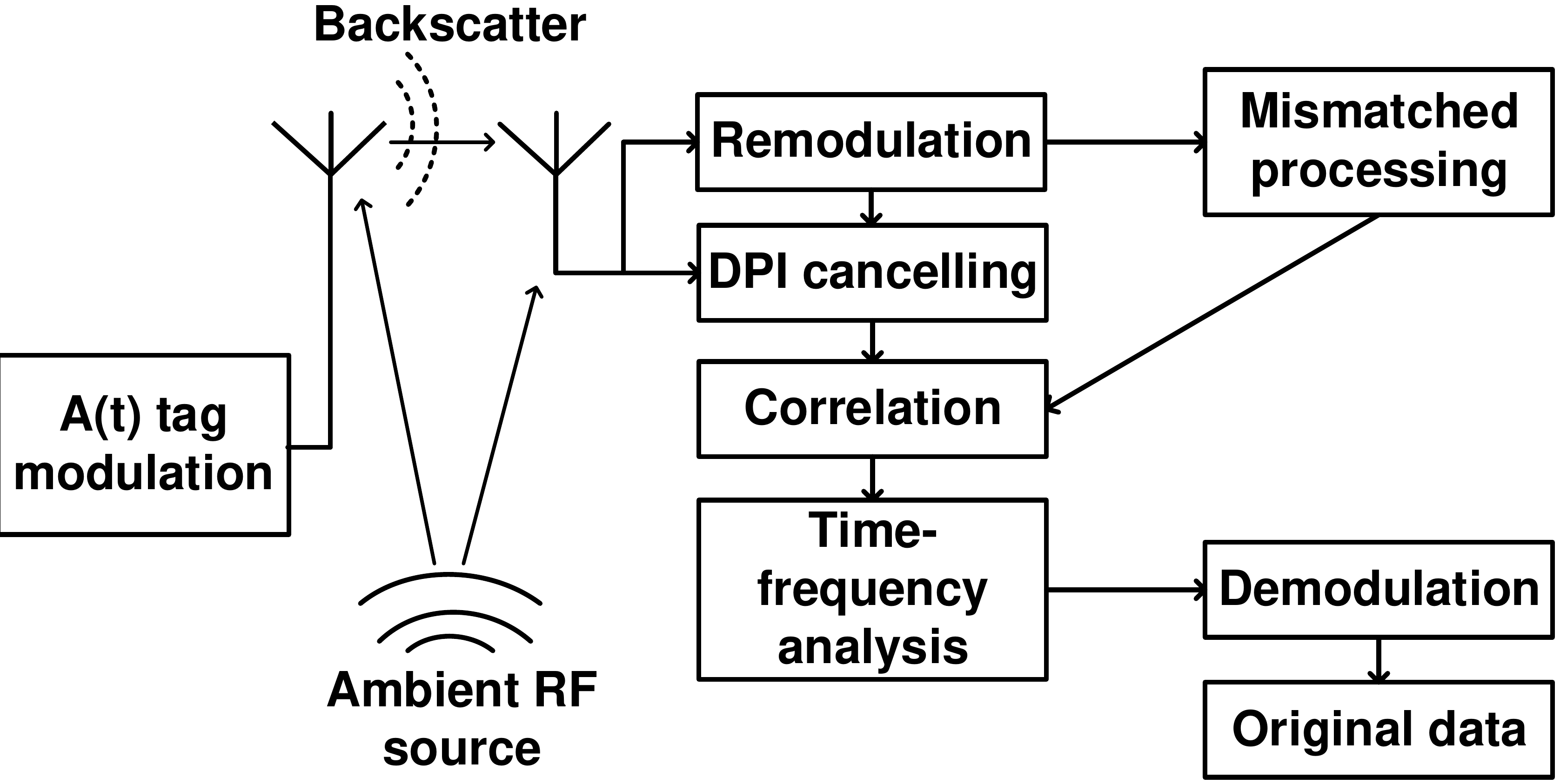}
	\caption{Block diagram of the signal processing for the passive backscatter receiver~\cite{Barott2014Coherent}.}
	\label{fig:chap6_Barott2014Coherent}
\end{figure}

In~\cite{Barott2014Coherent}, the authors first indicate that received signals $s(t)$ at the backscatter receiver consist of ambient RF signals $r(t)$, reflected signals $A(t)$, and noise $n(t)$. Thus, it is able to recover $A(t)$ from $s(t)$ by cross-correlation $s(t) \star r(t)$. The cross-correlation measures the similarity between two signals and is represented by the notation $\star$. To do so, the passive coherent processing is proposed with four stages as shown in Fig.~\ref{fig:chap6_Barott2014Coherent}. First, the received signals are passed to the remodulation stage to recover $r(t)$. In particular, the remodulation stage isolates a copy of $r(t)$ from the received signals $s(t)$ by using modulators and a demodulator. In general, the remodulation process generates two output waveforms. The first waveform is a clutter-free and noise-free reference signals used in adaptive clutter cancellation in the ambient RF signals, i.e., the direct path interference (DPI) cancellation. The second waveform is a mismatched reference signal used in cross-correlation, i.e., the mismatched processing. The principles of the remodulation can be found in~\cite{Palmer2013DVBT}. Then, the noise, i.e., DPI, is eliminated in DPI cancelling stage by using the Wiener-Hopf filtering~\cite{WienerHopffiltering} and the extensive cancellation algorithm~\cite{ECA}. Finally, the originally-transmitted signals are recovered from the noiseless signals in the correlation processing and time-frequency analysis stages. Then, the data sent by a backscatter transmitter is extracted through a demodulator. Theoretical analyses demonstrate that the passive coherent processing can achieve a bitrate of 1 kbps at the range of 100 meters with the TV tower operating at 626-632 MHz.

In~\cite{Daskalakis2017AmbientFM}, the authors introduce an ambient backscatter communications system that utilizes broadcast FM signals. In particular, the backscatter transmitter adopts OOK modulation and FM0 encoding on the ambient signals from an FM station to transmit data. At the backscatter receiver, an algorithm is employed to derive original data sent from the backscatter transmitter. The key idea of this algorithm is reducing the difference between frequencies at the backscatter transmitter and backscatter receiver, i.e., CFO correction. Then, a matched filter and a downsampling component are applied to remove noise and interference of the received signals to improve system performance. From the experimental results, the proposed backscatter system can achieve a bitrate of 2.5 kbps over a distance of 5 meters between the backscatter transmitter and backscatter receiver.

\begin{figure}[htb]
	\centering
	\includegraphics[scale=0.23]{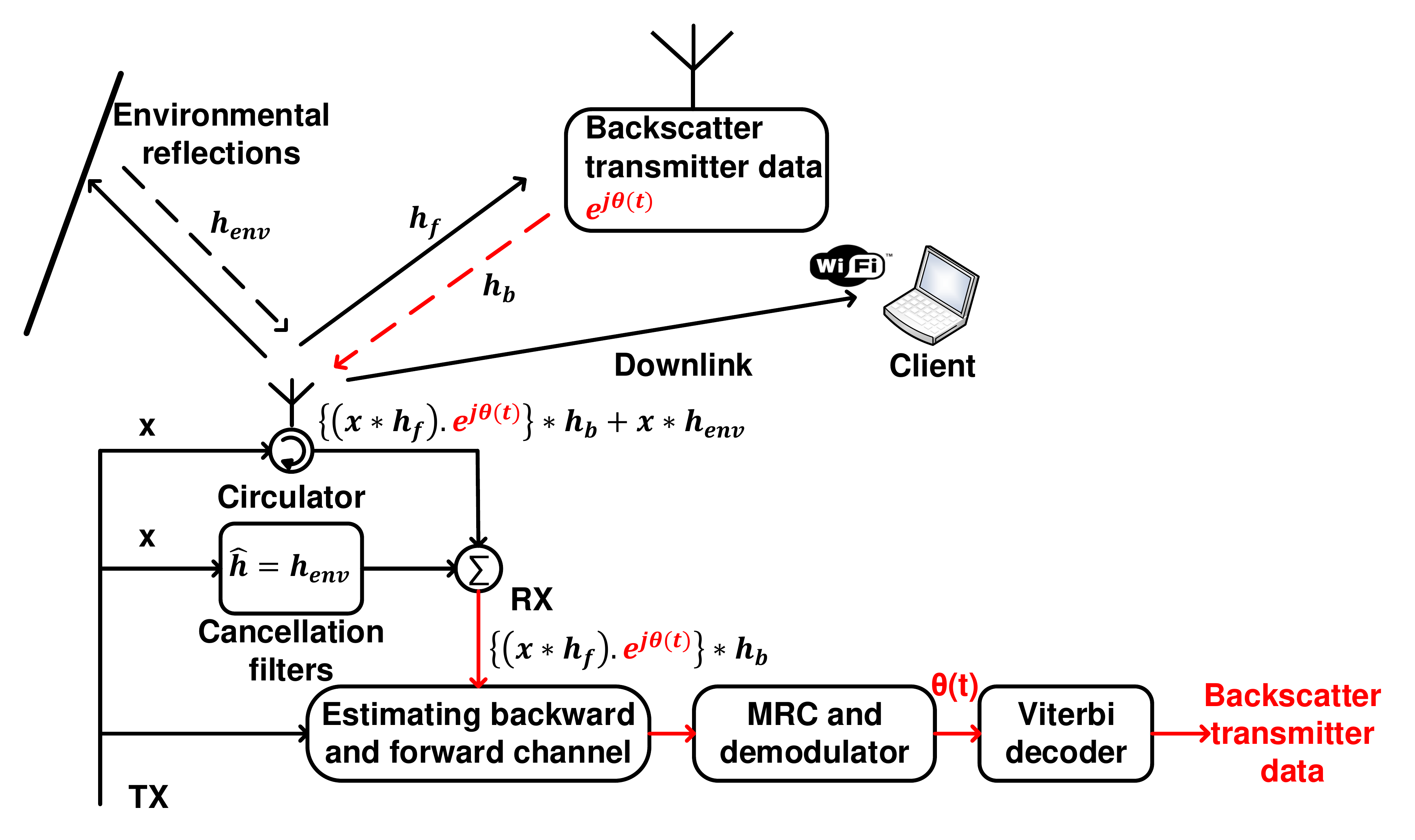}
	\caption{Architecture of the backscatter receiver used in BackFi~\cite{Bharadia2015BackFi}.}
	\label{fig:chap6_BackFi_reader}
\end{figure}

In~\cite{Bharadia2015BackFi}, the authors introduce BackFi which offers high bitrate and long-range communication between backscatter sensor transmitters. Unlike~\cite{Liu2013Ambient}, BackFi uses a Wi-Fi AP as an ambient RF source as well as a backscatter receiver. Thus, the backscatter sensor transmitters are able to not only communicate with each other, but also connect to the Wi-Fi AP. BackFi is different from RFID systems since it reuses ambient signals from the Wi-Fi AP which is already deployed for standard wireless networks. In~\cite{Bharadia2015BackFi}, the authors focus on improving the performance of the uplink transmission from the backscatter sensor transmitter to the Wi-Fi AP, i.e., BackFi AP. An important finding is that self-interference at the backscatter receiver, i.e., BackFi AP, can significantly reduce the communication range and transmission rate of the system. The self-interference arises from two sources: (i) signals from the Wi-Fi AP and (ii) reflected signals from non-transmitter objects in the environment. Then, a self-interference cancellation technique is proposed for the backscatter receiver as shown in Fig.~\ref{fig:chap6_BackFi_reader}. 
The Wi-Fi signals $x$, which are sent to a client, e.g., a laptop, are reflected by the environment and by a backscatter sensor transmitter. First, the reflected signals by the environment, i.e., $h_{env}$, are extracted from the received signals by using digital and analog finite impulse response filters, i.e., cancellation filters. The remaining signals after cancellation are used to estimate backward and forward channels, i.e., $h_b$ and $h_f$, respectively. However, $h_b$ and $h_f$ are in a cascaded form, i.e., $h_b*h_f$, where $*$ represents the convolution of two signals. Therefore, the authors use the maximal-ratio combining (MRC) technique~\cite{MRC} to recover the data from the backscatter sensor transmitter, i.e., $\theta (t)$, from $h_b*h_f$ signals. Then, the Viterbi decoder~\cite{Viterbi} is adopted to extract useful information. The authors then implement an experiment in an indoor environment with multi-path reflections. The experimental results show that BackFi can achieve the throughput of 5 Mbps at a range of 1 meter and the throughput of 1 Mbps at a range of 5 meters with a 2.4 GHz Wi-Fi AP.

Another technique which aims to reduce self-interference at the backscatter receiver is introduced in~\cite{Zhang2016Enabling}, named frequency-shifted backscatter. The key idea of this technique is that the backscatter transmitters shift the ambient RF signals, i.e., Wi-Fi signals, to an adjacent frequency band before reflecting. As such, the backscatter receiver can decode data from the reflected signals without self-interference. To do so, the authors use an oscillator at the backscatter transmitter to shift the RF signals by 20 MHz. The experimental results demonstrate that frequency-shifted backscatter can achieve a bitrate of 50 kbps at a range of 3.6 meters with BER of $10^{-3}$.

In~\cite{Park2014Turbocharging}, the authors design a multi-antenna backscatter transmitter, i.e., $\mu mo$, and a low-power coding mechanism, i.e., $\mu code$, to improve communication performance in terms of data rates and transmission ranges.
\begin{figure}[htb] 
	\centering
	\includegraphics[scale=0.3]{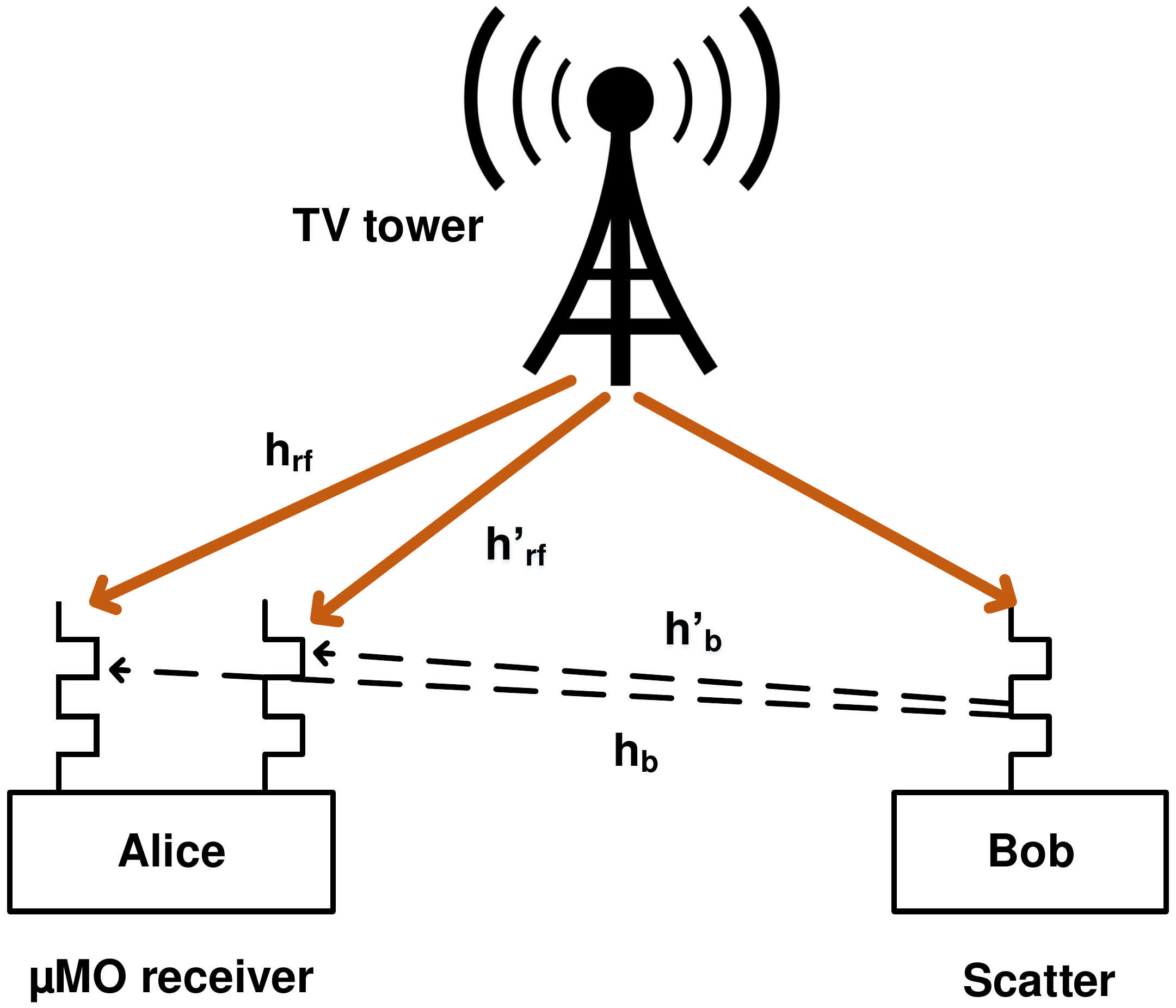}
	\caption{$\mu mo$ decoding~\cite{Park2014Turbocharging}.}
	\label{fig:chap6_Turbocharging_mo}
\end{figure}
By using multiple antennas, we can eliminate interference from ambient RF signals, e.g., TV signals, thereby increasing the bitrate between the backscatter transmitters. The main design principle of $\mu mo$ is shown in Fig.~\ref{fig:chap6_Turbocharging_mo}. Let $s(t)$ be RF signals from a TV tower and Bob transmits data by reflecting and absorbing $s(t)$ to convey bit `1' and bit `0', respectively. The received signals at two antennas of Alice are expressed as follows:
\begin{equation} \label{eq:chap6_received_signals}
\begin{split}
y_1(t) = h_{rf}s(t) + h_bB(t)s(t), \\
y_2(t) = h_{rf}^{'}s(t) + h_b^{'}B(t)s(t),
\end{split}
\end{equation}
where $h_{rf}$, $h_{rf}^{'}$ and $h_b$, $h_b^{'}$ are the channels from the TV tower and Bob to the two antennas of Alice, respectively. In addition, $B$ takes a value of `0' or `1' depending on non-reflecting or reflecting state, respectively. By dividing (\ref{eq:chap6_received_signals}) together, we have the following fraction:
\begin{equation}
	\label{eq:chap6_dividing}
	\frac{|y_1(t)|}{|y_2(t)|} = \frac{|h_{rf} + h_bB(t)|}{|h_{rf}^{'} + h_b^{'}B(t)|}.
\end{equation}
From~(\ref{eq:chap6_dividing}), this fraction is independent of the TV signals, i.e., $s(t)$. Since the value of $B$ is either `0' or `1', the fraction results in two levels, i.e., $\frac{|h_{rf}|}{|h_{rf}^{'}|}$ and $\frac{|h_{rf} + h_b|}{|h_{rf}^{'} + h_b^{'}|}$, corresponding to the non-reflecting and reflecting states, respectively. Therefore, Alice can decode data sent from Bob without estimating channel parameters.

Moreover, a low-power coding scheme based on CDMA, is proposed to increase the communication ranges between the backscatter transceivers. In this scheme, the backscatter transmitter encodes bit `0' and `1' into different chip sequences, and the backscatter receiver correlates the received signals with the chip sequence patterns to decode the data. Longer chip sequences for encoding can be used at both the backscatter transmitter and the backscatter receiver to increase the SNR. The authors then implement $\mu mo$ and $\mu code$ on a circuit board to evaluate their performance. The experimental results show that $\mu mo$ increases the bitrate up to 1 Mbps at distances from 4 feet to 7 feet and $\mu code $ increases the communication ranges up to 80 feet at 1 kbps by backscattering signals from a TV tower operating at 539 MHz.

In~\cite{Penichet2016Do}, the authors propose a solution to improve the bitrate for ABCSs by modifying the encoding technique $\mu code$. The authors in~\cite{Penichet2016Do} highlight that encoding multiple bits per symbol can significantly increase the bitrate. However, this encoding scheme may make the transmission more sensitive to noise and interference. Thus, the authors use simulation to investigate the trade-off between the bitrate and robustness of the proposed encoding scheme. Note that the proposed scheme encodes two bits per symbol. Instead, $\mu code$ encodes only one bit per symbol. The simulation results show that the energy per chip over noise spectral density ($E_c$/$N_0$) of the proposed scheme is higher than that of $\mu code$ with the same number of chips per symbol. This means that applying multiple bits per symbol can increase the bitrate. However, with the same value of $E_c$/$N_0$, $\mu code$ shows better robustness than that of the scheme from~\cite{Penichet2016Do}. In other words, the transmission is more likely to be corrupted by noise and interference when the number of bits per symbol is increased. The authors then conclude that (i) longer chip sequences are more robust and increase the communication range but result in low bitrates, and (ii) encoding more bits per symbol increases the bitrate, but reduces the robustness.

In~\cite{Liu2014Enabling}, the authors introduce a full-duplex technique to improve the performance of ABCSs. In this technique, after receiving reflected signals, the backscatter receiver can send a feedback to the backscatter transmitter to inform any error. The authors indicate that the challenge when designing the full-duplex system is that the amplitudes of the received signals at the backscatter receiver can change considerably when the receiver backscatters to send feedback signals. This issue arises due to the fact that the backscatter receiver uses the same antenna to transmit and receive signals. Therefore, the authors change the impedance of the antenna at the backscatter receiver to create phase shifts to the received signals, and thus the amplitudes of the received signals at the backscatter receiver are maintained. The authors then introduce a protocol with two steps for the feedback channel. First, as soon as the backscatter receiver receives signals sent from the backscatter transmitter, the backscatter receiver begins to transmit preamble bits on the feedback channel. Then, the backscatter receiver divides the received signals into chunks of $b$ bits and computes $c$-bit checksum for each group of $b$ bits. Second, the backscatter receiver transmits the checksum back to the backscatter transmitter. The values of $b$ and $c$ are determined by a ratio between the transmission rate of data channel, i.e., transmitter-to-receiver channel, and that of feedback channel, i.e., receiver-to-transmitter channel. In this way, the transmission times of both data and feedback are approximately equal. By using the feedback data, the backscatter transmitter can detect errors and collision, and is able to adjust its bitrates based on the channel condition. Additionally, by calculating the $c$-bit checksum for each chunk of $b$ bits, the full-duplex technique allows the backscatter transmitter to re-transmit a subset of the bits rather than the whole chunk when an error is detected.

However, in~\cite{Liu2017FullDuplex}, the authors study that this full-duplex technique focuses on mixed transmissions of data and feedback signals, thereby requiring asymmetric rates for transmissions in the opposite directions. This is not feasible for future wireless applications, e.g., IoT, in which communication links among the tremendous number of devices exist. Therefore, the authors in~\cite{Liu2017FullDuplex} propose a novel multiple-access scheme, namely time-hopping full-duplex backscatter communication (BackCom), to simultaneously mitigate interference and enable asymmetric full-duplex communications. In particular, the proposed scheme includes two components, i.e., a sequence-switch modulation and full-duplex BackCom. The key idea of the sequence-switch modulation is that bits are transmitted by switching between a pair of time-hopping spread-spectrum sequences with different nonzero chips to represent bits `0' and `1'. By doing this, the interference produced by time-hopping spread-spectrum is reduced. The numerical and simulation results demonstrate that the proposed full-duplex BackCom achieves higher performance in terms of BER, energy-transfer rates, and supporting symmetric full-duplex data rates. However, this system occupies a large spectrum bandwidth since it adopts time-hopping spread-spectrum.

In~\cite{Shen2016Phase}, a multi-phase backscatter modulator is introduced to circumvent the phase cancellation problem at the backscatter transmitters. The authors indicate that the phase difference between ambient RF signals and reflected signals at the backscatter receiver can significantly impact the amplitude of received signals. Thus, during the cancellation phase, the backscatter receiver cannot extract data from the received signals. To address this problem, the authors propose a modulator for the backscatter transmitter which enables multi-phase backscattering. Under this scheme, the backscatter transmitter backscatters its data in two successive intervals with different phases. Thus, if there is a cancellation phase during one of the intervals, the other interval, which operates at the different phase, will be immune to the cancellation phase. To further improve the transmission performance, the authors propose a hybrid scheme which combines the backscattered signals in these intervals. With an envelope detector, the backscatter receiver can identify four amplitude differences, which helps to differentiate between the amplitudes of the ambient RF signals and the reflected signals more accurately. The simulation and experimental results show that the proposed solutions successfully avoid the phase cancellation problem, and thus improve communication ranges and robustness for ABCSs.

In~\cite{Kim2017Optimum}, the authors propose an optimum modulation and coding scheme to maximize the network capacity of ABCSs. This scheme finds an optimal value of the reflection coefficient $\alpha$ and the code rate $\rho$. The authors then formulate a joint optimization problem of $\alpha$ and $\rho$ and use line search algorithms such as Golden section method to find the solution. The simulation results demonstrate that the network capacity can be improved by 90\% higher compared to the conventional modulations, e.g., BPSK. The authors also note that there is a trade-off in a choice of the variables $\alpha$ and $\rho$. For small $\alpha$, the backscatter transmitter harvests more energy and reflects fewer signals to the backscatter receiver. Consequently, this may lead to an information outage. For large $\alpha$, the backscatter transmitter harvests less energy and reflects more signals, and thus possibly resulting in a power outage. Likewise, for large $\rho$, the bitrate is increased but the reliability of the transmission deteriorates. In contrast, for small $\rho$, the reliability increases while the bitrate is reduced.

Different from all aforementioned schemes, several works focus on signal detection techniques to improve BER performance of ABCSs. In~\cite{Wang2016Ambient}, the authors introduce an ML detector to minimize BER without requiring channel state information. The authors indicate that the probability density functions of the conditional random variables vary at different transmission slots. Additionally, as the channel state information is unknown, the backscatter receiver cannot distinguish which energy level corresponds to which state. Thus, it is difficult to detect and extract data at the backscatter receiver. Therefore, the ML detector uses an approximate threshold to measure the difference between two adjacent energy levels. If there is a significant change between two successive energy levels, the detector can decode binary symbols sent from the backscatter transmitter. The simulation results show that the proposed ML detector can achieve high BER performance at around $10^{-1}$ and $10^{-2}$ with 5 dB and 30 dB of transmit SNR, respectively.

In~\cite{Yang2016Backscatter}, the authors propose a backscatter transceiver design to cancel out the direct-link interference for backscatter communications over an ambient orthogonal frequency division multiplexing (OFDM) carrier without increasing hardware complexity. This is a novel joint design for backscatter transmitter waveform and the detector of the backscatter receiver. The time duration of each backscatter transmitter symbol is set to one OFDM symbol period. Thus, for bit `1', there is an additional state transition in the middle of each OFDM symbol period within one backscatter transmitter symbol duration. Therefore, the designed waveform can be easily implemented in low-cost backscatter devices since it has similar characteristics to FM waveform. Additionally, a cyclic prefix is added at the beginning of the OFDM signals to create a repeating structure. With this design, the backscatter receiver can remove the direct-link interference by using an ML detector which exploits the repeating structure of ambient OFDM signals. The simulation results show that the proposed solution can achieve the BER of $9\times10^{-4}$ with 24 dB of transmit SNR.

In~\cite{YangModulation}, the authors extend the work in~\cite{Yang2016Backscatter} by considering a transceiver design for multi-antenna backscatter receivers. The authors propose an optimal detector to decode bits from the backscatter transmitter by using a linear combination of the received signals at each antenna. Similar to~\cite{Yang2016Backscatter}, this detector also exploits the repeating structure of ambient OFDM signals to obtain interference-free signals. The simulation results demonstrate that the multi-antenna backscatter receiver can achieve higher BER performance than that of the single-antenna backscatter receiver. In particular, the BER decreases quickly as the number of the backscatter receiver's antennas increases, i.e., from $0.5\times10^{-2}$ to about $10^{-6}$ at the SNR of 9 dB when the number of antennas increases from 1 to 6.

Although ML detectors can achieve a decent detection performance, their computational complexity may not be suitable for low-power backscatter receivers. Hence, the authors in~\cite{Qian2016Noncoherent} introduce a low-complexity detector which is able to maintain considerable detection performance. Similar to the ML detector in~\cite{Wang2016Ambient}, this detector also uses a detection threshold to determine `0' and `1' bits. Nevertheless, the threshold is computed using statistic variances of the received signals which are easy to derive. Thus, the computational complexity of the proposed detector is significantly reduced. Intuitively, with $M$ transmitted symbols, and the sampling number of the received signals corresponding to one single symbol is $N$, the ML detector requires at least $(18M + N)$ complex multiplier and adder (CMA) units and $4M$ exponent arithmetic units for the calculation of two probability density functions. Instead, the proposed detector needs just 4 CMAs. The simulation results also demonstrate that the BER performance of the proposed detector is as good as the ML detector in~\cite{Wang2016Ambient}.

\begin{figure}[htb] 
	\centering
	\includegraphics[scale=0.32]{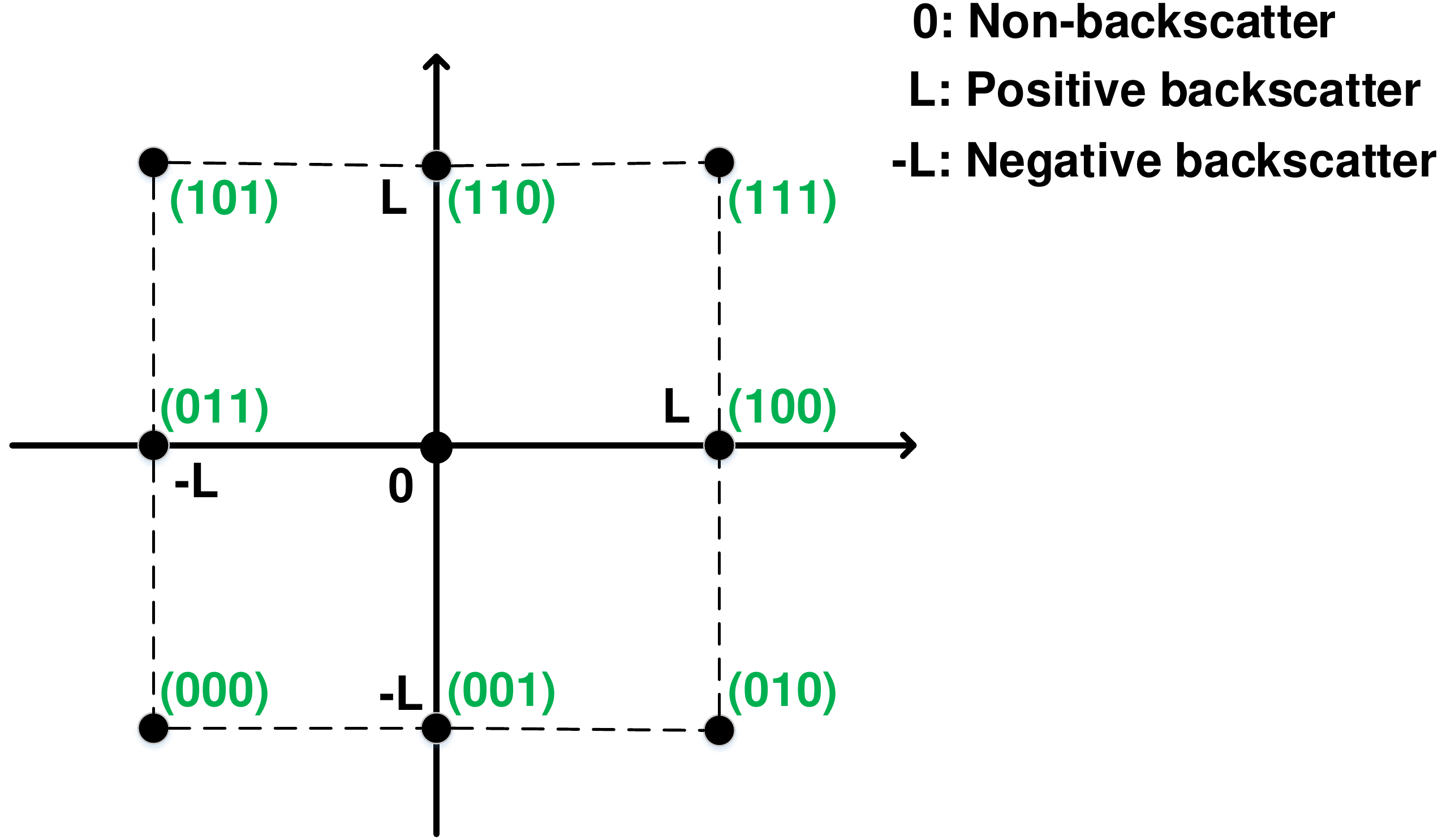}
	\caption{Signal constellation and coding scheme~\cite{Liu2017Coding}.}
	\label{fig:chap6_Coding_scheme}
\end{figure}

In~\cite{Liu2017Coding}, the authors introduce a coding scheme to increase throughput of backscatter communication. In the proposed coding scheme, three states, i.e., reflecting, non-reflecting, and negative-reflecting are used. The reflecting and non-reflecting states are the same as in conventional ABCSs. In the negative-reflecting state, the backscatter transmitter adjusts its antenna impedance to reflect RF signals in an inverse phase. With this three states, there are nine points in the signal constellation, and each point represents three-bit symbols as shown in Fig.~\ref{fig:chap6_Coding_scheme} where $L$ is the unit distance between two adjacent constellation points. Coding theory shows that it is possible to evaluate the BER performance of coding schemes by using the average Euclidean distance~\cite{Viterbi}. Thus, the coding scheme without point $(0,0)$ has low BER. As such, the proposed coding scheme removes point $(0,0)$ in the signal constellation to minimize the BER. Based on this coding scheme, the authors then design a maximum a posteriori (MAP) detector to detect signals at the backscatter receiver. Both the simulation and theoretical results demonstrate that the proposed solutions can reduce the BER to $10^{-3}$ with 15 dB of transmit SNR and increase the throughput up to $10^{-1}$ bits/s/Hz with 20 dB of transmit SNR.

\subsubsection{Power Reduction}
As the amount of energy harvested from ambient RF signals is usually small, backscatter transmitters may not have sufficient power for their operations. Thus, several solutions are proposed to deal with this problem. In fact, these solutions share the same idea as those in BBCSs, i.e., using low-power components in backscatter transmitter circuits.

In~\cite{Kellogg2016Passive}, the authors introduce a passive Wi-Fi backscatter transmitter which harvests energy from a Wi-Fi AP. The backscatter transmitter is designed by using low-power analog devices to reduce its energy consumption. The backscatter modulator of the backscatter transmitter consists of an HMC190BMS8 RF switch~\cite{HMC190BMS8} to modulate data by adjusting the antenna impedance. Additionally, for baseband processing, the authors use a 65 nm LP CMOS node~\cite{CMOS_TSMC} to save power. The authors then implement a prototype on a DE1 Cyclone II FPGA development board by Altera~\cite{FPGA} to measure the power consumption of the backscatter transmitter. The experimental results demonstrate that the passive Wi-Fi backscatter transmitter consumes as low as 14.5 $\mu W$ at 1 Mbps.

Similarly, in~\cite{Wang2017FMBackscatter}, the authors design a low-power backscatter transmitter which uses off-the-shelf components such as Tektronix 3252 arbitrary waveform generator~\cite{Tektronix3252} as a modulator, ADG902~\cite{ADG902} as an RF switch, and a 65 nm LP CMOS node~\cite{CMOS_TSMC} as a baseband processing. With this design, the backscatter transmitter consumes only 11.7 $\mu W$ of power. In~\cite{Liu2014Enabling}, a low-power backscatter transmitter is implemented on a four-layer printed circuit board using off-the-shelf components such as ADG919 RF switch~\cite{ADG919} connected directly to the antenna of the backscatter transmitter and the STMicroelectronics TS881~\cite{TS881} as an ultra-low power comparator. Furthermore, with a technique that re-transmits a subset of bits rather than the whole packet when an error occurs, the backscatter transmitter can save a significant amount of energy. The authors demonstrate that the proposed backscatter transmitter consumes around 0.25 $\mu W$ for TX and 0.54 $\mu W$ for RX.

\subsubsection{Multiple Access}
In ABCSs, there can be several backscatter transmitters operating simultaneously. Therefore, multiple access schemes are needed to achieve optimal network performance.

\begin{figure}[htb] 
	\centering
	\includegraphics[scale=0.25]{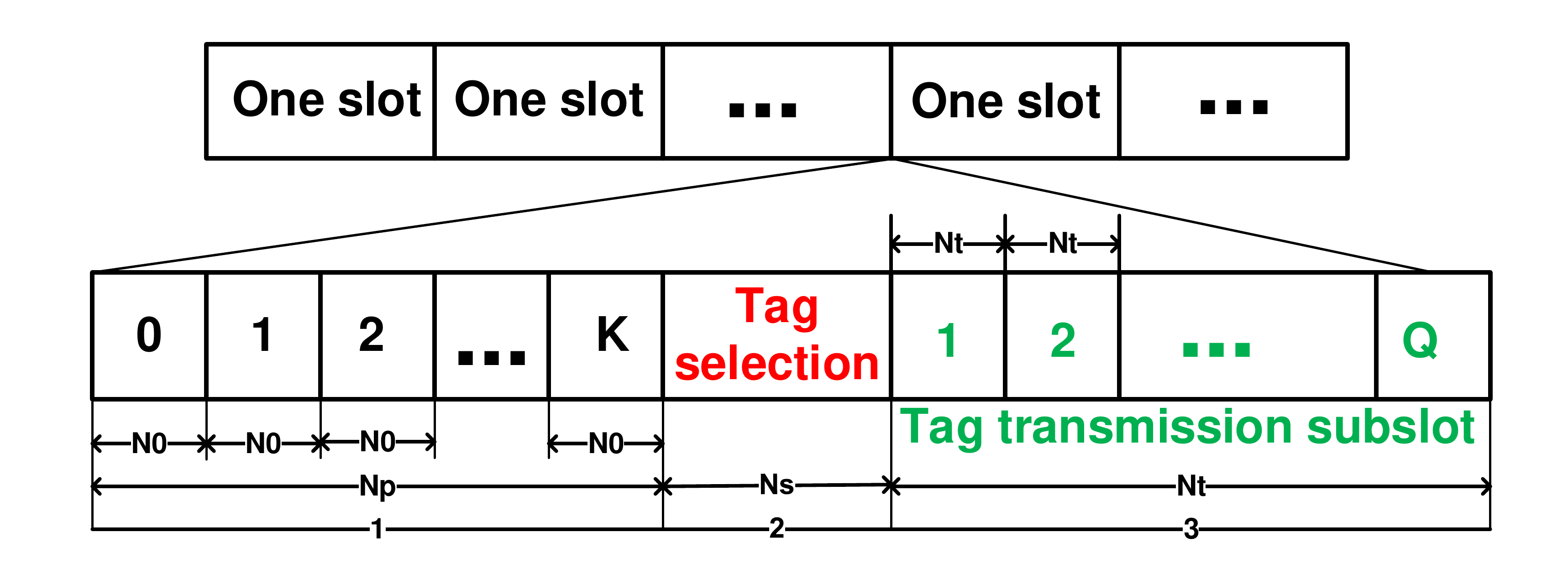}
	\caption{Slotted structure of the communication process between the backscatter receiver and $K$ backscatter transmitters~\cite{Zhou2017An}.}
	\label{fig:chap6_Slotted}
\end{figure}

In~\cite{Zhou2017An}, the authors propose a backscatter transmitter selection technique which allows $K$ backscatter transmitters to communicate with a backscatter receiver. The transmission process is divided into slots, and each slot consists of three sub-slots as shown in Fig.~\ref{fig:chap6_Slotted}. 
The first sub-slot contains $(K+1)N_0$ symbols. Note that the value of $N_0$ is not fixed. In the first $N_0$ symbols, the backscatter transmitters do not backscatter RF signals. In the following $KN_0$ symbols, each backscatter transmitter backscatters RF signals sequentially. In other words, in the $k$-th $N_0$ symbols, only the $k$-th backscatter transmitter backscatters RF signals for its own data transmission. In the second sub-slot, the backscatter receiver selects a backscatter transmitter with the best transmission condition based on the energy levels of received signals in the first sub-slot. In the third sub-slot, the selected backscatter transmitter is able to transmit data to the backscatter receiver while the other backscatter transmitters remain silent. In this way, the backscatter receiver can handle transmissions from backscatter transmitters without any interference. The simulation results demonstrate that the backscatter transmitter selection technique can allow the backscatter receiver to successfully receive data from 8 backscatter transmitters.

In~\cite{Liu2017MultipleAccess}, the authors introduce a multiple-access scheme to reduce the direct-link interference at the backscatter receiver. Their work considers an ambient backscatter multiple-access system, e.g., for smart-home applications, which allows the backscatter receiver to detect both the signals sent from the RF source and backscatter transmitter instead of adopting cancellation techniques as in most existing works. Specifically, this multiple-access system is different from conventional linear additive multiple-access systems since backscatter transmitters adopt multiplicative operations, and thus a multiplicative multiple-access channel (M-MAC) is also deployed. The numerical results show that the achievable rate region of the M-MAC is larger than that of the conventional TDMA scheme. Moreover, the rate performance of the system in the range of 0-30 dB of the direct-link SNR is significantly improved.

\subsection{Potential Applications}
The development of ambient backscatter techniques opens considerable opportunities for D2D communications. Thus, ABCSs can be adopted in many applications such as smart life, logistics, and medical biology~\cite{Liu2013Ambient},~\cite{Liu2017Backscatter},~\cite{Wang2017FMBackscatter},~\cite{Gollakota2013Emergence},~\cite{Huang2016Batteryfree},~\cite{DeylePatent}. ABCSs allow devices, i.e., backscatter transmitters, to operate independently with minimal human intervention.

\subsubsection{Smart world}

ABCSs can be deployed in many areas to improve quality of life. For example, in a smart home, a large number of passive backscatter sensor transmitters can be placed at flexible locations, e.g., inside walls, ceilings, and furniture~\cite{Liu2017Backscatter}. These backscatter sensor transmitters can operate for a long period of time without additional power sources and maintenance. The applications include detection of toxic gases, e.g., gas, smoke, and CO, monitoring movements, and surveillance. Furthermore, backscatter transmitters can be embedded inside things in our daily life, i.e., IoT. A proof-of-concept of ABCSs, i.e., smart card applications, is first introduced in~\cite{Liu2013Ambient}. The authors implement a simple scenario, i.e., a smart card transmits texts ``Hello World'' to another smart card. The experimental results show that the texts ``Hello World'' can be transmitted at a bitrate of 1 kbps and a range of 4 inches with 94\% of successful ratio without any retries. In~\cite{Wang2017FMBackscatter}, the authors deploy a backscatter transmitter inside a poster. By using ambient signals from a local FM station operating at 94.9 MHz, the poster can transmit data with texts and audio to its receiver, e.g., a smart phone, to show its supplementary contents. The experimental results show that the prototype can achieve a bitrate of 100 bps at distances of up to ten feet.

\subsubsection{Biomedical Applications}

Biomedical applications such as wearable and implantable health monitoring require small and long-lasting communication devices. Ambient backscatter transmitters can meet these requirements. Some biomedical prototypes have been implemented. For example, in~\cite{Huang2016Batteryfree}, the authors design a battery-free platform for wearable devices, e.g., smart shoes, through backscattering ambient RF signals. A pair of shoes is implemented with sensors and ambient backscatter modules. the sensor in each shoe performs separate tasks, e.g., counting steps and heart rate, and two shoes are coordinated by using the ambient backscatter modules. The experimental results demonstrate that the proposed platform successfully operates in real-life scenarios. However, the bitrate may significantly reduce when moving speeds are high. Another interesting application is introduced in~\cite{Wang2017FMBackscatter}, i.e., smart fabric. The authors embed a backscatter module inside a shirt to monitor vital signs such as heart and breathing rates. The bitrates between the backscatter module and its receiver, i.e., smart phone, are set to 100 bps and 1.6 kbps. The experimental results show that at a bitrate of 1.6 kbps, the BER is roughly 0.02. However, at a low bitrate, i.e., 100 bps, the BER is less than 0.005. 

\subsubsection{Logistics}

ABCSs can also be adopted in logistics applications because of its low cost. In~\cite{Liu2013Ambient}, ABCS is implemented to remind when an item is out of place in a grocery store. Each item is equipped with a backscatter transmitter, and has a specific identification number. The backscatter transmitter then broadcasts its identification number in an interval of 5 seconds. Furthermore, all backscatter transmitters in the network periodically listen and store their neighbors' backscatter transmitters. In this way, a backscatter transmitter can indicate whether it is out of place or not by comparing its identification number with those of its neighbors. The experimental results show that the backscatter transmitter just needs less than 20 seconds to successfully detect its location, i.e., out of place or not.

\subsection{Discussion}

In this section, we have presented a common architecture of ABCSs as well as its limitations. The designs and solutions are reviewed as summarized in Table~\ref{table_sec6_sum}. Similar to BBCSs, in ABCSs, the existing works mainly focus on improving the communication range and bitrate. A few critical issues such as multiple accesses and security are less studied. A majority of the proposed designs are deployed with two backscatter transmitters communicating with each other. However, in reality, more than two backscatter transmitters may send data to the backscatter receiver simultaneously. Thus, multiple access is an important issue which needs to be further studied. As the backscatter transmitters are designed to to have low-complexity and low-cost, simple multiple access schemes such as TDMA and FDMA can be adopted. Random access schemes are yet to be adopted. Furthermore, the backscatter receiver can receive the signals coming from multiple backscatter transmitters by using different antennas to improve performance. Nevertheless, this may not be feasible in the real system due to higher cost, complexity, and size. Similar to BBCSs, ABCSs can suffer from security threats as the backscatter transmitters are simple devices. Thus, designing security schemes is a critical research direction for the ABCSs. Additionally, most of the existing works only adopt theoretical analysis and numerical simulations which may not be sufficient to prove practicability for real applications. Hence, there is a need for future research to conduct real experiments.

\section{Emerging Backscatter Systems}
\label{sec:Emerging}
Due to its attractive features, many new backscatter communication systems have been developed and introduced lately. In this section, we first review emerging RFID systems with tag-to-tag communication capability which is similar to bistatic systems. Then, the integration of backscatter communications with wireless-powered communication networks (WPCNs)~\cite{WPCNs} is discussed. Finally, RF-powered backscatter cognitive radio networks (CRNs), i.e., an integration of RF-powered CRNs and ABCSs are reviewed.

\subsection{Tag-to-tag Communication RFID Systems}

RFID systems are mainly used for tracking and identifying objects, e.g., in supply chain applications. However, RFID systems are only reliable at a range of few meters as the tags rely on RFID readers to backscatter their data. The studies have shown that even if a dense infrastructure of RFID readers is deployed, still 20-80\% of RFID tags may be located in blind spots~\cite{Loo2009Experimental}. This is stems from the fact that the communication between the tags and the reader can be adversely affected by interference or orientation misalignment~\cite{Loo2009Experimental}. Consequently, RFID readers need to be carefully deployed around the areas, e.g., in warehouses, to collect information from the tags. This is a major challenge for Amazon and Walmart nowadays~\cite{Ma2017Drone}. Therefore, novel RFID systems are proposed to deal with this problem.

\begin{figure}[htb]
	\centering
	\includegraphics[scale=0.22]{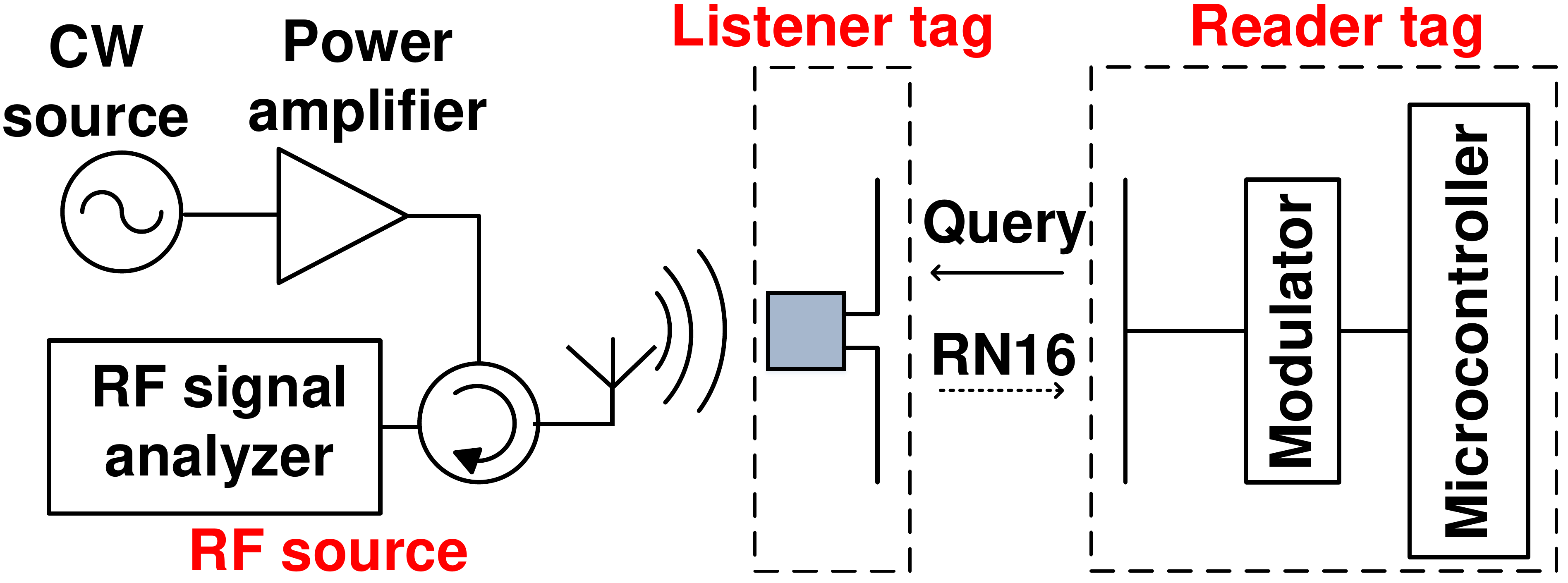}
	\caption{Block diagram of the proof-of-concept passive tag-to-tag communication system~\cite{Nikitin2012Passive}.}
	\label{fig:chap7_PassiveT2T}
\end{figure}

In~\cite{Nikitin2012Passive}, the authors propose a passive RFID system in which tags can communicate with each other directly. To do so, the tags backscatters signals from an RF source. If these signals are strong, the tags can be completely passive. Otherwise, the tags are equipped with batteries, i.e., semi-passive, but they still communicate with each other by backscattering and require no active RF transmitter. In~\cite{Nikitin2012Passive}, the authors then introduce a proof-of-concept passive tag-to-tag communication system as shown in Fig.~\ref{fig:chap7_PassiveT2T}. This system works in a \textit{master-slave} mode, which is compatible with existing Gen2 tags~\cite{RFIDGen2}. The master tag backscatters commands, i.e., queries, to the slave tags around it, and receives and decodes tag identification number and other data simultaneously. The slave tags are simply Gen2 tags which respond to the master tag's commands through their RN16 messages~\cite{RFIDGen2}. The RN16 is a 16-bit random number generated by the tag, and used for tag identification. The RF signal analyzer is deployed to ensure that the Gen2 tags correctly respond to the master tag. The experimental results demonstrate that the proposed tag-to-tag communication system is feasible. However, the authors note that the maximum reliable tag-to-tag communication distance, i.e., the distance between the master and a slave tag, is below 1 inch, since the communication is very sensitive to the positions of the tag.

Being inspired by~\cite{Nikitin2012Passive}, the authors in~\cite{Niu2014Cross} propose a cross-layer design to improve the performance of tag-to-tag communication systems. This approach consists of two protocols, i.e., a multiple access protocol in the data link layer and a routing protocol in the network layer. For the multiple access protocol, similar to Ethernet or 802.11, a carrier sense multiple access (CSMA) scheme is adopted, i.e., network allocation vector. However, this scheme requires timers to run precisely and consistently, and thus it needs more computational resources as well as memories at the tags. Therefore, the authors propose a \textit{Dual-ACK} virtual carrier-sensing method. The key idea is that the network allocation vector table is updated only when request-to-send, clear-to-send, and acknowledgment messages are detected. In addition, to deal with the \textit{hidden-node} problem, two acknowledgment messages are used to ensure that the states of the network allocation vector table are correct. As such, the tag does not need to rely on its timers, and the access to the transmission medium is reduced accordingly. For the routing protocol, the authors design an optimal link cost multi-path routing (OLCMR) protocol based on modulation depth, i.e., the ratio of the higher voltage level to the lower voltage level of the demodulated signals. Similar to the optimum link state routing (OLSR) protocol~\cite{OLSR}, OLCMR also constructs the routing table in which the destination address, next-hop address, the number of hops, and cost to that destination are included. However, unlike OLSR, the cost in OLCMR is computed by using the modulation depth. The bigger modulation depth is, the more energy the tags can harvest. The simulation results demonstrate that the proposed cross-layer design improves the performance of tag-to-tag communication networks in terms of end-to-end (E2E) delay, E2E cost, i.e., the modulation depth, and packet delivery ratio. In particular, the E2E delay is around 15 ms when the number of tags is 160 in the case without collision and around 70 ms with 140 tags in the case with a collision. Moreover, the delivery ratio significantly increases up to 98\% with 150 tags in the field. However, it is noted that there is a trade-off among the delivery ratio, E2E hops, and E2E cost. In particular, the higher delivery ratio requires more hops and incurs more cost.

\subsection{RF-Powered Cognitive Radio Networks and Backscatter Communication}

In RF-powered CRN~\cite{Lee2013Opportunistic}, a secondary transmitter (ST) can harvest energy from signals of a primary transmitter (PT) and uses the harvested energy to directly transmit data to the secondary receiver (SR) when the primary receiver is not transmitting or is sufficiently far away. This is known as harvest-then-transmit (HTT) method.
\begin{figure*}[tbh]
	\centering
	\includegraphics[scale=0.35]{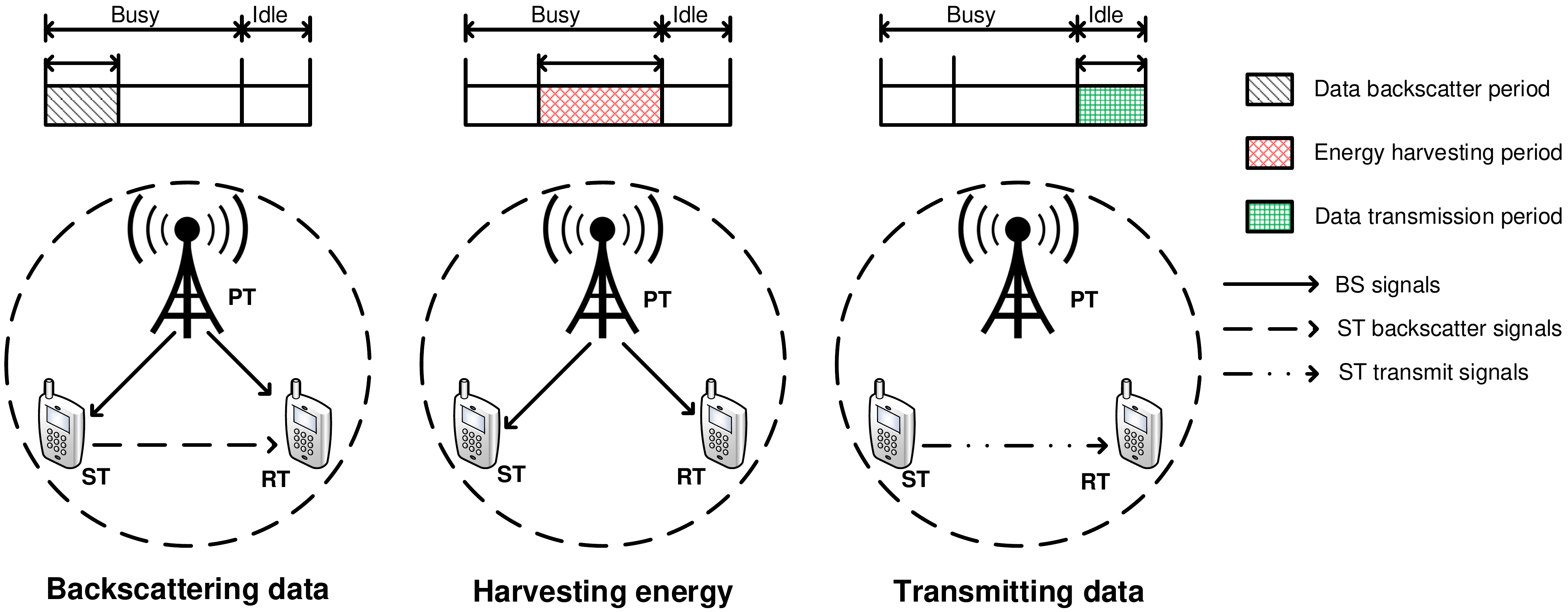}
	\caption{RF-powered cognitive radio network with ambient backscatter communication~\cite{Hoang2016Tradeoff}.}
	\label{fig:chap7_overlay}
\end{figure*}
In~\cite{Hoang2016Tradeoff}, the authors indicate that the performance of RF-powered CRNs depends greatly on the amount of harvested energy and the condition of the primary channels. For example, when the amount of harvested energy is too small and/or the channel idle probability is low, i.e., the PT frequently accesses the channel, the total transmitted bits will be reduced. Thus, the authors propose a combination of RF-powered CRN and ambient backscatter communications system, namely RF-powered backscatter CRN, which allows the ST not only to harvest energy from primary signals, but also to transmit data to the SR by backscattering primary signals. Note that the backscatter communications and the energy harvesting cannot efficiently be performed at the same time. The reason is that the amount of harvested energy will be significantly reduced if the ST backscatters data, and thus the harvested energy is not enough for the active transmission. The authors define three different subperiods corresponding to three activities, i.e., backscattering data, harvesting energy, and transmitting data as shown in Fig.~\ref{fig:chap7_overlay}. In particular, when the PT transmits data, i.e., the channel is busy, the ST can transmit data by using backscatter communication or harvest energy from the RF signals. Otherwise, when the channel is idle, if there is enough energy, the ST directly transmits data to the SR. Therefore, there is a trade-off between backscatter and HTT time to achieve optimal network throughput in which an optimization problem is formulated. It is proved that the network throughput is a convex function, and thus there always exists the globally optimal network throughput. The numerical results show that the solution of the proposed optimization problem can achieve significantly better performance than that of using either backscatter communications or HTT protocol alone.

In~\cite{Hoang2017Overlay}, the authors consider the case in which the SR charges a price/fee to the ST if the ST backscatters data to the SR. The Stackelberg game model for the RF-powered backscatter CRNs is introduced. In the first stage of the game, the SR, i.e., the leader, offers a price, i.e., for the backscatter time, to the ST, i.e., the follower, such that the expected SR's profit is maximized. Then, in the second stage, given the offered price, the ST chooses its optimal backscatter time to maximize its utility. To find the Stackelberg solution, the authors adopt the backward induction. The simulation results demonstrate that the solution of the Stackelberg game can maximize the profit of the SR as well as the utility of the ST.

In both~\cite{Hoang2016Tradeoff} and~\cite{Hoang2017Overlay}, the authors just consider overlay CRNs in which the ST can harvest energy when the channel is busy and transmit data when the channel is idle. Instead, in~\cite{Hoang2017Ambient}, the authors extend the work in~\cite{Hoang2016Tradeoff} by considering both overlay and underlay CRNs. Different from an overlay CRN, in an underlay CRN, the primary channel is always busy. Thus, the transmit power of the ST needs to be controlled to avoid interference to the primary receiver (PR). The authors define a threshold value for the transmit power of the ST to ensure that the interference at the PR is acceptable. Moreover, to maximize the transmission rate, the authors determine an optimal trade-off among backscatter, energy harvesting, and transmitting times, under the transmit power constraint of the ST. The simulation results suggest that the proposed solution can provide a solution for RF-powered CRN nodes to choose the best mode to operate, thereby improving the performance of the system.

In~\cite{Hoang2017Optimaltime}, RF-powered CRNs with multiple STs are taken into account. Similar to~\cite{Hoang2016Tradeoff}, the authors formulate an optimization problem to find the trade-off between data backscatter time and energy harvesting time to maximize network throughput. The authors demonstrate that the objective function, i.e., the network throughput, is convex. Thus, there exists a globally optimal trade-off between data backscatter and energy harvesting time, and also time sharing among STs. 

\begin{figure}[htb]
	\centering
	\includegraphics[scale=0.23]{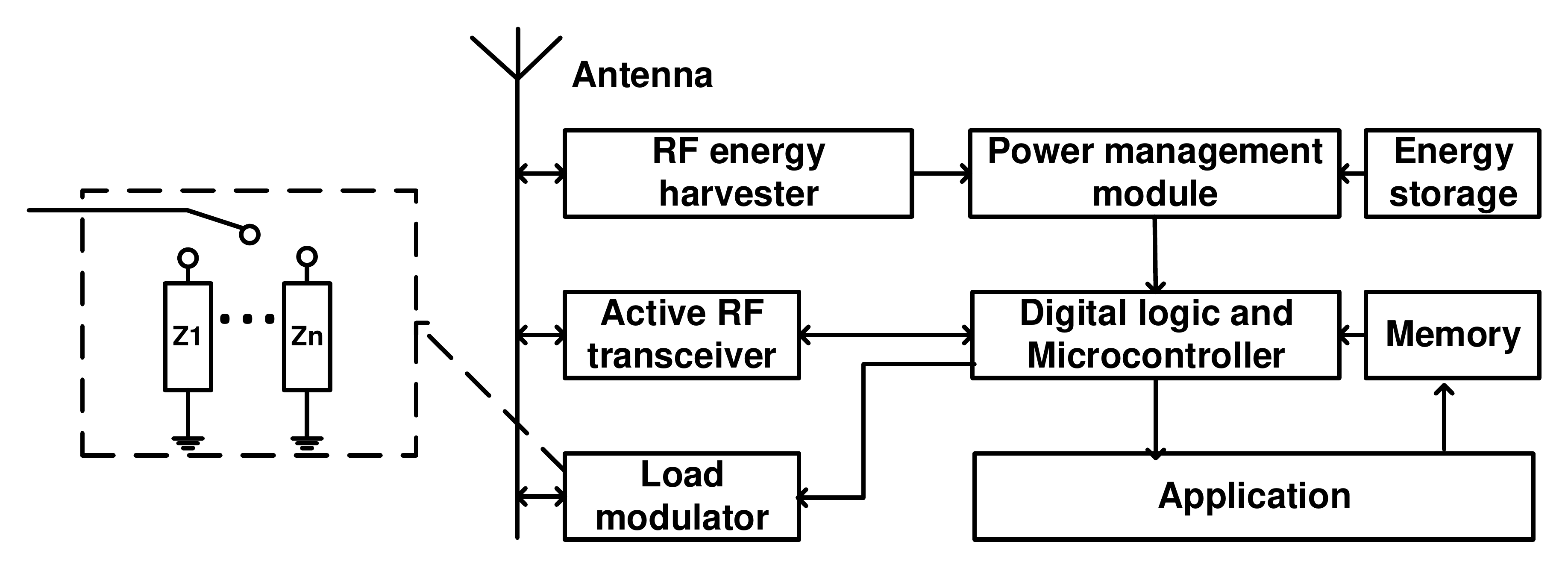}
	\caption{The structure of the hybrid transmitter~\cite{Lu2017Ambient}.}
	\label{fig:chap7_hybrid_transmitter}
\end{figure}

In~\cite{Lu2017Ambient}, the authors introduce a hybrid transmitter that integrates ambient backscatter with wireless-powered communication capability to improve transmission performance. The structure of this hybrid transmitter is shown in Fig.~\ref{fig:chap7_hybrid_transmitter}.
The transmitter consists of the following main components:
\begin{itemize}
	\item Antenna: shared by an RF energy harvester, a load modulator, and an active RF transceiver,
	\item RF energy harvester: to harvest energy from RF signals, 
	\item Load modulator: to modulate data for ambient backscatter communication, and 
	\item Active RF transceiver: to transmit or receive active RF signals for wireless-powered communication. 
\end{itemize}
In comparison with an ambient backscatter transmitter or a wireless-powered transmitter alone, this hybrid transmitter has many advantages such as supporting long duty-cycle and large transmission range. Additionally, the authors propose a multiple access scheme for the ambient backscatter-assisted WPCNs to maximize the sum of throughput of all STs in a CRN. The key idea is similar to the work in~\cite{Hoang2017Optimaltime}. Through the numerical results, the authors demonstrate the superiority of the proposed hybrid transmitter compared to traditional designs.

\subsection{Wireless-Powered Communication Networks and Backscatter Communication}

WPCNs allow devices to use energy from dedicated or ambient RF sources for their own data transmission. However, in the WPCNs, a wireless-powered transmitter may require a long time to acquire enough energy for active transmissions, and thus the performance of the system is significantly sub-optimal. Therefore, backscatter communication systems, i.e., BBCSs and ABCSs, are integrated with the WPCNs. The motivation is from the fact that the backscatter communications can transmit data by backscattering RF signals without requiring any external power source. 

\begin{figure}[htb]
	\centering
	\includegraphics[scale=0.25]{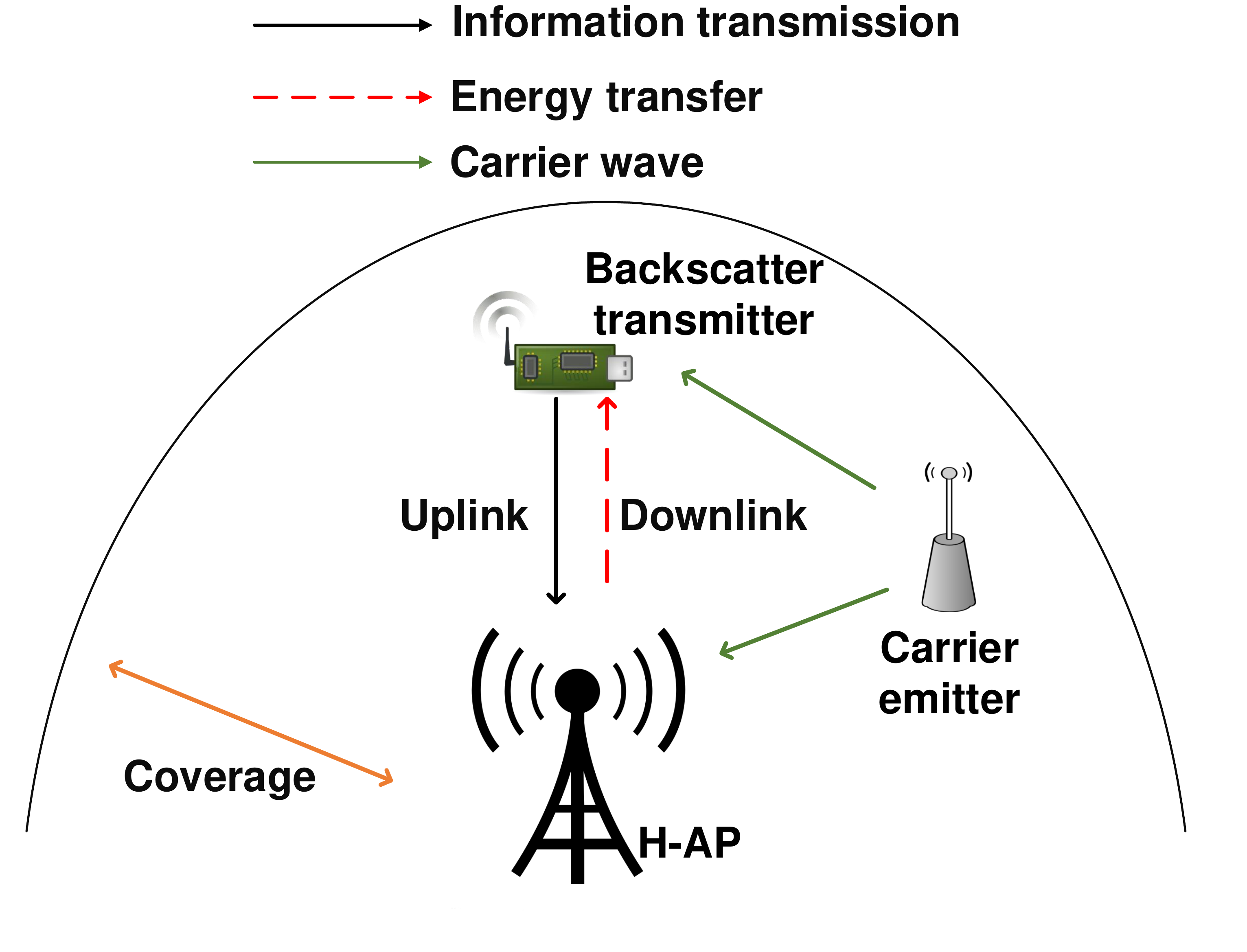}
	\caption{A backscatter radio based wireless-powered communication network~\cite{Choi2015Backscatter}.}
	\label{fig:chap7_hybrid_bistatic}
\end{figure}

In~\cite{Choi2015Backscatter}, the authors introduce an RF-powered bistatic backscatter design aiming to achieve a long-range coverage. The authors propose a solution combining backscatter radios and WPCNs. It is observed that the backscatter transmitters far away from the backscatter receiver can harvest less energy than that of the nearby backscatter transmitters. The authors propose a bistatic backscatter system which is composed of a carrier emitter and a hybrid access point (H-AP)~\cite{Ju2014HAP} as shown in Fig.~\ref{fig:chap7_hybrid_bistatic}. The H-AP not only broadcasts RF signals to the backscatter transmitters, but also receives backscattered signals. Therefore, the far backscatter transmitters can harvest energy from RF signals of both the carrier emitter and the backscatter receiver, i.e., the H-AP, to improve network performance. 

The authors also propose a two-phase transmission protocol for the H-AP. In the first phase, the H-AP uses the downlink to transfer wireless energy to the backscatter transmitters. The backscatter transmitters then reflect data by using FSK modulation on the uplink in the second phase. In contrast, the carrier emitter, which is deployed close to the backscatter transmitters, can always transmit RF signals. As a result, the far backscatter transmitters can derive sufficient energy for their operations. The results show that this network design can extend system coverage range up to 120 meters with 25 dBm and 13 dBm of transmit power at the H-AP and carrier emitter operating at 868 MHz, respectively.

\begin{figure}[htb]
	\centering
	\includegraphics[scale=0.1]{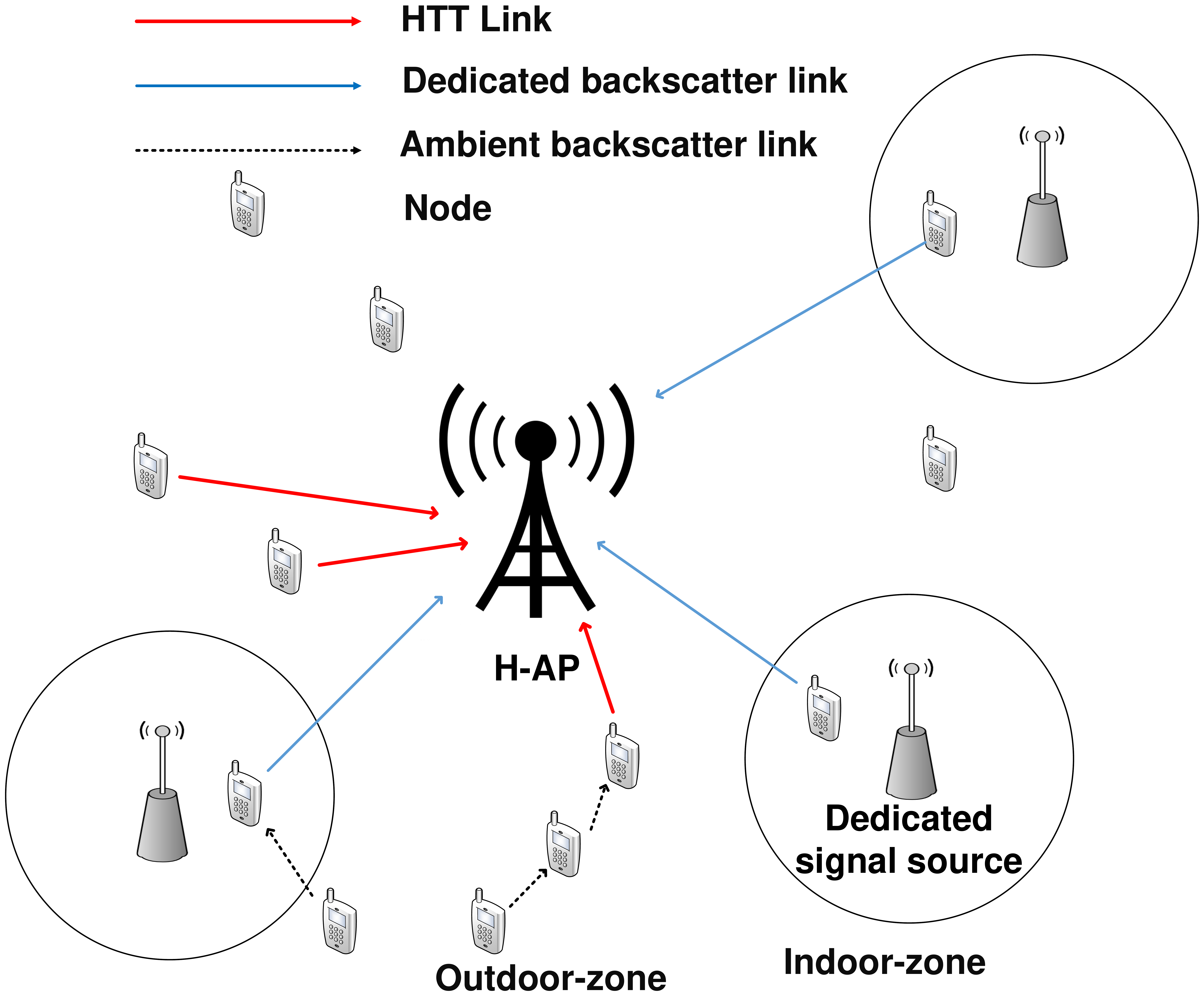}
	\caption{The wireless-powered heterogeneous network (WPHetNet) model with hybrid backscatter communication~\cite{Kim2016Hybrid},~\cite{Kim2017Hybrid}.}
	\label{fig:chap7_hybrid_ambient}
\end{figure}

In~\cite{Kim2016Hybrid}, the authors propose a hybrid backscatter communications for WPCNs to improve transmission range and bitrate. Different from~\cite{Choi2015Backscatter}, this system adopts dual mode operation of bistatic backscatter and ambient backscatter depending on indoor and outdoor zones, respectively. In particular, the proposed WPCN includes ambient RF sources, i.e., TV towers or high-power base stations, e.g., macrocells, and dedicated RF sources, leading to a wireless-powered heterogeneous network (WPHetNet) as shown in Fig.~\ref{fig:chap7_hybrid_ambient}.
If the ST is in the coverage of the carrier emitter, i.e., \textit{indoor-zone}, it can use both the ambient backscatter and bistatic backscatter, i.e., the dual mode operation. Otherwise, in \textit{outdoor-zone}, the ST can only adopt ambient backscatter. The authors note that the ST can flexibly select between \textit{HTT with bistatic backscatter protocol} and \textit{HTT with ambient backscatter protocol} based on its location, i.e., \textit{indoor-zone} and \textit{outdoor-zone}, and energy status. Similar to~\cite{Hoang2016Tradeoff}, the authors also define the harvesting time, backscatter time, and data transmission time to formulate an optimal time allocation problem. The objective is to maximize the throughput of the hybrid backscatter communications in \textit{indoor-zone}. However, in this work, the energy harvesting and backscatter communication processes can be performed at any time while the data transmission is only performed during the channel idle period to protect the PU's signals. The authors show that the optimal time allocation is a concave problem and can be solved by using KKT conditions. The numerical results demonstrate that the proposed hybrid communication can significantly increase the system throughput. In particular, with 25 W of transmit power at the H-AP and 23 dBm of transmit power at the dedicated carrier signals, the HTT with bistatic backscatter protocol and the HTT with ambient backscatter protocol can achieve throughput of up to 2.5 kbps and 115 kbps, respectively.

\begin{figure}[htb]
	\centering
	\includegraphics[scale=0.25]{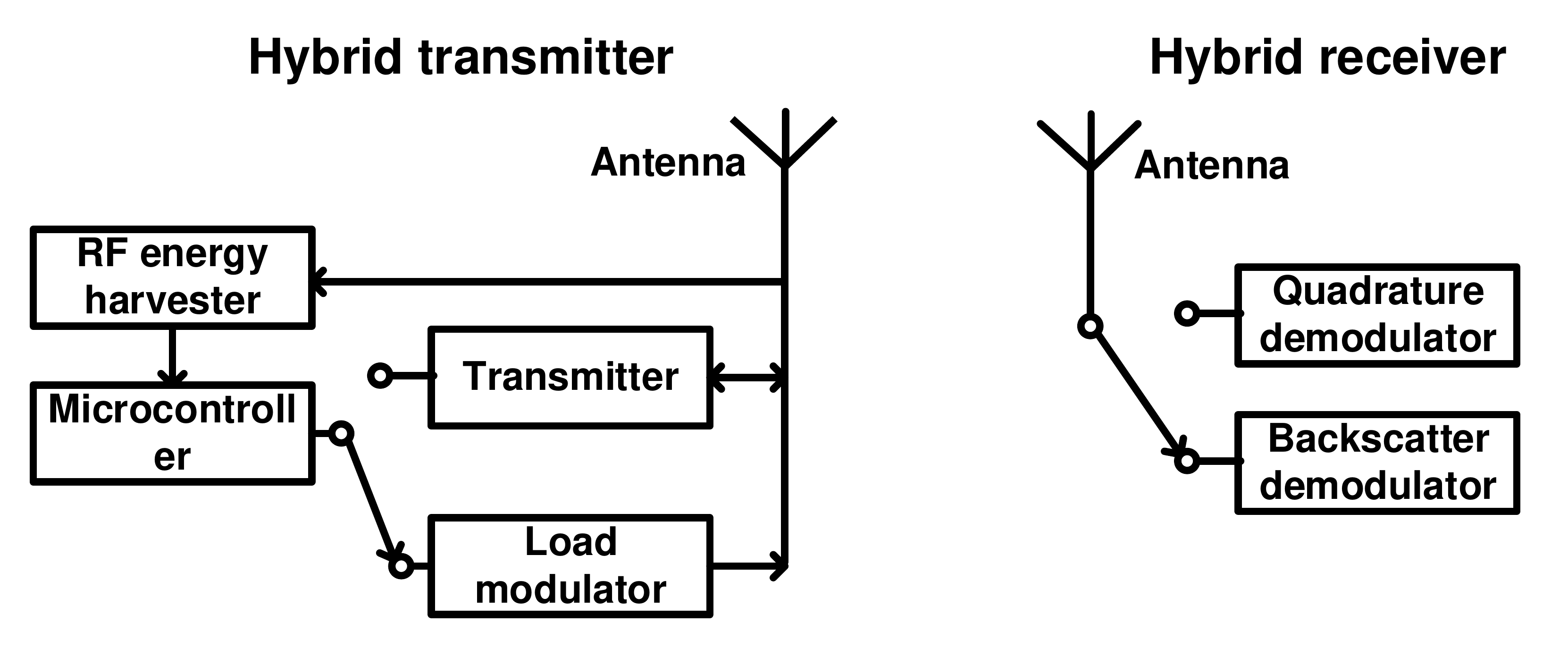}
	\caption{The structure of the hybrid transmitter and hybrid receiver~\cite{Lu2017Wireless}.}
	\label{fig:chap7_hybrid_trans_recei}
\end{figure}

In~\cite{Lu2017Wireless}, the authors propose hybrid D2D communications by integrating ambient backscatter and wireless-powered communications to improve the performance of the system. The authors then design a hybrid transmitter and a hybrid receiver as shown in Fig.~\ref{fig:chap7_hybrid_trans_recei}.
Similar to~\cite{Lu2017Ambient}, the authors introduce a hybrid receiver which can receive and decode data from both the modulated backscatter and active RF transmission. The structure of the hybrid receiver consists of two sub-blocks. The first block adopts a conventional quadrature demodulator, a phase shift module and a phase detector to decode the data from active RF transmission. The second block is a simple circuit composed of three main components, i.e., an envelope average circuit, a threshold calculator, and a comparator, to decode the modulated signals. By such, the hybrid receiver can decode both the ambient backscatter and wireless-powered transmission from the hybrid transmitter. As both ambient backscattering and wireless-powered transmission are based on ambient RF energy harvesting which requires no internal power source, the performance of the hybrid transmitter greatly depends on the environment factors, e.g., density of ambient transmitters and their spatial distribution. Therefore, the authors design a two-mode selection protocol for hybrid D2D communications, i.e., power threshold-based protocol and SNR threshold-based protocol. Under the power threshold-based protocol, the hybrid transmitter first detects the available energy harvesting rate. If this rate is lower than the power threshold which needs to power the active RF transmission, the ambient backscatter mode will be used. Otherwise, the HTT mode will be adopted. Under the SNR threshold-based protocol, the hybrid transmitter first tries to transmit data by backscattering. If the SNR of the backscattered signals at the receiver is lower than the threshold to decode information correctly, the transmitter will switch to the HTT mode. The authors then analyze the hybrid D2D communications in terms of energy outage probability, coverage probability, and average throughput. Through the stochastic geometry analysis, it is shown that the D2D communications benefit from larger geographical repulsion among energy sources, transmission load and density of ambient transmitters. Additionally, the power threshold-based protocol is more suitable for the scenarios with a high density of ambient transmitters and low interference level. On the contrary, the SNR threshold-based protocol is more suitable for the scenarios where the interference level and density of ambient transmitters are both low or both high.

\subsection{Backscatter Relay Networks}

Although many designs and solutions are introduced to improve the performance of backscatter networks, the single-hop communication range is still limited. One of the practical solutions recently proposed is to use relay nodes.
\begin{figure}[htb]
	\centering
	\includegraphics[scale=0.12]{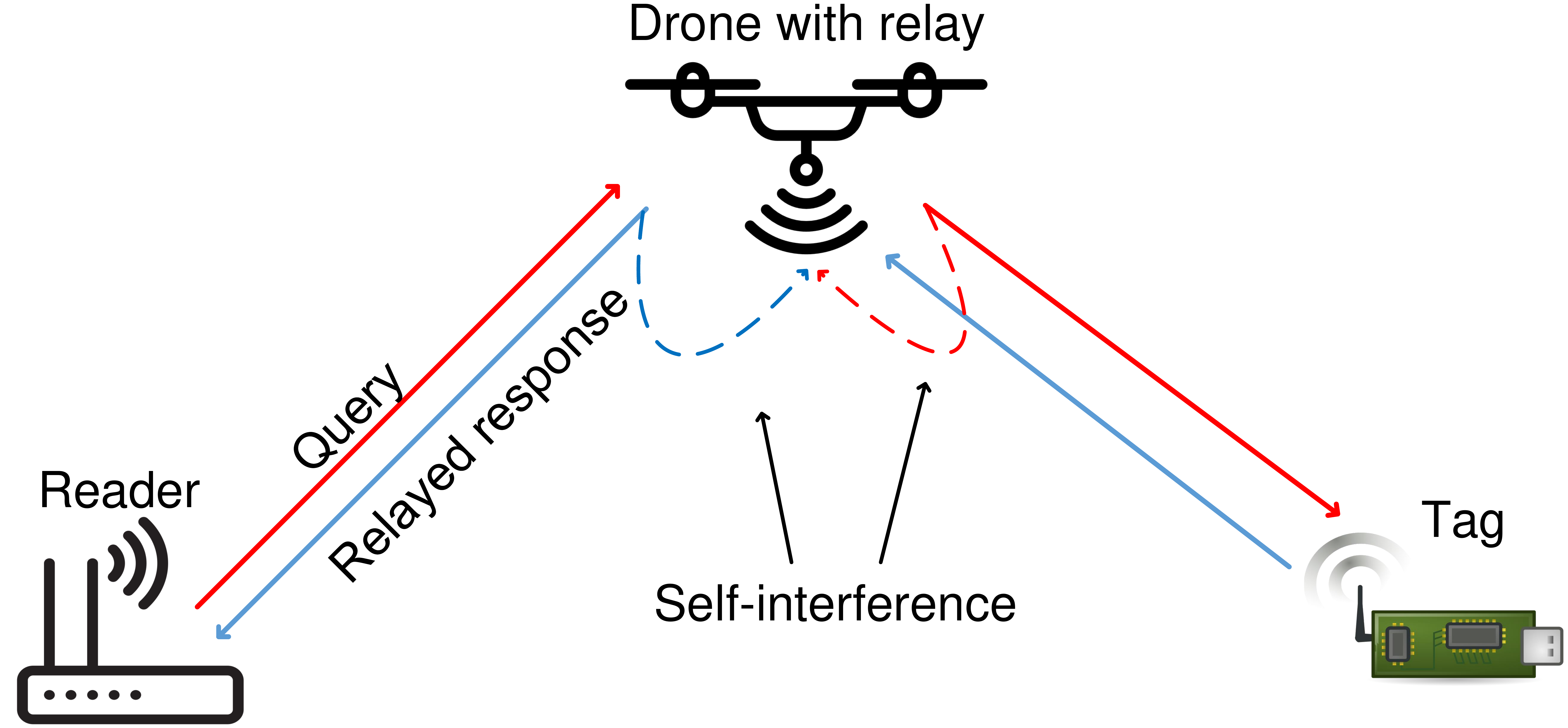}
	\caption{The system model of RFly~\cite{Ma2017Drone}.}
	\label{fig:chap7_RFly}
\end{figure}

In~\cite{Ma2017Drone}, the authors introduce an RFID system, named ``RFly'', that leverages drones as relays to extend the communication range as shown in Fig.~\ref{fig:chap7_RFly}. The key idea of RFly is that the drone is configured to collect queries from a reader, forward it to the tag, i.e., backscatter transmitter, and send the tag's reply back to the reader. However, the signals received at the drone's antennas may be affected by interference, i.e., inter-link self-interference and intra-link self-interference. The inter-link self-interference is from the uplink, i.e., from the tag to the reader, and the downlink, i.e., from the reader to the tag, operating at the same frequency. The intra-link self-interference is the leakage between the drone's receive and transmit antennas. To address the self-interference, the authors adopt a \textit{downconvert-upconvert} approach and a baseband filter. For the inter-link self-interference, RFly first \textit{downconverts} the received signals to baseband, low-pass filters for the downlink and bandpass filters for the uplink, and then \textit{upconverts} before sending. Through this filtering, RFly prevents the relay's self-interference from leaking into the uplink and downlink channels. For the intra-link self-interference, RFly leverages the downconvert-upconvert approach by using different frequencies in the upconvert stage. As such, the frequencies of reader-relay half-link and relay-transmitter half-link are different, and thus the intra-link self-interference is avoided. Through the experiments, the authors demonstrate that RFly can enable the communication between the reader and the tags at over 50 meters in LOS scenarios.

In~\cite{Munir2017Lowpower}, the authors introduce a relaying technique for full-duplex backscatter devices~\cite{Liu2014Enabling} to extend the communication ranges. Specifically, the authors assume a model in which a source backscatter transmitter, i.e., $ST$, wants to transmit data to a destination backscatter transmitter, i.e., $DT_s$, but the channel conditions between $ST$ and $DT_s$ are not feasible for the transmission. Hence, another backscatter transmitter, i.e., $RT$, which is located close to $ST$, is used as a relay between $ST$ and $DT_s$. However, the relay $RT$ may have data which needs to transmit to its own destination, i.e., $DT_r$. Therefore, the authors propose a protocol including two cases, i.e., $RT$ with and without data to transmit. For each case, the transmission time is divided into two phases. $ST$ transmits data to $RT$ in the first phase, and $RT$ transmits the received data to $DT_s$ in the second phase. If the relay $RT$ has its own data to transmit in the first phase, $RT$ receives data sent from $ST$ and transmits its data to $DT_r$ simultaneously. The authors then set up a simulation test to evaluate the performance of the relaying technique. A TV tower operating at 539 MHz with 10 kW of transmit power is used as an RF source. The $ST$-to-$RT$ and $RT$-to-$DT_s$ distances are 1 meter. The simulation results show that $ST$ can successfully send data to $DT_s$ through the support of $RT$. Additionally, the $ST$-to-$RT$ and $RT$-to-$DT_s$ bitrates can be up to 2 kbps and 1 kbps, respectively.

\subsection{Visible Light Backscatter Communications}
\emph{Visible light backscatter communications system (VLBCS)} is proposed to enable efficient data transmissions in RF limited environments, e.g., in hospitals or on planes. In general, the principles of VLBCSs are similar to backscatter RF systems. Specifically, the authors in~\cite{Li2015RetroVLC} design a backscatter transmitter, namely ViTag, that transmits data by using ambient visible light. ViTag first harvests energy from ambient light through solar cells to support its internal operations. Then, ViTag adopts a liquid crystal display (LCD) shutter to modulate, i.e., block or pass, the light carrier reflected by a retro-reflector. At its backscatter receiver, the modulated signals are amplified, demodulated, digitized, and finally decoded. In other words, ViTag can send its data to the backscatter receiver by backscattering visible light. The experimental results demonstrate that ViTag can achieve a downlink rate of 10 kbps and an uplink rate of 0.5 kbps over a distance of up to 2.4 meters.

However, as VLBCSs usually use single-carrier pulsed modulation scheme, i.e., OOK, their throughput is limited. Thus, in~\cite{Shao2017Pixelated}, the authors extend the idea in~\cite{Li2015RetroVLC} by using 8-pulse amplitude modulation (8-PAM) scheme to increase the throughput. The experimental results show that by using 8-PAM, a bitrate of 600 bps can be achieved at a distance of 2 meters compared to 200 bps when using OOK scheme. To further improve the bitrate of VLBCSs, the authors in~\cite{Xu2017PassiveVLC} propose a trend-based modulation scheme. In OOK modulation, a symbol is modulated once the LCD completely changes its on/off state, and thus the interval for modulation is not minimized, e.g., 4 ms with ViTag. The authors observe that as soon as the LCD changes its states, even if incompletely, the level of its transparency will change over short time, i.e., 1 ms. This time is long enough to produce a distinguishable decreasing trend on the backscatter receiver side. This means that 1 ms can be used as a minimum modulation interval in the trend-based modulation. As a result, the proposed modulation scheme can achieve a bitrate of up to 1 kbps and 4 times higher than that of the ViTag.

\subsection{Long-range LoRa Backscatter Communications}
Nowadays, as wireless applications are dynamically expanding their scale, there is a demand for wide area backscatter communications. Therefore, in~\cite{Talla2017LoRa}, the authors introduce a backscatter communication system enabling long-range transmissions, namely LoRa. Specifically, LoRa uses the chirp spread spectrum (CSS) modulation which represents a bit `0' as a continuous chirp that increases linearly with frequency while a bit `1' is a chirp that is cyclically shifted in time. The CSS modulation has several advantages for long-range communications such as achieving high sensitivity and resilient to fading, Doppler, and interference. However, the CSS modulation requires continuously changing the frequency as a function of time. Thus, the authors propose a hybrid digital-analog backscatter design which uses digital components to create a frequency plan for the continuously varying CSS signals and map it to analog components by using a low-power DAC. Furthermore, the authors introduce a backscatter harmonic cancellation mechanism to reduce the interference and improve the system performance. The key idea of this mechanism is adding voltage levels to approximate the sinusoidal signals and obtain a cleaner frequency spectrum. The experimental results show that LoRa can operate at the distance between the RF source and the backscatter receiver up to 475 meters. Additionally, the authors deploy LoRa in different scenarios, i.e., a 446 $m^2$ house spread across three floors, a 1210 $m^2$ office area covering 41 rooms, and a one-acre 4046 $m^2$ vegetable farm, and demonstrate that LoRa backscatter can achieve reliable coverage.

\section{Open Issues and Future Research Directions}
\label{sec:ChallengesandFuture}

In this section, we discuss a few open research problems that have not been fully studied in the literature and require further research attention.

\subsection{Heterogeneity of Ambient Signals}

Almost all the ambient backscatter transmitter designs are specific to a certain signal source, e.g., Wi-Fi or TV signals. However, in many cases, the signal from the specific source is not available. Therefore, UWB backscatter techniques are introduced~\cite{Yang2017UWB}. The techniques allow the backscatter transmitter to use a wide variety of ambient sources operating in the 80 MHz to 900 MHz range such as FM radios, digital TVs, and cellular networks. However, different signals from different sources have different characteristics and patterns. It is important to optimize the backscatter transmission strategies, e.g., using FM bands and digital TV jointly to optimize network throughput.

\subsection{Interference to Licensed Systems}

Many ABCSs rely on ambient signals from licensed sources. Therefore, backscattering can cause interference to licensed users. In~\cite{Liu2013Ambient}, through experiments, the authors show that if the ambient backscatter rates are less than 10 kbps, the TV receiver does not see any noticeable glitches for distances greater than 7.2 inches. However, in general, this may not be the case especially for the high bitrate ABCSs. As a result, modeling interference for ambient backscatter transmitters need to be done carefully. In particular, stochastic geometry models and spatial analysis can be applied to investigate and evaluate the interference to licensed systems.

\subsection{Standards and Network Protocols}

So far, testbeds and network protocols used in ABCSs have been developed for particular purposes and have proprietary features. For example, in~\cite{Liu2013Ambient} FM0 encoding is used to reduce energy consumption on backscatter devices, while in~\cite{Penichet2016Do} multi-bit encoding mechanism is adopted to increase the data rate for ambient backscatter communications. This makes backscatter devices less interoperable or even totally incompatible. Therefore, there is an urgent need for the development of communication standards and network protocols, e.g., packet format, network stack, and MAC protocol, for future ABCSs. 

\subsection{Security and Jamming Issues}

Due to the simple coding and modulation schemes adopted, backscatter communications are vulnerable to security attacks such as eavesdropping and jamming. The passive nature of backscatter communications making it challenging to secure backscatter secrecy. On one hand, any attacker that uses active RF transmitters can be more powerful to impair the modulated backscatter~\cite{Hong2017Jamming}. On the other hand, attacks on the signals sources, e.g., denial-of-service attack, can also jeopardize backscatter communications. Moreover, the resource constraints in backscatter transceivers make it impractical or even impossible to implement typical security solutions such as encryption and digital signature. Some existing research efforts mainly focus on physical-layer security approaches to protect secrecy. For example, references~\cite{Q.2016Yang,W.2014Saad} utilize artificial noise injection with the help of the reader to safeguard backscatter communications in RFID systems. However, this approach cannot be directly adopted in ABCSs as there are not dedicated readers. It is imperative to design simple, yet effective solutions to enable secure ambient backscatter communications.

\subsection{Millimeter-wave-based Ambient Backscatter}

Utilizing high-frequency millimeter waves (mmWave) for high speed communication has been deemed as one of the enabling technologies for the fifth-generation cellular networks. Due to different physical characteristics from UHF waves, mmWave requires LOS communication channels and miniaturized high-gain antennas and antenna arrays~\cite{S.2014Rangan}. The recent work in~\cite{J.2017Kimionis} demonstrates that the MBCSs working in mmWave bands can achieve a 4 Gigabit backscatter transmission rate with binary modulation. The ABCSs using mmWave are feasible to be developed.

\section{Conclusion}
\label{sec:Conclusion}

Ambient backscatter is a promising technology for today's large-scale self-sustainable wireless networks such as wireless sensor networks and Internet of Things. In this article, we have presented a comprehensive survey of ambient backscatter communications systems. We have first introduced the fundamentals of backscatter communications in different configurations. Common channel coding and modulation schemes are also reviewed. Then, we have provided literature reviews of bistatic backscatter communications systems regarding principles, advantages, limitations, and solutions to improve the system performance. Next, we have presented the general architecture as well as basics of ambient backscatter communications systems. Several state-of-the-art designs, solutions, and implementations in the literature are reviewed in detail. Furthermore, emerging applications in many areas of backscatter communications have been discussed. Finally, we have highlighted the practical challenges and future research directions.



\begin{thebibliography}{100}
\bibliographystyle{IEEEtranS}

\bibitem{StockmanModulated}
H.~Stockman, ``Communication by means of reflected power,'' in \emph{Proc. of I.R.E}, vol. 36, no. 10, Oct. 1948, pp. 1196-1204.
\bibitem{Vann2008}
G.~Vannucci, A.~Bletsas, and D.~Leigh, ``A software-defined radio system for backscatter sensor networks,'' \emph{IEEE Transactions on Wireless Communications}, vol. 7, no. 6, Jun. 2008, pp. 2170-2179.
\bibitem{Blet2008Anti}
A.~Bletsas, S.~Siachalou, and J.~N.~Sahalos, ``Anti-collision tags for backscatter sensor networks,'' in \emph{Proc. of 38th European Microwave Conference EuMC 2008}, Amsterdam, Netherlands, Oct. 2008, pp. 179-182.
\bibitem{Kimionis2013Bistatic}
J.~Kimionis, A.~Bletsas, and J.~N.~Sahalos, ``Bistatic backscatter radio for power-limited sensor networks,'' in \emph{Proc. of IEEE Global Communications Conference (GLOBECOM)}, Atlanta, GA, USA, Dec. 2013, pp. 353-358.
\bibitem{Blet2009Anti}
A.~Bletsas, S.~Siachalou, and J.~N.~Sahalos, ``Anti-collision backscatter sensor networks,'' \emph{IEEE Transactions on Wireless Communications}, vol. 8, no. 10, Oct. 2009, pp. 5018-5029.
\bibitem{Griffin2009Complete}
J.~D.~Griffin, and G.~D.~Durgin, ``Complete link budgets for backscatter-radio and RFID systems,'' \emph{IEEE Antennas and Propagation Magazine}, vol. 51, no. 2, Apr. 2009, pp. 11-25.
\bibitem{RFID2006survey}
A.~Juels, ``RFID security and privacy: A research survey,'' \emph{IEEE journal on selected areas in communications}, vol. 24, no. 2, Feb. 2006, pp. 381-394.
\bibitem{Lopez2006Survey}
P.~Peris-Lopez, J.~C.~Hernandez-Castro, J.~M.~Estevez-Tapiador, and A.~Ribagorda, ``RFID systems: A survey on security threats and proposed solutions,'' in \emph{Proc. of International Conference on Personal Wireless Communications}, Albacete, Spain, Sept. 2006, pp. 159-170.
\bibitem{Shih2006Survey}
D.~H.~Shih, P.~L.~Sun, D.~C.~Yen, and S.~M.~Huang, ``Taxonomy and survey of RFID anti-collision protocols,'' \emph{Computer communications}, vol. 29, no. 11, Jul. 2006, pp. 2150-2166.
\bibitem{Klair2010Survey}
D.~K.~Klair, K.~W.~Chin, and R.~Raad, ``A survey and tutorial of RFID anti-collision protocols,'' \emph{IEEE Communications Surveys \& Tutorials}, vol. 12, no. 3, Apr. 2010, pp. 400-421.
\bibitem{Zhang2012BLINK}
P.~Zhang, J.~Gummeson, and D.~Ganesan, ``BLINK: A high throughput link layer for backscatter communication,'' in \emph{Proc. of International conference on Mobile systems, applications, and services}, Low Wood Bay, Lake District, UK, Jun. 2012, pp. 99-112.
\bibitem{Liu2013Ambient}
V.~Liu, A.~Parks, V.~Talla, S.~Gollakota, D.~Wetherall, and J.~R.~Smith, ``Ambient backscatter: wireless communication out of thin air,'' in \emph{Proc. of ACM SIGCOMM 2013}, Hong Kong, China, Aug. 2013, pp. 39-50.
\bibitem{Kimionis2014Increased}
J.~Kimionis, A.~Bletsas, and J.~N.~Sahalos, ``Increased range bistatic scatter radio,'' \emph{IEEE Transactions on Communications}, vol. 62, no. 3, Mar. 2014, pp. 1091-1104.
\bibitem{Choi2015Backscatter}
S.~H.~Choi, and D.~I.~Kim, ``Backscatter radio communication for wireless powered communication networks,'' in \emph{Proc. of 21st Asia-Pacific Conference on Communications (APCC)}, Kyoto, Japan, Oct. 2015, pp. 370-374.
\bibitem{Lu2017Ambient}
X.~Lu, D.~Niyato, H.~Jiang, D.~I.~Kim, Y.~Xiao, and Z.~Han, ``Ambient Backscatter Networking: A Novel Paradigm to Assist Wireless Powered Communications'', {\em IEEE Wireless Communications}, to appear. 
\bibitem{Loo2008Chipimpedance}
C.~H.~Loo et al., ``Chip impedance matching for UHF RFID tag antenna design,'' \emph{Progress In Electromagnetics Research}, vol. 81, 2008, pp. 359-370.
\bibitem{Zhang2017Areview}
J.~Zhang, G.~Y.~Tian, A.~M.~Marindra, A.~I.~Sunny, and A.~B.~Zhao, ``A review of passive RFID tag antenna-based sensors and systems for structural health monitoring applications,'' \emph{Sensors}, vol. 17, no. 2, Feb. 2017, pp. 265.
\bibitem{Nikitin2008Antenna}
P.~V.~Nikitin, and K.~V.~S.~Rao, ``Antennas and propagation in UHF RFID systems,'' in \emph{Proc. of IEEE International Conference on RFID}, Las Vegas, NV, USA, May 2008, pp. 277-288.
\bibitem{ADC}
R.~Walden, ``Analog-to-digital converter survey and analysis,'' \emph{IEEE Journal on Selected Areas in Communications}, vol. 17, no. 4, Apr. 1999, pp. 539-550.
\bibitem{Rao2015Antenna}
K.~S.~Rao, P.~V.~Nikitin, and S.~F.~Lam, ``Antenna design for UHF RFID tags: A review and a practical application,'' \emph{IEEE Transactions on antennas and propagation}, vol. 53, no. 12, Dec. 2015, pp. 3870-3876.
\bibitem{RFID_fre_regulations}
RFID frequency ranges [Online]. Available: \url{http://www.centrenational-rfid.com/rfid-frequency-ranges-article-16-gb-ruid-202.html}
\bibitem{Lehpamer2012RFID}
H.~Lehpamer, \emph{RFID design principles,} Artech House, 2012.
\bibitem{Shirane2015RFpowered}
A.~Shirane, Y.~Fang, H.~Tan, T.~Ibe, H.~Ito, N.~Ishihara, and K.~Masu, ``RF-powered transceiver with an energy-and spectral-efficient IF-based quadrature backscattering transmitter,'' \emph{IEEE Journal of Solid-State Circuits}, vol. 50, no. 12, Aug. 2015, pp. 2975-2987.
\bibitem{Ensworth2015Every}
J.~F.~Ensworth, and M.~S.~Reynolds, ``Every smart phone is a backscatter reader: Modulated backscatter compatibility with bluetooth 4.0 low energy (ble) devices,'' in \emph{Proc. of 2015 IEEE International Conference on RFID}, San Diego, CA, USA, Apr. 2015, pp. 78-85.
\bibitem{Ensworth2015BLE}
J.~F.~Ensworth, M.~S.~Reynolds, ``BLE-Backscatter: Ultralow-Power IoT Nodes Compatible With Bluetooth 4.0 Low Energy (BLE) Smartphones and Tablets,'' \emph{IEEE Transactions on Microwave Theory and Techniques}, vol. 65, Sept. 2017, pp. 3360-3368.
\bibitem{Kim2003Measurments}
D.~Kim, M.~A.~Ingram, and W.~W.~Smith, ``Measurements of small-scale fading and path loss for long range RF tags,'' \emph{IEEE Transactions on Antennas and Propagation}, vol. 51, no. 8, Aug. 2003, pp. 1740-1749.
\bibitem{Kellogg2014WiFiIn}
B.~Kellogg, A.~Parks, S.~Gollakota, J.~R.~Smith, and D.~Wetherall, ``Wi-fi backscatter: Internet connectivity for rf-powered devices,'' in \emph{Proc. of 2014 ACM conference on SIGCOMM}, Chicago, Illinois, USA, Aug. 2014, pp. 607-618.
\bibitem{Griffin2009HighDiss}
J.~D.~Griffin, ``High-frequency modulated-backscatter communications using multiple antennas,'' Ph.D. dissertation, Georgia Institute Technology, Georgia, Mar. 2009.
\bibitem{Dardari2010Ultrawide}
D.~Dardari, R.~Derrico, C.~Roblin, A.~Sibille, and M.~Z.~Win, ``Ultrawide Bandwidth RFID: The Next Generation?,'' \emph{Proc. of the IEEE}, vol. 98, no. 9, Sept. 2010, pp. 1570-1582.
\bibitem{Guidi2010Backscatter}
F.~Guidi, D.~Dardari, C.~Roblin, and A.~Sibille, ``Backscatter communication using ultrawide bandwidth signals for RFID applications,'' \emph{The Internet of Things}, Springer New York, Jan. 2010, pp. 251-261.
\bibitem{UWB}
Federal Communications Commission, \emph{Revision of Part 15 of the Commissions Rules Regarding Ultra-Wideband Transmission Systems}, First Rep. Order (ET Docket 98-153), Adopted Feb. 14, 2002. Released Apr. 22, 2002.
\bibitem{Yeoman2014Impedance}
M.~S.~Yeoman, and M.~A.~O'Neill, ``Impedance Matching of Tag Antenna to Maximize RFID Read Ranges \& Design Optimization,'' \emph{2014 COMSOL Conference}, Cambridge, UK, Sept. 2014.
\bibitem{Grosinger2012Diss}
J.~Grosinger, ``Backscatter Radio Frequency Systems and Devices for Novel Wireless Sensing Applications,'' Ph.D. dissertation, Vienna University of Technology, Vienna, Austria, Aug. 2012.
\bibitem{AntennaDesign}
Antenna Design and RF Layout Guidelines [Online]. Available: \url{http://www.cypress.com/file/136236/download}
\bibitem{AntennaGain}
Antenna Gain [Online]. Available: \url{http://www.antenna-theory.com/basics/gain.php}
\bibitem{6Factors}
6 Factors that Affect RFID Read Range [Online]. Available: \url{https://blog.atlasrfidstore.com/improve-rfid-read-range}
\bibitem{AntennaTheory}
C.~A.~Balanis, \emph{Antenna Theory: Analysis and Design}, John Wiley \& Sons, Inc., 4th edition, 2016.
\bibitem{Wu2013circularpolarization}
T.~Wu, H.~Su, L.~Gan, H.~Chen, J.~Huang, and H.~Zhang, ``A compact and broadband microstrip stacked patch antenna with circular polarization for 2.45-GHz mobile RFID reader,'' \emph{IEEE Antennas and Wireless Propagation Letters}, vol. 12, May 2013, pp. 623-626.
\bibitem{Deavours2009circularpolarization}
D.~D.~Deavours, ``A circularly polarized planar antenna modified for passive UHF RFID,'' in \emph{Proc. of IEEE International Conference on RFID}, Orlando, FL, USA, May 2009, pp. 265-269.
\bibitem{Nikitin2006Reply}
P.~V.~Nikitin, and K.~V.~S.~Rao, ``Reply to" Comments on'Antenna Design for UHF RFID Tags: A Review and a Practical Application'",'' \emph{IEEE Transactions on Antennas and Propagation}, vol. 54, no. 6, Jun. 2006, pp. 1906-1907.
\bibitem{RFIDHandbook}
K.~Finkenzeller, \emph{RFID handbook: fundamentals and applications in contactless smart cards, radio frequency identification and near-field communication}, John Wiley \& Sons, 3rd edition, Jun. 2010.
\bibitem{RFIDFundamentals}
A.~Lozano-Nieto, \emph{RFID design fundamentals and applications}, CRC press, 2010.
\bibitem{Griffin2009Fundamentals}
J.~Griffin, ``The Fundamentals of Backscatter Radio and RFID Systems,'' \emph{Disney Research Pittsburgh}, Disney Research, Pittsburgh, Pennsylvania, United States, Jun. 2009.
\bibitem{Lalitha2014CodingReview}
V.~Lalitha, and S.~Kathiravan, ``A Review of Manchester, Miller, and FM0 Encoding Techniques,'' \emph{Smart Computing Review}, vol. 4, no. 6, Dec. 2014, pp. 481-490.
\bibitem{Kim2016Selfpowered}
Y.~H.~Kim, H.~S.~Ahn, C.~S.~Yoon, Y.~S.~Lim, and S.~O.~Lim, ``Self-powered and Backscatter-communicated IoT Device System for Zero-energy Wireless Communication in a Wi-Fi Environment,'' in \emph{Proc. of 6th International Conference on the Internet of Things}, Stuttgart, Germany, Nov. 2016, pp. 163-164.
\bibitem{Kim2017Implementation}
Y.~H.~Kim, H.~S.~Ahn, C.~S.~Yoon, Y.~S.~Lim, and S.~O.~Lim, and M.~H.~Yoon, ``Implementation of Bistatic Backscatter Wireless Communication System Using Ambient Wi-Fi Signals,'' \emph{KSII Transactions on Internet \& Information Systems}, vol. 11, no. 2, Feb. 2017, pp. 1250-1264.
\bibitem{Boyer2013Spacetime}
C.~Boyer, and S.~Roy, ``Space time coding for backscatter RFID,'' \emph{IEEE Transactions on Wireless Communications}, vol. 12, no. 5, Mar. 2013, pp. 2272-2280.
\bibitem{Durgin2017Improved}
G.~D.~Durgin, and B.~P.~Degnan, ``Improved Channel Coding for Next-Generation RFID,'' \emph{IEEE Journal of Radio Frequency Identification}, vol.1, no. 1, Sept. 2017, pp. 68-74.
\bibitem{Nikos2015Coherent}
N.~Fasarakis-Hilliard, P.~N.~Alevizos, and A.~Bletsas. ``Coherent detection and channel coding for bistatic scatter radio sensor networking,'' in \emph{Proc. of IEEE International Conference on Communications (ICC)}, London, UK, Jun. 2015, pp. 4895-4900.
\bibitem{cycliccode}
R.~J.~McEliece, \emph{The Theory of Information and Coding}, 2nd edition, Cambridge Univ. Press, 2001.
\bibitem{Park2014Turbocharging}
A.~N.~Parks, A.~Liu, S.~Gollakota, and J.~R.~Smith, ``Turbocharging ambient backscatter communication,'' in \emph{Proc. of 2014 ACM SIGCOMM}, Chicago, Illinois, USA , Aug. 2014, pp. 619-630.
\bibitem{Durgin2017Abetter}
G.~D.~Durgin, and B.~P.~Degnan, ``A better channel code than FM0 for next-generation RFID,'' \emph{IEEE International Conference on RFID}, Phoenix, AZ, USA, Jun. 2017, pp. 1-5.
\bibitem{ASK_advantges_disadvantages}
Advantages and disadvantages of ASK [Online]. Available \url{http://www.rfwireless-world.com/Terminology/Advantages-and-Disadvantages-of-ASK.html}
\bibitem{Molina2013BAT}
A.~M.~Markham, S.~S.~Clark, B.~Ransford and K.~Fu, ``BAT: Backscatter anything-to-tag communication,'' \emph{Wirelessly Powered Sensor Networks and Computational RFID}, Jan. 2013, pp. 131-142.
\bibitem{Kuester2012Baseband}
D.~G.~Kuester, D.~R.~Novotny, and J.~R.~Guerrieri, ``Baseband signals and power in load-modulated digital backscatter,'' \emph{IEEE Antennas and Wireless Propagation Letters}, vol. 11, Nov. 2012, pp. 1374-1377.
\bibitem{Kimionis2012Bistatic}
J.~Kimionis, A.~Bletsas, and J.~N.~Sahalos, ``Bistatic backscatter radio for tag read-range extension,'' in \emph{Proc. of IEEE International Conference on RFID-Technologies and Applications (RFID-TA)}, Nice, France, Nov. 2012, pp. 356-361.
\bibitem{Zhang2016HitchHike}
P.~Zhang, D.~Bharadia, K.~Joshi, and S.~Katti, ``HitchHike: Practical backscatter using commodity wifi,'' in \emph{Proc. of 14th ACM Conference on Embedded Network Sensor Systems CD-ROM }, Stanford, CA, USA, Nov. 2016, pp. 259-271.
\bibitem{EkhoNet2014Zhang}
P.~Zhang, P.~Hu, V.~Pasikanti, and D.~Ganesan, ``EkhoNet: high speed ultra low-power backscatter for next generation sensors,'' in \emph{Proc. of 20th annual international conference on Mobile computing and networking}, Maui, Hawaii, USA, Sept. 2014, pp. 557-568.
\bibitem{Tountas2015Bistatic}
K.~Tountas, P.~N.~Alevizos, A.~Tzedaki, and A.~Bletsas, ``Bistatic architecture provides extended coverage and system reliability in scatter sensor networks,'' in \emph{Proc. of 2015 International EURASIP Workshop on RFID Technology (EURFID)}, Rosenheim, Germany, Oct. 2015, pp. 144-151.
\bibitem{Kampi2013Backscatter}
E.~Kampianakis, J.~Kimionis, K.~Tountas, C.~Konstantopoulos, E.~Koutroulis, and A.~Bletsas, ``Backscatter sensor network for extended ranges and low cost with frequency modulators: Application on wireless humidity sensing,'' \emph{2013 IEEE SENSORS}, Baltimore, MD, USA, Nov. 2013, pp. 1-4.
\bibitem{Vannucci2007Implementing}
G.~Vannucci, A.~Bletsas, and D.~Leigh, ``Implementing backscatter radio for wireless sensor networks,'' in \emph{Proc. of IEEE 18th International Symposium on Personal, Indoor and Mobile Radio Communications}, Athens, Greece, Dec. 2007, pp. 1-5.
\bibitem{Alevizos2014Channel}
P.~N.~Alevizos, N.~F.~Hilliard, K.~Tountas, N.~Agadakos, N.~Kargas, and A.~Bletsas, ``Channel coding for increased range bistatic backscatter radio: Experimental results,'' in \emph{Proc. of IEEE RFID Technology and Applications Conference (RFID-TA)}, Tampere, Finland, Sept. 2014, pp. 38-43.
\bibitem{Varshney2016Loera}
A.~Varshney, O.~Harms, C.~P.~Penichet, C.~Rohner, F.~Hermans, T.~Voigt, ``LoRea: A Backscatter Reader for Everyone!,'' [Online]. Available: arXiv:1611.00096.
\bibitem{Vougiou2016Could}
G.~Vougioukas, S.~N.~Daskalakis, and A.~Bletsas, ``Could battery-less scatter radio tags achieve 270-meter range?,'' in \emph{Proc. of IEEE Wireless Power Transfer Conference (WPTC)}, Aveiro, Portugal, May 2016, pp. 1-3.
\bibitem{Alevizos2017Noncoherent}
P.~Alevizos, A.~Bletsas, and G.~N.~Karystinos, ``Noncoherent short packet detection and decoding for scatter radio sensor networking,'' \emph{IEEE Transactions on Communications}, vol. 65, no. 5, May 2017, pp. 2128 - 2140.
\bibitem{Alevizos2015Noncoherent}
P.~N.~Alevizos, and A.~Bletsas, ``Noncoherent composite hypothesis testing receivers for extended range bistatic scatter radio WSNs,'' in \emph{Proc. of IEEE International Conference on Communications (ICC)}, London, UK, Jun. 2015, pp. 4448-4453.
\bibitem{Alevizos2017Scatter}
P.~N.~Alevizos, A.~Bletsas, ``Scatter Radio Receivers for Extended Range Environmental Sensing WSNs,'' [Online]. Available: arXiv:1703.07139.
\bibitem{Liu2017Backscatter}
W.~Liu, K.~Huang, X.~Zhou, and S.~Durrani, ``Next Generation Backscatter Communication: Theory and Applications,'' [Online]. Available: arXiv:1701.07588.
\bibitem{Wang2017FMBackscatter}
A.~Wang, V.~Iyer, V.~Talla, J.~R.~Smith, and S.~Gollakota, ``FM Backscatter: Enabling Connected Cities and Smart Fabrics,'' in \emph{Proc. of 14th USENIX Symposium on Networked Systems Design and Implementation (NSDI 17)}, 2017, pp. 243-258.
\bibitem{Griffin2008Gains}
J.~D.~Griffin, and G.~D.~Durgin, ``Gains for RF tags using multiple antennas,'' \emph{IEEE Transactions on Antennas and Propagation}, vol. 56, no. 2, Feb. 2008, pp. 563-570.
\bibitem{Darsena2017Modeling}
D.~Darsena, G.~Gelli, and F.~Verde, ``Modeling and Performance Analysis of Wireless Networks With Ambient Backscatter Devices,'' \emph{IEEE Transactions on Communications}, vol. 65, no. 4, Apr. 2017, pp. 1797-1814.
\bibitem{Iyer2016Inter}
V.~Iyer, V.~Talla, B.~Kellogg, S.~Gollakota, and J.~Smith, ``Inter-technology backscatter: Towards internet connectivity for implanted devices,'' in \emph{Proc. of 2016 Conference on ACM SIGCOMM}, Florianopolis, Brazil, Aug. 2016, pp. 356-369.
\bibitem{Bharadia2015BackFi}
D.~Bharadia, K.~R.~Joshi, M.~Kotaru, and S.~Katti, ``BackFi: High throughput wifi backscatter,'' in \emph{Proc. of ACM Conference on Special Interest Group on Data Communication}, London, United Kingdom, Aug. 2015, pp. 283-296.
\bibitem{Qian2016Noncoherent}
J.~Qian, F.~Gao, G.~Wang, S.~Jin, and H.~B.~Zhu, ``Noncoherent Detections for Ambient Backscatter System,'' \emph{IEEE Transactions on Wireless Communications}, vol. 6, no. 3, Apr. 2016, pp. 1412-1422.
\bibitem{Kim2017Optimum}
T.~Y.~Kim and D.~I.~Kim,``Optimum MCS for High-Throughput Long-Range Ambient Backscatter Communication Networks,'' (invited) \emph{IEEE SPAWC 2017 Special Session on Signal Processing for Wireless Powered Communications}, Sapporo, Japan, July 2017.
\bibitem{Wang2012Efficient}
J.~Wang, H.~Hassanieh, D.~Katabi, and P.~Indyk, ``Efficient and reliable low-power backscatter networks,'' in \emph{Proc. of ACM SIGCOMM 2012 conference on Applications, Technologies, Architectures, and Protocols for computer communication}, Helsinki, Finland, Aug. 2012, pp. 61-72.
\bibitem{Boyer2012codedQAM}
C.~Boyer, and S.~Roy, ``Coded QAM backscatter modulation for RFID,'' \emph{IEEE Transactions on Communications}, vol. 60, no. 7, May 2012, pp. 1925-1934.
\bibitem{Correia2016Design}
R.~Correia, and N.~B.~Carvalho, ``Design of high order modulation backscatter wireless sensor for passive IoT solutions,'' in \emph{Proc. of IEEE Wireless Power Transfer Conference (WPTC)}, Aveiro, Portugal, Jun. 2016, pp. 1-3.
\bibitem{Lee2017Determination}
W.~S.~Lee, C.~H.~Kang, Y.~K.~Moon, and H.~K.~Song, ``Determination Scheme for Detection Thresholds Using Multiple Antennas in Wi-Fi Backscatter Systems,'' \emph{IEEE Access}, no. 99, Oct. 2017, pp. 22159-22165.
\bibitem{FSK_advantages_disadvantages}
Advantages and disadvantages of FSK [Online]. Available: \url{http://www.radio-electronics.com/info/rf-technology-design/fm-frequency-modulation/advantages-disadvantages.php}
\bibitem{Darsena2016Performance}
D.~Darsena, G.~Gelli, and F.~Verde, ``Performance analysis of ambient backscattering for green Internet of Things,'' in \emph{Proc. of 27th Annual International Symposium on Personal, Indoor, and Mobile Radio Communications (PIMRC)}, Valencia, Spain, Sept. 2016, pp. 1-6.
\bibitem{Shen2016Phase}
Z.~Shen, A.~Athalye, and P.~M.~Djuric, ``Phase Cancellation in Backscatter-Based Tag-to-Tag Communication Systems,'' \emph{IEEE Internet of Things Journal}, vol. 3, no. 6, Dec. 2016, pp. 959-970.
\bibitem{Boyer2014Backscatter}
C.~Boyer, and S.~Roy, ``Backscatter communication and rfid: Coding, energy, and mimo analysis,'' \emph{IEEE Transactions on Communications}, vol. 62, no. 3, Mar. 2014, pp. 770-785.
\bibitem{Noncoherent}
S.~Hussain, and S.~K.~Barton, ``Noncoherent detection of FSK signals in the presence of oscillator phase noise in an AWGN channel,'' in \emph{Proc. of European Conference on Mobile and Personal Communications}, Brighton, UK, Dec. 1993, pp. 95-98.
\bibitem{Nikos2015Trans}
N.~Fasarakis-Hilliard, P.~N.~Alevizos, and A.~Bletsas, ``Coherent detection and channel coding for bistatic scatter radio sensor networking,'' \emph{IEEE Transactions on Communications}, vol. 63, no. 5, May 2015, pp. 1798-1810.
\bibitem{Proakis2007DigitalcommunicationBook}
J.~G.~Proakis, \emph{Digital Communications}, New York: McGraw-Hill, 5th edition, 2007.
\bibitem{Wang2015Uplink}
G.~Wang, F.~Gao, Z.~Dou, and C.~Tellambura, ``Uplink detection and BER analysis for ambient backscatter communications systems,'' in \emph{Proc. of IEEE Global Communications Conference (GLOBECOM)}, San Diego, CA, USA, Dec. 2015, pp. 1-6.
\bibitem{Qian2017SemiCoherent}
J.~Qian, F.~Gao, G.~Wang, S.~Jin, and H.~Zhu, ``Semi-coherent Detection and Performance Analysis for Ambient Backscatter System,'' \emph{IEEE Transactions on Communications}, Aug. 2017.
\bibitem{Lu2015Signal}
K.~Lu, G.~Wang, F.~Qu, and Z.~Zhong, ``Signal detection and BER analysis for RF-powered devices utilizing ambient backscatter,'' in \emph{Proc. of 2015 International Conference on Wireless Communications \& Signal Processing (WCSP)}, Nanjing, China, Oct. 2015, pp. 1-5.
\bibitem{Ingram2001Transmite}
M.~A.~Ingram, M.~F.~Demirkol, and D.~Kim, ``Transmit diversity and spatial multiplexing for RF links using modulated backscatter,'' \emph{ISSSE 2001}, Tokyo, Japan, Jul. 2001.
\bibitem{Griffin2007Reduced}
J.~D.~Griffin, and G.~D.~Durgin, ``Reduced fading for RFID tags with multiple antennas,'' in \emph{Proc. of IEEE Antennas and Propagation Society International Symposium}, Honolulu, HI, USA, Jun. 2007, pp. 1201-1204.
\bibitem{Griffin2007Link}
J.~D.~Griffin, and G.~D.~Durgin, ``Link envelope correlation in the backscatter channel,'' \emph{IEEE Communications Letters}, vol. 11, no. 9, Sept. 2007, pp. 735-737.
\bibitem{Kim2001Small}
D.~Kim, M.~A.~Ingram, and W.~W.~Smith, ``Small-scale fading for an indoor wireless channel with modulated backscatter,'' in \emph{Proc. of 54th Vehicular Technology Conference}, Atlantic City, NJ, USA, USA, Oct. 2001, pp. 1616-1620.
\bibitem{Griffin2011Fading}
J.~D.~Griffin, and G.~D.~Durgin, ``Fading Statistics for Multi-Antenna RF Tags,'' in \emph{Handbook of Smart Antennas for RFID Systems}. New York: Wiley, 2010.
\bibitem{Griffin2010Multipath}
J.~D.~Griffin, and G.~D.~Durgin, ``Multipath fading measurements at 5.8 GHz for backscatter tags with multiple antennas,'' \emph{IEEE Transactions on Antennas and Propagation}, vol. 58, no. 11, Nov. 2010, pp. 3693-3700.
\bibitem{Griffin2009Multipath}
J.~D.~Griffin, and G.~D.~Durgin, ``Multipath fading measurements for multi-antenna backscatter RFID at 5.8 GHz,'' in \emph{Proc. of IEEE International Conference on RFID}, Orlando, FL, USA, Apr. 2009, pp. 322-329.
\bibitem{Kang2017Signaldetection}
C.~H.~Kang, W.~S.~Lee, Y.~H.~You, and H.~K.~Song, ``Signal Detection Scheme in Ambient Backscatter System With Multiple Antennas,'' \emph{IEEE Access}, vol. 5, Jul. 2017, pp. 14543-14547.
\bibitem{Smietanka2012Modeling}
G.~Smietanka, and J.~Gotze, ``Modeling and simulation of MISO diversity for UHF RFID communication,'' in \emph{Proc. of Federated Conference on Computer Science and Information Systems (FedCSIS)}, Wroclaw, Poland, Sept. 2012, pp. 813-820.
\bibitem{He2010Gains}
C.~He, and Z.~J.~Wang, ``Gains by a space-time-code based signaling scheme for multiple-antenna RFID tags,'' in \emph{Proc. of Canadian Conference on Electrical and Computer Engineering (CCECE)}, Calgary, AB, Canada, Sept. 2010, pp. 1-4.
\bibitem{Kamp2014Wireless}
E.~Kampianakis, J.~Kimionis, K.~Tountas, C.~Konstantopoulos, E.~Koutroulis, and A.~Bletsas, ``Wireless environmental sensor networking with analog scatter radio and timer principles,'' \emph{IEEE Sensors Journal}, vol. 14, no. 10, Oct. 2014, pp. 3365-3376.
\bibitem{Konstan2016Converting}
C.~Konstantopoulos, E.~Koutroulis, N.~Mitianoudis, and A.~Bletsas, ``Converting a Plant to a Battery and Wireless Sensor with Scatter Radio and Ultra-Low Cost,'' \emph{IEEE Transactions on Instrumentation and Measurement}, vol. 65, no. 2, Feb. 2016, pp. 388-398.
\bibitem{HMC190BMS8}
Analog Devices. \emph{HMC190BMS8.} [Online]. Available: \url{http://www.analog.com/media/en /technical-documentation/data-sheets/hmc190b.pdf/}
\bibitem{MSP430}
Texas Instruments. \emph{MSP430 FR5969.} [Online]. Available: \url{http://www.ti.com/product/MSP430FR5969}
\bibitem{SI1064Datasheet}
SI 1064 Data sheet [Online]. Available: \url{https://www.silabs.com/documents/public/data-sheets/Si106x-8x.pdf}
\bibitem{SpeedwayR420}
Speedway R420 Data Sheet [Online]. Available: \url{http://www.ptsmobile.com/impinj/impinj-speedway-datasheet.pdf}
\bibitem{N6841A}
N6841A RF Sensor [Online]. Available: \url{http://literature.cdn.keysight.com/litweb/pdf/5990-3839EN.pdf}
\bibitem{SensaphoneWSG30}
Sensaphone WSG30 Gateway [Online]. Available: \url{http://shared.sensaphone.com/pdfs/FGD-WSG30.pdf}
\bibitem{RFIDPrice}
How much does an RFID tag cost today? [Online]. Available: \url{https://www.rfidjournal.com/faq/show?85}
\bibitem{CC2420Data}
CC2420 Data Sheet [Online]. Available: \url{http://www.ti.com/lit/ds/symlink/cc2420.pdf}
\bibitem{CC2500Data}
CC2500 Data Sheet [Online]. Available: \url{http://www.ti.com/lit/ds/symlink/cc2500.pdf}
\bibitem{CC2420}
CC2420 [Online]. Available: \url{http://www.ti.com/product/CC2420/samplebuy}
\bibitem{CC2500}
CC2500 [Online]. Available: \url{http://www.ti.com/product/CC2500/samplebuy}
\bibitem{RFIDReaderCost}
RFID Reader [Online]. Available: \url{https://www.rfidjournal.com/faq/show?86}
\bibitem{Hoang2017Optimal}
D.~T.~Hoang, D.~Niyato, P.~Wang, D.~I.~Kim, and L.~B.~Le, ``Optimal Data Scheduling and Admission Control for Backscatter Sensor Networks,'' \emph{IEEE Transactions on Communications}, vol. 65, no. 5, May 2017, pp. 2062-2077.
\bibitem{Kampianakis2014Wireless}
E.~Kampianakis, J.~Kimionis, K.~Tountas, C.~Konstantopoulos, E.~Koutroulis, and A.~Bletsas, ``Wireless environmental sensor networking with analog scatter radio \& timer principles,'' \emph{IEEE Sensors Journal}, vol. 14, no. 10, Oct. 2014, pp. 3365-3376.
\bibitem{Schater2011Savitzky}
R.~W.~Schafer, ``What is a Savitzky-Golay filter? [Lecture Notes],'' \emph{EEE Signal Process. Mag.}, vol. 28, no. 4, Jul. 2011, pp. 111-117.
\bibitem{Forney2005BlockCode}
G.~D.~Forney Jr., ``Introduction to binary block codes,'' \emph{Principles of Digital Communication II}, MIT, 2005.
\bibitem{Daskalakis2016Soil}
S.~N.~Daskalakis, S.~D.~Assimonis, E.~Kampianakis, and A.~Bletsas, ``Soil Moisture Scatter Radio Networking With Low Power,'' \emph{IEEE Transactions on Microwave Theory and Techniques}, vol. 64, no. 7, Jul. 2016, pp. 2338-2346.
\bibitem{Daskalakis2014Soil}
S.~N.~Daskalakis, S.~D.~Assimonis, E.~Kampianakis, and A.~Bletsas, ``Soil moisture wireless sensing with analog scatter radio, low power, ultra- low cost and extended communication ranges,'' \emph{IEEE SENSORS}, Valencia, Spain, Nov. 2014, pp. 122-125.
\bibitem{Smith1997dutycycle}
S.~W.~Smith, \emph{The Scientist and Engineer's Guide to Digital Signal Processing}, California Tech. Publishing, San Diego, CA, USA, 1997.
\bibitem{Gollakota2013Emergence}
S.~Gollakota, M.~S.~Reynolds, J.~R.~Smith, and D.~J.~Wetherall, ``The emergence of RF-powered computing,'' \emph{Computer}, vol. 47, no. 1, Nov. 2013, pp. 32-39.
\bibitem{Penichet2016PhD}
C.~P.~Penichet, ``Ph. D. Forum Abstract: Ambient Backscatter Communication,'' in \emph{Proc. of 15th ACM/IEEE International Conference on Information Processing in Sensor Networks (IPSN)}, Vienna, Austria, Apr. 2016, pp. 1-2.
\bibitem{Xiao2015RFHarvest}
X.~Lu, P.~Wang, D.~Niyato, D.~I.~Kim, and Z.~Han, ``Wireless networks with RF energy harvesting: A contemporary survey,'' \emph{IEEE Communications Surveys \& Tutorials}, vol. 17, no. 2, Second Quarter 2015, pp. 757-789.
\bibitem{ADG902}
Analog Devices. (2005). \emph{ADG902 RF Switch, Product Manual.} [Online]. Available: \url{http://www.analog.com/media/en/technical-documentation/data-sheets/ADG901\_902.pdf/}
\bibitem{WISP}
A.~Sample, D.~Yeager, P.~Powledge, A.~Mamishev, and J.~Smith, ``Design of an RFID-based battery-free programmable sensing platform,'' \emph{IEEE Transactions on Instrumentation and Measurement}, vol. 57, no. 11, Nov. 2008, pp. 2608-2615.
\bibitem{Wang2016Ambient}
G.~Wang, F.~Gao, R.~Fan, and C.~Tellambura, ``Ambient backscatter communications systems: detection and performance analysis,'' \emph{IEEE Transactions on Communications}, vol. 64, no. 11, Nov. 2016, pp. 4836-4846.
\bibitem{Barott2014Coherent}
W.~C.~Barott, ``Coherent backscatter communications using ambient transmitters and passive radar processing,'' in \emph{Proc. of National Wireless Research Collaboration Symposium (NWRCS)}, Idaho Falls, ID, USA, May. 2014, pp. 15-20.
\bibitem{WienerHopffiltering}
J.~E.~Palmer and S.~J.~Searle, ``Evaluation of adaptive filter algorithms for clutter cancellation in passive bistatic radar,'' in \emph{Proc. of IEEE Radar Conference}, Atlanta, GA, USA, May 2012, pp. 493-498.
\bibitem{ECA}
F.~Colone, D.~W.~O'Hagan, P.~Lombardo, and C.~J.~Baker, ``A multistage processing algorithm for disturbance removal and target detection in passive bistatic radar,'' \emph{IEEE Transactions on Aerospace and Electronic Systems}, vol. 45, no. 2, Apr. 2009, pp. 698-722.
\bibitem{Daskalakis2017AmbientFM}
S.~N.~Daskalakis, J.~Kimionis, A.~Collado, G.~Goussetis, M.~M.~Tentzeris, and A.~Georgiadis, ``Ambient Backscatterers Using FM Broadcasting for Low Cost and Low Power Wireless Applications,'' \emph{IEEE Transactions on Microwave Theory and Techniques}, no. 99, Nov. 2017, pp. 1-12.
\bibitem{Palmer2013DVBT}
J.~E.~Palmer, H.~A.~Harms, S.~J.~Searle, and L.~Davis, ``DVB-T passive radar signal processing,'' \emph{IEEE transactions on Signal Processing}, vol. 61, no. 8, Apr. 2013, pp. 2116-2126.
\bibitem{MRC}
D.~Brennan, ``Linear diversity combining techniques,'' in \emph{Proc. of The IEEE}, vol. 91, no. 2, Feb. 2003, pp. 331-356.
\bibitem{Viterbi}
J.~Proakis, \emph{Digital Communications}, McGraw-Hill, 2001.
\bibitem{Zhang2016Enabling}
P.~Zhang, M.~Rostami, P.~Hu, and D.~Ganesan, ``Enabling Practical Backscatter Communication for On-body Sensors,'' in \emph{Proc. of ACM SIGCOMM}, Florianopolis, Brazil, Aug. 2016, pp. 370-383.
\bibitem{Penichet2016Do}
C.~P.~Penichet, A.~Varshney, F.~Hermans, C.~Rohner, and T.~Voigt, ``Do Multiple Bits per Symbol Increase the Throughput of Ambient Backscatter Communications?,'' in \emph{Proc. of 2016 International Conference on Embedded Wireless Systems and Networks}, Graz, Austria, Feb. 2016, pp. 355-360.
\bibitem{Liu2014Enabling}
V.~Liu, V.~Talla, and S.~Gollakota, ``Enabling instantaneous feedback with full-duplex backscatter,'' in \emph{Proc. of 20th annual international conference on Mobile computing and networking}, Maui, Hawaii, USA, Sept. 2014, pp. 67-78.
\bibitem{Liu2017FullDuplex}
W.~Liu, K.~Huang, X.~Zhou, and S.~Durrani, ``Full-Duplex Backscatter Interference Networks Based on Time-Hopping Spread Spectrum,'' \emph{IEEE Transactions on Wireless Communications}, vol. 16, no. 7, Jul. 2017, pp. 4361-4377.
\bibitem{Yang2016Backscatter}
G.~Yang, and Y.~C.~Liang, ``Backscatter communications over ambient OFDM signals: transceiver design and performance analysis,'' in \emph{Proc. of Global Communications Conference (GLOBECOM)}, Washington, DC, USA, Dec. 2016, pp. 1-6.
\bibitem{YangModulation}
G.~Yang, Y.~C.~Liang, R.~Zhang, and Y.~Pei, ``Modulation in the Air: Backscatter Communication over Ambient OFDM Carrier,'' \emph{IEEE Transactions on Communications}, no. 99, Nov. 2017.
\bibitem{Liu2017Coding}
Y.~Liu, G.~Wang, Z.~Dou, and Z.~Zhong, ``Coding and Detection Schemes for Ambient Backscatter Communication Systems,'' \emph{IEEE Access}, vol. 5, Mar. 2017, pp. 4947-4953.
\bibitem{Kellogg2016Passive}
B.~Kellogg, ``Passive Wi-Fi: Bringing Low Power to Wi-Fi Transmissions,'' in \emph{Proc. of 13th Usenix Conference on Networked Systems Design and Implementation}, Santa Clara, CA, Mar. 2016, pp. 151-164.
\bibitem{CMOS_TSMC}
65 nm LP CMOS node by TSMC [Online]. Available: \url{http://www.tsmc.com/english/dedicatedFoundry/technology/65nm.htm}
\bibitem{FPGA}
Altera de1 fpga development board [Online]. Available: \url{https://www.terasic.com.tw/cgi-bin/page/archive.pl?No=83}
\bibitem{Tektronix3252}
Tektronix 3252 arbitrary waveform generator [Online]. Available: \url{http://www.tek.com/datasheet/afg3000-series}
\bibitem{ADG919}
ADG919 RF Switch [Online]. Available: \url{http://www.analog.com/media/en/technical-documentation/data-sheets/ADG918_919.pdf}
\bibitem{TS881}
TS881 Comparator [Online]. Available: \url{http://www.st.com/en/amplifiers-and-comparators/ts881.html}
\bibitem{Zhou2017An}
X.~Zhou, G.~Wang, Y.~Wang, and J.~Cheng, ``An Approximate BER Analysis for Ambient Backscatter Communication Systems with Tag Selection'', \emph{IEEE Access}, vol. 5, Jul. 2017, pp. 22552-22558.
\bibitem{Liu2017MultipleAccess}
W.~Liu, Y.~C.~Liang, Y.~Li, and B.~Vucetic, ``Backscatter Multiplicative Multiple-Access Systems: Fundamental Limits and Practical Design,'' [Online]. Available: arXiv:1711.10694.
\bibitem{Huang2016Batteryfree}
Q.~Huang, Y.~Mei, W.~Wang, and Q.~Zhang, ``Battery-free sensing platform for wearable devices: The synergy between two feet,'' in \emph{Proc. of IEEE International Conference on Computer Communications}, San Francisco, CA, USA, Apr. 2016, pp. 1-9.

\bibitem{DeylePatent}
T.~Deyle, ``Streaming display data from a mobile device using backscatter communications,'' U.S. Patent 9,792,082, issued Oct. 2017.
\bibitem{WPCNs}
X.~Lu, P.~Wang, D.~Niyato, D.~I.~Kim, and Z.~Han, ``Wireless Charging Technologies: Fundamentals, Standards, and Network Applications,'' \emph{IEEE Communications Surveys and Tutorials}, vol. 18, no. 2, Second Quarter, 2016, pp. 1413-1452.
\bibitem{Loo2009Experimental}
C.~H.~Loo, A.~Z.~Elsherbeni, F.~Yang, and D.~Kajfez, ``Experimental and simulation investigation of RFID blind spots,'' \emph{Journal of Electromagnetic Waves and Applications}, vol. 23, no. 5-6, 2009, pp. 747-760.
\bibitem{Ma2017Drone}
Y.~Ma, N.~Selby, and F.~Adib, ``Drone Relays for Battery-Free Networks,'' in \emph{Proc. of ACM Special Interest Group on Data Communication}, Los Angeles, CA, USA, Aug. 2017, pp. 335-347.
\bibitem{Nikitin2012Passive}
P.~V.~Nikitin, S.~Ramamurthy, R.~Martinez, K.~V.~S.~Rao, ``Passive Tag-to-Tag Communication,'' in \emph{Proc. of IEEE International Conference on RFID}, Orlando, FL, USA, Apr. 2012, pp. 177-184.
\bibitem{RFIDGen2}
Radio-Frequency Identity Protocols Generation-2 UHF RFID [Online]. Available: \url{https://www.gs1.org/sites/default/files/docs/epc/uhfc1g2_2_0_0_standard_20131101.pdf}
\bibitem{Niu2014Cross}
H.~Niu, and S.~Jagannathan, ``A cross layer routing scheme for passive RFID tag-to-tag communication,'' in \emph{Proc. of IEEE Conference on Local Computer Networks}, Edmonton, AB, Canada, Sept. 2014, pp. 438-441.
\bibitem{OLSR}
P.~Jacquet, P.~Muhlethaler, T.~Clausen, A.~Laouiti, A.~Qayyum and L.~Viennot, ``Optimized link state routing protocol for ad hoc networks,'' in \emph{Proc. of IEEE INMIC}, Lahore, Pakistan, Pakistan, Dec. 2001, pp. 62-68.

\bibitem{Lee2013Opportunistic}
S.~Lee, R.~Zhang, and K.~Huang, ``Opportunistic wireless energy har- vesting in cognitive radio networks,'' \emph{IEEE Transactions on Wireless Communications}, vol. 12, no. 9, Sept. 2013, pp. 4788-4799.
\bibitem{Hoang2016Tradeoff}
D.~T.~Hoang, D.~Niyato, P.~Wang, D.~I.~Kim, and Z.~Han, ``The tradeoff analysis in RF-powered backscatter cognitive radio networks,'' in \emph{Proc. of Global Communications Conference (GLOBECOM)}, Washington, DC, USA, Dec. 2016, pp. 1-6.
\bibitem{Hoang2017Ambient}
D.~T.~Hoang, D.~Niyato, P.~Wang, D.~I.~Kim, and Z.~Han, ``Ambient backscatter: A new approach to improve network performance for RF-powered cognitive radio networks,'' \emph{IEEE Transactions on Communications}, vol. 65, no. 9, Jun. 2017, pp. 3659-3674.
\bibitem{Hoang2017Overlay}
D.~T.~Hoang, D.~Niyato, P.~Wang, D.~I.~Kim, and L.~B.~Long, ``Overlay RF-powered backscatter cognitive radio networks: A game theoretic approach,'' in \emph{Proc. of IEEE International Conference on Communications (ICC)}, Paris, France, May 2017, pp. 1-6.
\bibitem{Hoang2017Optimaltime}
D.~T.~Hoang, D.~Niyato, P.~Wang, D.~I.~Kim, ``Optimal time sharing in RF-powered backscatter cognitive radio networks,'' in \emph{Proc. of IEEE International Conference on Communications (ICC)}, Paris, France, May 2017, pp. 1-6.

\bibitem{Ju2014HAP}
H.~Ju and R.~Zhang, ``Throughput maximization in wireless powered communication networks,'' \emph{IEEE Trans. Wireless Commun.}, vol. 13, no. 1, Jan. 2014, pp. 418-428.
\bibitem{Kim2016Hybrid}
S.~H.~Kim, and D.~I.~Kim, ``Hybrid backscatter communications for wireless powered communication networks,'' in \emph{Proc. of 2016 International Symposium on Wireless Communication Systems (ISWCS)}, Poznan, Poland, Sept. 2016, pp. 265-269.
\bibitem{Kim2017Hybrid}
S.~H.~Kim, and D.~I.~Kim, ``Hybrid Backscatter Communication for Wireless-Powered Heterogeneous Networks,'' \emph{IEEE Transactions on Wireless Communications}, vol. 16, no. 10, Oct. 2017, pp. 6557-6570.
\bibitem{Lu2017Wireless}
X.~Lu, H.~Jiang, D.~Niyato, D.~I.~Kim, and Z.~Han, ``Analysis of Wireless-Powered Device-to-Device Communications with Ambient Backscattering'', {\em IEEE Transactions on Wireless Communications}, to appear. 
\bibitem{Munir2017Lowpower}
D.~Munir, S.~T.~Shah, W.~J.~Lee, and M.~Y.~Chung, ``Low-power backscatter relay network,'' \emph{11th International Conference on Ubiquitous Information Management and Communication}, Beppu, Japan, Jan. 2017.
\bibitem{Li2015RetroVLC}
J.~Li, A.~Liu, G.~Shen, L.~Li, C.~Sun, and F.~Zhao, ``Retro-VLC: Enabling battery-free duplex visible light communication for mobile and iot applications,'' in \emph{Proc. of International Workshop on Mobile Computing Systems and Applications}, Santa Fe, New Mexico, USA, Feb. 2015, pp. 21-26.
\bibitem{Shao2017Pixelated}
S.~Shao, A.~Khreishah, and H.~Elgala, ``Pixelated VLC-backscattering for self-charging indoor IoT devices,'' \emph{IEEE Photonics Technology Letters}, vol. 29, no. 2, Jan. 2017, pp. 177-180.
\bibitem{Xu2017PassiveVLC}
X.~Xu, Y.~Shen, J.~Yang, C.~Xu, G.~Shen, G.~Chen, and Y.~Ni, ``PassiveVLC: Enabling Practical Visible Light Backscatter Communication for Battery-free IoT Applications,'' in \emph{Proc. of 23rd Annual International Conference on Mobile Computing and Networking }, Snowbird, Utah, USA, Oct. 2017, pp. 180-192.
\bibitem{Talla2017LoRa}
V.~Talla, M.~Hessar, B.~Kellogg, A.~Najafi, J.~R.~Smith, and S.~Gollakota, ``LoRa backscatter: Enabling the vision of ubiquitous connectivity,'' \emph{Proc. of ACM on Interactive, Mobile, Wearable and Ubiquitous Technologies}, vol. 1, no. 3, Sept. 2017.
\bibitem{Yang2017UWB}
Yang C, Gummeson J, and Sample A, ``Riding the airways: Ultra-wideband ambient backscatter via commercial broadcast systems,'' in \emph{Proc. of IEEE INFOCOM}, May 2017, pp. 1-9.
\bibitem{Hong2017Jamming}
S.~G.~Hong, Y.~M.~Hwang, S.~Y.~Lee, D.~I.~Kim, Y.~Shin, and J.~Y.~Kim, ``Game-Theoretic Modeling of Backscatter Wireless Sensor Networks under Smart Interference,'' accepted for publication, IEEE Commun. Letters, Nov. 2017.

\bibitem{Q.2016Yang} 
Q. Yang, H. Wang, Y. Zhang, and Z. Han, ``Physical Layer Security in MIMO Backscatter Wireless Systems,"
\emph{IEEE Transactions on Wireless Communications}, 
vol. 15, no. 11, Nov. 2016, pp. 7547-7560.
\bibitem{W.2014Saad}
W. Saad, X. Zhou, Z. Han, and H. V. Poor, ``On the physical layer security of backscatter wireless systems," \emph{IEEE Trans. Wireless Commun.},
vol. 13, no. 6, Jun. 2014, pp. 3442-3451.
\bibitem{S.2014Rangan}
S. Rangan, T. S. Rappaport, and E. Erkip, ``Millimeter-wave cellular wireless networks: Potentials and challenges," \emph{Proc. of the IEEE}, vol. 102, no. 3, Mar. 2014, pp.366-385.
\bibitem{J.2017Kimionis}
J. Kimionis, A. Georgiadis, M. M. Tentzeris, ``Millimeter-wave backscatter: A quantum leap for gigabit communication, RF sensing, and wearables,"
in \emph{Proc. of IEEE MTT-S International Microwave Symposium (IMS)}, Honololu, HI, USA, Jun. 2017, pp. 812-815.

\end{thebibliography}
\end{document}